\documentclass{article}

\usepackage{graphicx}

\usepackage{amsmath,amsfonts,amssymb,amsthm,epsfig}
\usepackage{mathrsfs}
\usepackage{subfig}
\usepackage{fullpage}
\usepackage{enumerate}
\usepackage{times}
\usepackage{multirow}
\usepackage{soul}
\usepackage{hyperref}
\usepackage[normalem]{ulem}

\newtheorem{thm}{Theorem}[section]
\newtheorem{remark}[thm]{Remark}

\numberwithin{equation}{section}
\usepackage{amscd}
\usepackage{tikz}
\usepackage{xcolor}
\usepackage[english]{babel}
\usepackage[utf8]{inputenc}
\usepackage{csquotes}
\usepackage{graphicx}
\usepackage{slidesec}
\usepackage{url}
\usepackage{graphics}
\usepackage{amsbsy}
\usepackage{algorithm}
\usepackage{algorithmic}
\usepackage[url=true,giveninits=true,maxnames=99,hyperref]{biblatex}
\usepackage[font=footnotesize,labelfont=bf]{caption}

\DeclareMathOperator*{\argmax}{argmax}
\DeclareMathOperator*{\argmin}{argmin}

\providecommand{\keywords}[1]
{
  \small	
  \textbf{Keywords:} #1
}

\providecommand{\ams}[1]
{
  \small	
  \textbf{AMS subject classifications:} #1
}

\title{The split Gibbs sampler revisited: improvements to its algorithmic structure and augmented target distribution}
\author{Marcelo Pereyra$^{2,3}$ \\ \href{mailto:m.pereyra@hw.ac.uk}{\small \bf m.pereyra@hw.ac.uk} \and Luis A. Vargas-Mieles$^{1,2,3}$ \\ \href{mailto:lvargas@ed.ac.uk}{\small \bf lv375@cam.ac.uk} \and Konstantinos C. Zygalakis$^{1,3}$ \\ \href{mailto:k.zygalakis@ed.ac.uk}{\small \bf k.zygalakis@ed.ac.uk}}

\begin{document}

\maketitle

\begin{abstract}
Developing efficient Bayesian computation algorithms for imaging inverse problems is challenging due to the dimensionality involved and because Bayesian imaging models are often not smooth. Current state-of-the-art methods often address these difficulties by replacing the posterior density with a smooth approximation that is amenable to efficient exploration by using Langevin Markov chain Monte Carlo (MCMC) methods. Such methods rely on gradient or proximal operators to exploit geometric information about the target posterior density and scale efficiently to large problems. An alternative approach is based on data augmentation and relaxation, where auxiliary variables are introduced in order to construct an approximate augmented posterior distribution that is amenable to efficient exploration by Gibbs sampling. This paper proposes a new accelerated proximal MCMC method called latent space SK-ROCK (ls SK-ROCK), which tightly combines the benefits of the two aforementioned strategies. Additionally, instead of viewing the augmented posterior distribution as an approximation of the original model, we propose to consider it as a generalisation of this model. Following on from this, we empirically show that there is a range of values for the relaxation parameter for which the accuracy of the model improves, and propose a stochastic optimisation algorithm to automatically identify the optimal amount of relaxation for a given problem. In this regime, ls SK-ROCK converges faster than competing approaches from the state of the art, and also achieves better accuracy since the underlying augmented Bayesian model has a higher Bayesian evidence. The proposed methodology is demonstrated with a range of numerical experiments related to image deblurring and inpainting, as well as with comparisons with alternative approaches from the state of the art. An open-source implementation of the proposed MCMC methods is available from \url{https://github.com/luisvargasmieles/ls-MCMC}.
\end{abstract}

\keywords{Bayesian inference, inverse problems, image processing, Markov chain Monte Carlo methods, mathematical imaging, proximal algorithms, uncertainty quantification.}

\vspace{1mm}
\ams{62F15, 65C40, 65C60, 65J22, 68U10, 68W25}

\footnotetext[1]{School of Mathematics, University of Edinburgh, James Clerk Maxwell Building, Edinburgh, EH9 3FD, Scotland, UK.}
\footnotetext[2]{School of Mathematical and Computer Sciences, Heriot-Watt University, Edinburgh, EH14 4AS, Scotland, UK.}
\footnotetext[3]{Maxwell Institute for Mathematical Sciences, The Bayes Centre, 47 Potterrow, EH8 9BT, Edinburgh, Scotland, UK.}

\section{Introduction}
The problem of estimating an unknown image from noisy and/or incomplete data is central to imaging sciences \cite{chambolle_pock_2016, arridge_maass_oktem_schonlieb_2019}. Canonical examples include, for example, noise removal \cite{doi:10.1137/120874989}, image inpainting \cite{NIPS2012_6cdd60ea}, image deblurring \cite{5701777}, medical imaging \cite{medicalImaging01,8271999,6168272}, astronomical imaging \cite{10.1093/mnras/sty2004,10.1093/mnras/sty2015}. Estimation by direct inversion of the foward model relating the unknown image to the data is not usually possible, inasmuch the inverse problem is often severely ill-conditioned or ill-posed. The literature describes a range of mathematical frameworks to incorporate regularisation and formulate well-posed solutions (see, e.g., \cite{KaipioJariP2005SaCI, variationalMethodsInImaging,arridge_maass_oktem_schonlieb_2019}).

We consider Bayesian statistical solutions to imaging inverse problems \cite{KaipioJariP2005SaCI}. In this case, the solution is a probability distribution characterising our knowledge about the value of the unknown image of interest given the observed data. This distribution can then be used to derive image estimators, calibrate unknown model parameters, perform uncertainty quantification analyses, or Bayesian model selection (see, e.g., \cite{doi:10.1137/16M1108340,vidal2019maximum,doi:10.1137/19M1283719,https://doi.org/10.48550/arxiv.2106.03646}). 

There are three main strategies to perform Bayesian computation in imaging inverse problems. One strategy relies on optimisation methods, which typically scale efficiently to large models and offer detailed convergence guarantees, but can only support maximum-a-posteriori (MAP) estimation and a very limited range of other inferences \cite{chambolle_pock_2016,10.1093/mnras/sty2015,doi:10.1137/18M1173629}. Another strategy is to iteratively construct an approximation of the distribution of interest by fitting a tractable surrogate model. Prominent examples of this strategy include variational Bayesian inference \cite{doi:10.1080/01621459.2017.1285773,7314898,5256324} and expectation propagation \cite{10.5555/2074022.2074067,doi:10.1137/21M1427541}. This approach can be very powerful for some models, but it often requires a careful model-specific implementation, it has weaker guarantees, and in some cases it can exhibit local convergence issues resulting in poor inferences. The third approach, which we adopt in this paper, is to directly approximate the distribution of interest by using a Markov chain Monte Carlo (MCMC) sampling method \cite{RobertChristianP.2004MCsm}. This allows computing the expectations and probabilities of interest in a highly reliable way.  However, the application of MCMC methods comes at the expense of a higher computational cost. Reducing the cost of MCMC Bayesian computation in imaging has been a focus of significant efforts recently (see, e.g.,\cite{doi:10.1137/16M1108340,doi:10.1137/19M1283719,8625467}). Leaving computation strategies aside, it is possible to gain valuable insights about a Bayesian imaging model through mathematical analysis. A prominent example is Gaussian denoising under a total variation prior (see \cite{doi:10.1137/120902276,6952578}).

There are two main challenges in designing efficient MCMC algorithms for Bayesian imaging: the dimensionality of the problem and the fact that imaging models are often not smooth (this makes it difficult to directly apply gradient-based Markov Chain Monte Carlo methods). A first attempt to address these problems was the proposal of proximal MCMC methods \cite{PereyraMarcelo2016PMcM}, such as the so-called Moreau Yosida unadjusted Langevin algorithm (MYULA) \cite{doi:10.1137/16M1108340} that combines ideas from Langevin gradient-based sampling with ideas from the field of non-smooth  convex optimization. MYULA and its variants represent a significant improvement in computational efficiency and have good theoretical convergence guarantees. However, they are computationally inefficient for problems that are severely ill-conditioned because of step-size restrictions that leads to a slow exploration of the solution space. Recently, two different approaches were proposed in order to accelerate the convergence of proximal MCMC algorithms: the proximal stochastic Runge-Kutta-Chebyshev method (SK-ROCK) \cite{doi:10.1137/17M1145859,doi:10.1137/19M1283719}, which carefully combines $s$ gradient evaluations to achieve an $s^2$-increase in the step-size, and the Split Gibbs Sampler (SGS) \cite{8625467,vono2020efficient}, which is based on an augmentation and relaxation scheme that can significantly improve convergence speed at the expense of some estimation bias.

This paper explores two natural questions. First, how do SGS and SK-ROCK compare methodologically and empirically. Second, if the two methods can be combined in order to yield even more efficient MCMC methods. We address these questions in the following way:
\begin{enumerate}
\item Rather than viewing the model augmentation and relaxation strategy of \cite{doi:10.1080/10618600.2020.1811105,8625467,vono2020efficient} as an approximation, we propose to regard the augmented model as a generalisation of the original model. We show empirically that there is a range of relaxation values for which the accuracy of the model improves. In this regime, relaxation leads to better convergence properties and better accuracy. Beyond this regime, the accuracy of the relaxed model deteriorates rapidly. 
\item Given the critical role of the amount of relaxation, we build on  \cite{vidal2019maximum} to propose an empirical Bayesian method to automatically estimate the value of the relaxation parameter by maximum marginal likelihood estimation.
\item We formally identify a relationship between SGS and MYULA by re-expressing SGS as a discrete-time approximation of a Langevin stochastic differential equation (SDE) closely related to MYULA. 
\item Having connected SGS and MYULA at the level of the SDE, we propose two novel MCMC methods for Bayesian imaging: 1) an integration of SGS and MYULA that improves on both SGS and MYULA; and 2) an integration of SGS and SK-ROCK that outperform SK-ROCK, the previously fastest method in the literature.
\end{enumerate}

The remainder of the paper is organised as follows:  In  Section \ref{sec:problem_statement} we introduce the models considered in this work, and recall the state-of-the-art MCMC methods to sample from them. In Section \ref{sec:augModelBetterModel} we revisit the augmented model and show empirically that there is a subset of hyperparemeter values that enhances this model, while we also present a method to  automatically compute  an optimal choice of these hyperparameters. In Section \ref{sec:SGS_noisy_MYULA_new_MCMC} we establish a formal connection between MYULA and SGS which allows us to propose two novel and more efficient sampling algorithms for the augmented model. Section \ref{sec:experiments} illustrates the proposed methodologies with two experiments related to image deblurring and image inpainting, where we report detailed comparisons with state-of-the-art algorithms. Conclusions and perspectives for future work are reported in Section \ref{sec:conclusions}.

\section{Problem statement}
\label{sec:problem_statement}
\subsection{Bayesian inference and imaging inverse problems}
\label{subsec:BayInf_ImInvProb}
Let $x \in \mathbb{R}^d$ be the image we are interested in estimating and $y$ the available observation related to $x$ by a statistical model with likelihood function
\[
p(y|x) \propto \mathrm{e}^{-f_y(x)}.
\]
In this work, we pay special attention to problems where the estimation of $x$ given $y$ is ill-posed or ill-conditioned\footnote{Either the problem does not admit a unique solution that changes continuously with $y$, or there exists a unique solution but it is not stable w.r.t. small perturbations in $y$.}. We address this difficulty by considering the Bayesian framework, where we regularize the estimation problem by specifying a prior distribution, given for any $x$ and $\theta \in (0,+\infty)^{d^{\prime}}$ by
\[
p(x|\theta ) \propto \mathrm{e}^{-\theta^{\mathsf{T}} g(x) },
\]
for some vector of statistics $g: \mathbb{R}^d \rightarrow \mathbb{R}^{d^{\prime}}$, where $\theta$ parametrises this prior distribution and controls the level of imposed regularity. Using Bayes' theorem, we derive the posterior distribution
\begin{equation}
\label{eqn:posterior_dist}
p(x|y,\theta )=\frac{p(y|x)p(x|\theta)}{p(y|\theta )} = \frac{\exp \left[-f_{y}(x)-\theta^{\mathsf{T}} g(x)\right]}{\displaystyle \int_{\mathbb{R}^{d}} \exp \left[-f_{y}(x^{\prime})-\theta^{\mathsf{T}} g(x^{\prime})\right] \mathrm{d}x^{\prime}} \, ,
\end{equation}
which models the knowledge we have about $x$ given the observed data $y$.

We focus on Bayesian computational methodology for log-concave models of the form \eqref{eqn:posterior_dist}, where $f_{y}$ and $g$ satisfy the following conditions:

\begin{enumerate}
	\item $f_{y}: \mathbb{R}^{d} \rightarrow \mathbb{R}$ is convex and Lipschitz continuously differentiable with constant $L_f$
	\item $(g_i)_{i \in \{1,\ldots,d^{\prime}\}}: \mathbb{R}^{d} \rightarrow \mathbb{R}$ is proper, convex, and lower semi continuous, but potentially non-smooth.
\end{enumerate}
Models with these characteristics are widely adopted in the imaging community due to the variety of currently available Bayesian optimization tools that exploit the convexity properties of $p(x|y,\theta )$ \cite{doi:10.1137/18M1173629,doi:10.1137/16M1071249}, such as the computation of the MAP estimate, which can be formulated as a convex optimization problem \cite{doi:10.1137/18M1174076} given by
\begin{equation}
\label{eqn:MAP_estimate}
\hat{x}_{\mathrm{MAP}} = \argmax_x p(x|y,\theta) = \argmin_x f_y(x) + \theta^{\mathsf{T}} g(x) ,
\end{equation}
and can be solved by using state-of-the-art convex optimization algorithms \cite{chambolle_pock_2016}.  However, there are other more complicated Bayesian analyses beyond point estimation, such as model calibration, Bayesian model selection and hypothesis testing \cite{Robert2007}, which cannot be addressed by using optimization algorithms and typically require the calculation of probabilities and expectations w.r.t $p (x|y,\theta)$. This is challenging in imaging problems since it requires calculating intractable integrals on $\mathbb{R}^d$. In this case, the application of MCMC methods \cite{RobertChristianP.2004MCsm, AndrieuChristophe2003AItM} and, in particular, proximal MCMC methods \cite{PereyraMarcelo2016PMcM,doi:10.1137/16M1108340,doi:10.1137/19M1283719,8625467,vono2020efficient}, specialised for non-smooth log-concave distributions, is the preferred approach.

\subsection{Sampling via the Langevin diffusion}
We consider proximal MCMC methods derived from the discretization of the overdamped Langevin diffusion process, which we discuss below. Assume that we are interested in calculating probabilities w.r.t. a smooth distribution with density $\pi(x)$, and consider the stochastic differential equation (SDE)
\begin{equation}\label{eqn:SDE_Langevin}
\mathrm{d}X_t = \nabla \log \pi(X_t) \mathrm{d}t + \sqrt{2} \mathrm{d} W_t \, ,
\end{equation}
where $W_t$ is a $d$-dimensional Brownian motion. Under mild assumptions on $\pi(x)$, this SDE has a unique strong solution and admits $\pi(x)$ as its unique invariant density \cite[Theorem 2.1]{roberts1996}. However, in imaging applications it is usually not possible to solve (\ref{eqn:SDE_Langevin}) exactly, so a numerical approximation needs to be employed. Most works consider the Euler-Maruyama (EM) scheme given by
\begin{equation}\label{eqn:EM_Langevin}
X_{n+1} = X_n + \delta \nabla  \log \pi(X_n) + \sqrt{2 \delta} Z_{n+1},
\end{equation}
where $\delta >0$ is a given step-size and $(Z_{n+1})_{n \geq 0}$ is an i.i.d. sequence of $d$-dimensional standard Gaussian random vectors. This recursion is known as the unadjusted Langevin algorithm (ULA) and has been shown to be a highly efficient method for high-dimensional Bayesian inference when $\pi(x)$ is log-concave and smooth with $\nabla \log \pi (x)$ $L$-Lipschitz continuous and $\delta< 1/L$ \cite{durmus2017, durmus2019}. However, in many imaging models, $\pi(x)$ is not smooth and hence appropriate adjustments need to be made to ULA.

\subsubsection{Moreau-Yosida unadjusted Langevin algorithm (MYULA)}
Proximal MCMC methods \cite{PereyraMarcelo2016PMcM} deal with the non-differentiability of  $ \pi (x)$ by replacing $\pi(x)$ with a smooth approximation $\pi^{\lambda}(x)$ which, by construction, satisfies the required conditions of ULA. In this case $ p (x|y,\theta)$ in \eqref{eqn:posterior_dist} is replaced by  $ p^{\lambda} (x|y,\theta)$ defined as
\begin{equation}
\label{eqn:postDist_w_MY}
p^{\lambda} (x|y,\theta) = \frac{p(y|x)p^{\lambda}(x|\theta)}{p^{\lambda} (y|\theta)} = \frac{\exp [-f_{y}(x) - \theta^{\mathsf{T}} g^{\lambda}(x)]}{ \displaystyle \int_{\mathbb{R}^{d}} \exp [-f_{y}(x^{\prime}) - \theta^{\mathsf{T}} g^{\lambda}(x^{\prime})] \mathrm{d}x^{\prime}} , 
\end{equation}
where
\[
g^{\lambda}(x) = [g_1^{\lambda}(x),\ldots,g_{d^{\prime}}^{\lambda}(x)],
\]
that is, each non-smooth term $g_i(x)$, $i \in \{1,\ldots,d^{\prime}\}$ is replaced by its Moreau-Yosida envelope\footnote{If the calculation of the proximal operator of the sum of some elements of $g$ is possible, it is not necessary to replace each of these elements of the vector $g$ with its corresponding Moreau-Yosida envelope. In addition, if there is some $g_k(x)$, $k \in \{1,\ldots,d^{\prime}\}$ that is Lipschitz differentiable, its gradient can be computed directly.}, defined as
\begin{equation}\label{eqn:moreauYosida}
g_i^{\lambda}(x)=\min\limits_{u\in\mathbb{R}^{d}} \left\lbrace g_i(u) + \frac{1}{2\lambda} \| x-u \|^{2} \right\rbrace .
\end{equation}
This leads to a smooth posterior \eqref{eqn:postDist_w_MY} which has the following properties:

\begin{itemize}
\item $p^{\lambda} (x|y,\theta)$ is log-concave and Lipschitz continuously differentiable with gradient
\begin{equation*}
\begin{split}
\nabla \log p^{\lambda} (x|y,\theta) =& -\nabla f_{y}(x) - \nabla (\theta^{\mathsf{T}} g^{\lambda}(x)) \, ,\\
= & - \nabla f_{y}(x) - \frac{1}{\lambda} \sum_{i=1}^p \left( x - \text{prox}_{\theta_i g_i}^{\lambda}(x) \right),
\end{split}
\end{equation*}
with Lipschitz constant $L= L_f + p/\lambda$, and for every $x \in \mathbb{R}^d$
\begin{equation*}
\mathrm{prox}_{g_i}^{\lambda}(x) = \argmin_{u\in\mathbb{R}^{d}} \left\lbrace g_i(u) + \frac{1}{2\lambda} \| x-u \|^{2} \right\rbrace.
\end{equation*}
\item $p^{\lambda} (x|y,\theta)$ converges in total variation to $p (x|y,\theta)$\cite[Proposition 3.1]{doi:10.1137/16M1108340}, i.e., 
\begin{equation*}
\lim_{\lambda \rightarrow 0} \| p^{\lambda} (x|y,\theta) - p (x|y,\theta) \|_{\mathrm{TV}} = 0.
\end{equation*}
\end{itemize}
Applying the ULA scheme to the smooth posterior approximation $p^{\lambda} (x|y,\theta)$ leads to the recursion
\begin{equation*}
X_{n+1} = X_n - \delta \nabla f_{y} (X_n) - \frac{\delta}{\lambda} \sum_{i=1}^{d^{\prime}} \left(X_n-\mathrm{prox}_{\theta_i g_i}^{\lambda}(X_n)\right)  + \sqrt{2\delta} Z_{n+1} ,
\end{equation*}
which is known as the Moreau-Yosida unadjusted Langevin algorithm (MYULA) \cite{doi:10.1137/16M1108340}. The main benefit of the MYULA is that now since $ p^{\lambda} (x|y,\theta)$ is smooth and preserves log-concavity, the results from \cite{durmus2017, durmus2019} apply hence providing an efficient method for imaging applications. 

One of the main computational bottlenecks of ULA and MYULA is the fact that, in order to converge, one needs to choose $\delta \leq 1/L$ where $L$ is the Lipschitz constant of $\nabla \log p^{\lambda} (x|y,\theta)$, given by $L= L_f + d^{\prime}/\lambda$ (see \cite[Theorem 3.2]{doi:10.1137/16M1108340}). This step-size restriction is not problematic for moderate values of $L$, for example in denoising problems. However, in problems of the form $y = Ax + \xi$ with $\xi \sim \mathcal{N}(0,\sigma^2 \mathbb{I}_m)$ and $\sigma>0$ where the forward operator $A$ is very poorly conditioned, the MYULA sampler will have poor mixing properties, particularly on the subspace of $\mathbb{R}^d$ where $x$ has high uncertainty (this is related to the \emph{slow} components of the Markov chain). A similar situation occurs when the problem requires a high level of accuracy in the approximated prior term $g_i^{\lambda}(x)$ (i.e., $\lambda$ is very small), for example in problems where one has to enforce domain constraints on the solution space.

\subsubsection{SK-ROCK}
\label{subsec:SK_ROCK}
A natural way of overcoming the step-size limitation of the MYULA is to adopt a discretization scheme for the Langevin Diffusion (\ref{eqn:SDE_Langevin}) with better numerical stability properties. In particular, an explicit stochastic Runge-Kutta-Chebyshev discretization of the Langevin SDE was proposed in \cite{doi:10.1137/17M1145859} called SK-ROCK, and a proximal variant suitable for computational imaging problems was recently proposed in \cite{doi:10.1137/19M1283719}. This highly advanced Runge-Kutta stochastic integration scheme extends the deterministic Chebyshev method \cite{Abdulle2015} to sampling processes and has been shown to require fewer gradient evaluations than ULA or MYULA to reach a given level of accuracy. In particular, it has been shown in \cite[Proposition 3.1]{doi:10.1137/19M1283719} that, in order to be $\varepsilon$ close to a multivariate Gaussian target distribution (in the 2-Wasserstein distance), SK-ROCK requires $\mathcal{O}(\sqrt{\kappa})$ gradient evaluations (where $\kappa$ is the condition number of $\nabla \log \pi(x)$) instead of $\mathcal{O}(\kappa)$ required by the ULA, similar to accelerated optimization methods. 

This is possible due to the fact that SK-ROCK uses $s$ gradient evaluations per iteration at carefully chosen extrapolated points, which allows using a much larger step-size than the MYULA and thus having a faster decorrelation of the Markov chain in a stable manner. This is one of the great advantages of SK-ROCK in very ill-posed and ill-conditioned models, compared to MYULA, accelerating its convergence while maintaining its stability, thus improving the quality of the generated samples.

The SK-ROCK scheme is shown in Algorithm \ref{alg:SKROCK}, where $T_s$ denotes the Chebyshev polynomial of order $s$ of the first kind, defined recursively by $T_{k+1} = 2xT_k(x) - T_{k-1}(x)$ with $T_0(x) = 1$ and $T_1(x) = x$. The two main parameters of the algorithm are the number of stages $s \in \mathbb{N}^*$ and the step-size $\delta \in (0,\delta_s^{max}]$, where the range of admissible values for $\delta$ is controlled by $s$: for any $s \in \mathbb{N}^*$, the maximum allowed step-size is given by $\delta_s^{max}=l_s / ( L_f + 1/\lambda)$ with $l_s=[(s-0.5)^2 (2 - 4/3 \eta) -1.5]$ and $\eta = 0.05$ \cite{doi:10.1137/17M1145859}. Please see Section \ref{subsec:implementation_guides} for guidelines for setting $s$ and $\delta$.

\begin{algorithm}
\caption{SK-ROCK}
\label{alg:SKROCK}
\begin{algorithmic}[1]
\STATE{Input: $X_0 \in \mathbb{R}^d$, $\lambda > 0$, $n, s \in \mathbb{N}$, $\eta = 0.05$}.
\STATE{Compute $l_s=(s-0.5)^2 (2 - 4/3 \eta) -1.5$},
\STATE{Compute $\omega_0=1 + \eta / s^2$, $\omega_1=T_s(\omega_0) / T_s' (\omega_0)$},
\STATE{Compute $\mu_1=\omega_1 / \omega_0$, $\nu_1=s \omega_1 /2$, $k_1 = s \omega_1 / \omega_0$},
\STATE{Choose $\delta  \in (0,\delta_s^{\max} ]$, where $\delta_s^{\max}=l_s / ( L_f + 1/\lambda)$},
\FOR{$i = 0:n-1$}
\STATE{Set $\text{\~{X}}_0  =  X_i$},
\STATE{Sample $\xi_{i+1} \sim \mathcal{N}(0,2 \delta\mathbb{I}_d)$},
\STATE{Compute $\text{\~{X}}_1 = \text{\~{X}}_0 + \mu_1 \delta \nabla \log p^{\lambda}(\text{\~{X}}_0 +\nu_1 \xi_{i+1}|y,\theta) + k_1 \xi_{i+1} $},
\FOR{$j = 2:s$}
\STATE{Compute $\mu_j = 2\omega_1 T_{j-1}(\omega_0) / T_j(\omega_0)$, $\nu_j = 2\omega_0 T_{j-1}(\omega_0) / T_j(\omega_0)$, $k_j = 1-\nu_j$},
\STATE{Compute $\text{\~{X}}_j  = \mu_j \delta \nabla \log p^{\lambda}(\text{\~{X}}_{j-1}|,y,\theta) + \nu_j \text{\~{X}}_{j-1} + k_j \text{\~{X}}_{j-2}$},
\ENDFOR
\STATE{Set $X_{i+1} = \text{\~{X}}_s$},
\ENDFOR
\STATE{Output: Samples $X_1,\ldots,X_n$}.
\end{algorithmic}
\end{algorithm}
	
\subsection{Sampling via augmentation: Split Gibbs sampler (SGS)}
\label{subsec:samplingAugmModel}
A separate line of research seeks to address the limitation of ULA (and MYULA) by introducing an auxiliary variable $z \in \mathbb{R}^d$ and operating on the augmented state-space $(x,z)$. This allows to relax the original model \eqref{eqn:posterior_dist} and instead uses the following augmented posterior
\begin{equation}\label{eqn:postDist_exp_augm}
\begin{split}
p (x,z|y,\theta,\rho^2) & = \frac{p(y|x)p(x,z|\theta,\rho^2)}{p(y|\theta,\rho^2)} = \frac{p(y|x)p(x|z,\rho^2)p(z|\theta)}{\displaystyle \int_{\mathbb{R}^d}\int_{\mathbb{R}^d} p(y|x)p(x|z,\rho^2)p(z|\theta) \mathrm{d}x \mathrm{d}z} \\ & = \frac{\exp\left[-f_{y}(x)-\theta^{\mathsf{T}} g(z)-\frac{1}{2 \rho^2} \Vert x-z \Vert^2 \right]}{ \displaystyle \int_{\mathbb{R}^d}\int_{\mathbb{R}^d} \exp\left[-f_{y}(x)-\theta^{\mathsf{T}} g(z)-\frac{1}{2 \rho^2} \Vert x-z \Vert^2 \right] \mathrm{d}x \mathrm{d}z}  , \quad \rho^2 > 0,
\end{split}
\end{equation}
where
\begin{equation}
\label{eqn:likelihood_and_prior_g}
p(y|x) = \frac{\exp(-f_{y}(x))}{\displaystyle\int_{\mathbb{R}^d} \exp(-f_{y}(x))\mathrm{d}x}, \quad p(z|\theta) = \frac{\exp (- \theta^{\mathsf{T}} g(z))}{\displaystyle\int_{\mathbb{R}^d} \exp (- \theta^{\mathsf{T}} g(z^{\prime}))\mathrm{d}z^{\prime}},
\end{equation}
\begin{equation}
\label{eqn:augmented_prior_full}
p(x|z,\rho^2) = \frac{\exp ( - \Vert x-z \Vert^2 / 2 \rho^2 )}{\displaystyle\int_{\mathbb{R}^d} \exp ( -\Vert x-z \Vert^2 / 2 \rho^2 ) \mathrm{d}x } = \frac{\exp ( - \Vert x-z \Vert^2 / 2 \rho^2)}{(2\pi \rho^2)^{d/2}},
\end{equation}
where $\rho^2$ controls the correlation between the variable of interest $x$ and the auxiliary variable $z$, and $f_{y}$, $g$ are the same as in Section \ref{subsec:BayInf_ImInvProb}. If we now consider the marginal posterior distribution
\begin{equation}
\label{eqn:marginal_x_augm}
p(x|y,\theta,\rho^2) = \int_{\mathbb{R}^d} p (x,z|y,\theta,\rho^2) \, \mathrm{d}z ,
\end{equation}
it is possible to show that it converges in total variation to the original posterior $p (x|y,\theta)$ as $\rho^2 \rightarrow 0$.

This approach was first introduced several decades ago as a way of calculating maximum likelihood estimates from incomplete data \cite{https://doi.org/10.1111/j.2517-6161.1977.tb01600.x}, and as an efficient method for sampling from posterior distributions \cite{doi:10.1080/01621459.1987.10478458} (see \cite{doi:10.1198/10618600152418584} for a review of these techniques). In the current literature, this model was revisited by \cite{doi:10.1080/10618600.2020.1811105} in the context of consensus Monte Carlo in distributed settings and applied to imaging inverse problems in \cite{8625467}, where its similarities to  the algorithmic structure of the Alternating Direction Method of Multipliers (ADMM) optimization algorithm \cite{10.1561/2200000016}  were also discussed.

From a computational point of view, as in the case of MYULA, because $g$ is not differentiable one needs to approximate $p (x,z|y,\theta,\rho^2)$ by
\begin{equation} \label{eqn:MY_augmented}
p^{\lambda} (x,z|y,\theta,\rho^2) \propto \exp\left[-f_{y}(x)-\theta^{\mathsf{T}} g^{\lambda}(z)-\frac{1}{2 \rho^2} \Vert x-z \Vert^2 \right], \quad \rho^2 > 0.
\end{equation}
To sample \eqref{eqn:MY_augmented}, \cite{8625467,vono2020efficient} proposed a Gibbs-like splitting strategy scheme, applied on the following conditional distributions
\begin{align}
\label{eqn:condDist_SP_x} p(x|y,z,\rho^2) & \propto \exp \left[ -f_{y}(x) -\frac{1}{2 \rho^2} \Vert x-z \Vert^2 \right], \\
\label{eqn:condDist_SP_z} p^{\lambda}(z|x,\theta,\rho^2) & \propto \exp \left[ -\theta^{\mathsf{T}} g^{\lambda}(z) -\frac{1}{2 \rho^2} \Vert x-z \Vert^2 \right].
\end{align}
This method is known as the split Gibbs sampler (SGS). See Algorithm \ref{alg:SP}.
\begin{algorithm}
\caption{SGS}
\label{alg:SP}
\begin{algorithmic}[1]
\STATE{Input: $X_0, Z_0 \in \mathbb{R}^d$, $\lambda, \rho^2 > 0$, $n \in \mathbb{N}$}.
\FOR {$i = 0:n-1$}
\STATE{Sample $X_{i+1} \sim p\left( x|y,Z_{i},\rho^2 \right)$ according to (\ref{eqn:condDist_SP_x})},
\STATE{Compute $Z_{i+1} = Z_i - \delta \sum_{k=1}^{d^{\prime}} [ Z_i -\mathrm{prox}_{\theta_k g_k}^{\lambda}(Z_i) ] / \lambda - \delta ( Z_i - X_{i+1} ) / \rho^2 +\sqrt{2 \delta} \zeta_{i+1}$; where $\zeta_{i+1} \sim \mathcal{N}(0,\mathbb{I}_d)$},
\ENDFOR
\STATE{Output: Samples $X_1,\ldots,X_n$}.
\end{algorithmic}
\end{algorithm}
In the case where the likelihood is Gaussian one can exactly sample from (\ref{eqn:condDist_SP_x}) \cite{6945861} (for a review and comparison of existing Gaussian sampling approaches, see \cite{doi:10.1137/20M1371026}). Additionally, when the iterates $Z_{i}$ are also sampled exactly from $p\left(z|y,X_{i},\rho^2 \right)$ (i.e., by replacing Step 4 by an exact sampler), the resulting scheme is provably ergodic and can be used for approximate inference w.r.t. $p\left(x|y\right)$ \cite{vono2020efficient}. Leaving exact sampling aside, a main benefit of this splitting approach is that the step-size one needs to set for the proximal MCMC method used for sampling (\ref{eqn:condDist_SP_z}) will be independent of the Lipschitz constant associated with the likelihood distribution, and will only depend on the parameters  $\lambda$ and $\rho^2$ (i.e., the step-size now depends on the Lipschitz constant indirectly via $\lambda$ and $\rho^2$, there is still some dependence w.r.t. the Lipschitz constant through the bias incurred). This can lead to faster sampling algorithms compared to MYULA for suitably chosen values of the parameter $\rho^2$ \cite{8625467}, albeit for a biased posterior distribution.

\section{Enhancing Bayesian imaging models by smoothing}
\label{sec:augModelBetterModel}

As discussed previously, the augmented model \eqref{eqn:postDist_exp_augm} was originally proposed  as a relaxation of \eqref{eqn:posterior_dist} that allows for  a faster exploration of the target distribution, at the expense of some additional bias when compared to the original model. One then might think that $\rho^2 = 0$ represents the best model for inference (at the expense of higher computing cost). However, we have found empirically that this is not the case.

As an illustration, Figure \ref{fig:cam_deb_mse_vs_rho2_w_SAPG}(a) shows the estimation mean-squared error (MSE) for a Bayesian image debluring problem (the details of this experiment will be explained in Section \ref{subsec:imageDeconvolution}). The error is computed w.r.t. the posterior mean, as estimated by an adaptation of the SK-ROCK method to target \eqref{eqn:MY_augmented} (see Section \ref{subsec:ls_SK_ROCK} for details), using a value of $\theta = 4.4 \times 10^{-2}$ estimated by \cite[Algorithm 1]{vidal2019maximum}, and by using different values for $\rho^2$. Recalling that increasing $\rho^2$ improves convergence speed, one can clearly identify a regime of small values of $\rho^2$ for which convergence speed improves without a deterioration in estimation accuracy (in fact, there is a mild improvement). Beyond this range, the estimation MSE deteriorates dramatically. This suggests the need for a method to automatically set the value of $\rho^2$.

We propose an empirical Bayesian method to estimate optimal values for $\theta$ and $\rho^2$ directly from $y$ by maximum marginal likelihood estimation (MMLE)
\begin{equation}
\label{eqn:argmax_p_y_given_rho_theta}
(\theta_*,\rho^2_*) = \argmax_{\theta \in \Theta, \rho^2 \in \Omega} p(y|\theta,\rho^2),
\end{equation}
where $\Theta \subset (0,+\infty)^{d^{\prime}}$, $\Omega \subset (0,+\infty)$ are compact convex sets, and $p(y|\theta,\rho^2)$ is defined in \eqref{eqn:postDist_exp_augm}. To solve \eqref{eqn:argmax_p_y_given_rho_theta}, we modify the stochastic approximation proximal gradient (SAPG) algorithm of \cite{vidal2019maximum}. By maximising the model evidence, \eqref{eqn:argmax_p_y_given_rho_theta} seeks to select the best model to perform inference within the class of posterior distributions parametrised by $\theta \in \Theta, \rho^2 \in \Omega$ \cite{10.2307/1912526}.

\subsection{Computing the optimal values for \texorpdfstring{$\theta$}{θ} and \texorpdfstring{$\rho^2$}{ρ²}}
We adopt the approach of \cite{vidal2019maximum} to solve \eqref{eqn:argmax_p_y_given_rho_theta} and estimate optimal values for $\theta$ and $\rho^2$ in (\ref{eqn:postDist_exp_augm}). The method \cite{vidal2019maximum} was proposed for models of the form \eqref{eqn:posterior_dist}, so we will now adapt it to the augmented model (\ref{eqn:postDist_exp_augm}).

We are interested in estimating the parameters $\theta \in \Theta$, $\rho^2 \in \Omega$ by MMLE (\ref{eqn:argmax_p_y_given_rho_theta}). If we had access to the gradients $\nabla_{\rho^2} \log p(y|\theta,\rho^2)$ and $\nabla_{\theta} \log p(y|\theta,\rho^2)$, then we could construct an interative algorithm that converges to the solution of \eqref{eqn:argmax_p_y_given_rho_theta} by using the projected gradient algorithm \cite{LEVITIN19661}
\begin{align*}
\rho^2_{n+1} & = \Pi_{\Omega} \left[ \rho^2_n + \gamma^{\prime}_n \nabla_{\rho^2} \log p(y|\theta_{n},\rho^2_n) \right] \\
\theta_{n+1} & = \Pi_{\Theta} \left[ \theta_n + \gamma_n \nabla_{\theta} \log p(y|\theta_n,\rho^2_{n}) \right],
\end{align*}
where $\Pi_{\Omega}$ and $\Pi_{\Theta}$ are the projection onto $\Omega$ and $\Theta$ respectively, and $(\gamma^{\prime}_n, \gamma_n)_{n \in \mathbb{N}}$ are sequences of non-increasing step-sizes such that $\sum_{n \in \mathbb{N}} \gamma^{\prime}_n \rightarrow + \infty$ and $\sum_{n \in \mathbb{N}} {\gamma^\prime}^2_n < \infty$, and similarly $\sum_{n \in \mathbb{N}} \gamma_n \rightarrow + \infty$ and $\sum_{n \in \mathbb{N}} \gamma^2_n < \infty$, (see Section \ref{subsec:implementation_guides} for details). However, due to the complexity of the model, $\nabla_{\rho^2} \log p(y|\theta,\rho^2)$ and $\nabla_{\theta} \log p(y|\theta,\rho^2)$ are intractable.

As shown in \cite{vidal2019maximum}, one can construct carefully designed stochastic estimates of these gradients that satisfy the conditions for the solution to converge to (\ref{eqn:argmax_p_y_given_rho_theta}). To build these stochastic estimators, we are going to express the gradients as expectations by applying the Fisher's identity \cite[Proposition D.4]{DoucRandal2014NTST}. More precisely, we have that
\begin{equation*}
\nabla_{\rho^2} \log p(y|\theta,\rho^2) = \int_{\mathbb{R}^d}\int_{\mathbb{R}^d} p(x,z|y,\theta,\rho^2) \nabla_{\rho^2} \log p(x,z,y|\theta,\rho^2)\mathrm{d}x \mathrm{d}z,
\end{equation*}
and
\begin{equation*}
\nabla_{\theta} \log p(y|\theta,\rho^2) = \int_{\mathbb{R}^d}\int_{\mathbb{R}^d} p(x,z|y,\theta,\rho^2) \nabla_{\theta} \log p(x,z,y|\theta,\rho^2)\mathrm{d}x \mathrm{d}z.
\end{equation*}
We can approximate these expectations by using MCMC. In fact, we will see that one MCMC sample will suffice to obtain an estimate of the gradient accurate enough to converge asymptotically to (\ref{eqn:argmax_p_y_given_rho_theta}). As $p(x,z,y|\theta,\rho^2) = p(y|x)p(x,z|\theta,\rho^2)=p(y|x)p(x|z,\rho^2)p(z|\theta)$, we have
\begin{equation*}
\nabla_{\rho^2} \log p(y|\theta,\rho^2) = \int_{\mathbb{R}^d}\int_{\mathbb{R}^d} p(x,z|y,\theta,\rho^2) \nabla_{\rho^2} \log p(x|z,\rho^2) \mathrm{d}x \mathrm{d}z,
\end{equation*}
and
\begin{equation*}
\nabla_{\theta} \log p(y|\theta,\rho^2) = \int_{\mathbb{R}^d}\int_{\mathbb{R}^d} p(x,z|y,\theta,\rho^2) \nabla_{\theta} \log p(z|\theta) \mathrm{d}x \mathrm{d}z.
\end{equation*}
Replacing (\ref{eqn:augmented_prior_full}) in  $p(x|z,\rho^2)$ we obtain 
\begin{equation*}
\nabla_{\rho^2} \log p(y|\theta,\rho^2) = A_{\theta,\rho^2}(y) - \frac{d}{2 \rho^2} ,
\end{equation*}
where
\begin{equation*}
A_{\theta,\rho^2}(y) = \mathbb{E}_{x,z | y,\theta,\rho^2} \left[ \frac{\Vert x-z \Vert^2}{2 (\rho^2)^2} \right] = \int_{\mathbb{R}^d}\int_{\mathbb{R}^d} p(x,z|y,\theta,\rho^2) \frac{\Vert x-z \Vert^2}{2 (\rho^2)^2} \mathrm{d}x \mathrm{d}z,
\end{equation*}
and similarly, replacing  (\ref{eqn:likelihood_and_prior_g}) in $p(z|\theta)$ gives
\begin{equation*}
\nabla_{\theta} \log p(y|\theta,\rho^2) = -B_{\theta,\rho^2}(y) - C_{\theta,\rho^2}(y),
\end{equation*}
where
\begin{equation*}
\begin{split}
B_{\theta,\rho^2}(y) & = \mathbb{E}_{x,z|y,\theta,\rho^2} [g(z)] = \int_{\mathbb{R}^d}\int_{\mathbb{R}^d} p(x,z|y,\theta,\rho^2) g(z) \mathrm{d}x \mathrm{d}z , \\
C_{\theta,\rho^2}(y) & = \mathbb{E}_{x,z|y,\theta,\rho^2} \left[ \nabla_{\theta} \log \left( \int_{\mathbb{R}^d} \exp(-\theta^{\mathsf{T}} g(z)) \mathrm{d}z \right) \right] \\ & = \int_{\mathbb{R}^d}\int_{\mathbb{R}^d} p(x,z|y,\theta,\rho^2) \nabla_{\theta} \log \left[ \int_{\mathbb{R}^d} \exp(-\theta^{\mathsf{T}} g(z)) \mathrm{d}z \right] \mathrm{d}x \mathrm{d}z.
\end{split}
\end{equation*}
Because of the complexity of the model, $A_{\theta,\rho^2}(y)$ and $B_{\theta,\rho^2}(y)$ are not available analytically and need to be approximated by MCMC computation (e.g., by using the methods we develop in Section \ref{sec:SGS_noisy_MYULA_new_MCMC}). With respect to $C_{\theta,\rho^2}(y)$, and more precisely, the integral between brackets, we can follow a similar procedure as in \cite[Section 3.2.1]{vidal2019maximum}. In particular, if we consider the case where each $g_i(z)$ is $\alpha_i$ positively homogeneous\footnote{$g(x)$ is $\alpha$ positively homogeneous if, for any $x \in \mathbb{R}^d$ and $t >0$, $g(tx) = t^{\alpha} g(x)$.}, which is the case for many regularizers such as $\ell_1$, $\ell_2$ or TV, we have that
\begin{equation*}
\frac{\partial \log p(y|\theta,\rho^2)}{\partial \theta^{(i)}} = \frac{d}{\alpha_i \theta^{(i)}} - \mathbb{E}_{x,z | y,\theta^{(i)},\rho^2} \left[ g_i(z) \right].
\end{equation*}
(See \cite{vidal2019maximum} for more details and \cite[Section 3.2]{vidal2019maximum} for the case of inhomogeneous regularizers).

Following on from this, and by using Monte Carlo approximations of $A_{\theta,\rho^2}(y)$ and $B_{\theta,\rho^2}(y)$, we construct an SAPG algorithm \cite{fort2019stochastic,vidal2019maximum} to solve \eqref{eqn:argmax_p_y_given_rho_theta} and produce optimal estimates of $\theta$ and $\rho^2$. This method is presented in Algorithm \ref{alg:SAPG_augm_model}. We refer the reader to Section \ref{subsec:implementation_guides} for guidelines on setting the step-size $(\gamma^{\prime}_n, \gamma_n)_{n \in \mathbb{N}}$ and the weights $(w_n)_{n \in \mathbb{N}}$ for the averages computed in the final step of Algorithm \ref{alg:SAPG_augm_model}. See also \cite{doi:10.1137/20M1339842,DeBortoli2021} for details about the convergence properties of this kind of SAPG algorithm.
\begin{algorithm}
\caption{SAPG algorithm for the augmented model (\ref{eqn:postDist_exp_augm})}
\label{alg:SAPG_augm_model}
\begin{algorithmic}[1]
\STATE{Input: $X_0^0, Z_0^0 \in \mathbb{R}^d$, $\theta_0, \rho^2_0, \gamma_0, \gamma_0^{\prime} \in \mathbb{R}$, $\lambda > 0$, $m, n \in \mathbb{N}$}.
\FOR{$i = 0:m-1$}
\IF{$i > 0$}
    \STATE{Set $X_i^{(0)} = X_{i-1}^{(n)}$},
\ENDIF
\FOR{$j = 0:n-1$}
\STATE{Sample $X_{i+1}^{(j+1)}$, $Z_{i+1}^{(j+1)}$ according to Algorithm \ref{alg:ls_MYULA}},
\ENDFOR
\FOR{$j = 1:d^{\prime}$}
\STATE{Set $\theta_{i+1}^{(j)} = \Pi_{\Theta} \left[ \theta_i^{(j)} + \frac{\gamma_{i+1}^{(j)}}{n} \sum_{k=1}^n \left\lbrace \frac{d}{\alpha_j \theta_i^{(j)}} - g_j(Z^{(k)}_{i+i}) \right\rbrace \right]$},
\ENDFOR
\STATE{Set $\rho^2_{i+1} = \Pi_{\Omega} \left[ \rho^2_i + \frac{\gamma_{i+1}^{\prime}}{n} \sum_{k=1}^n \left\lbrace \Vert X^{(k)}_{i+1} - Z^{(k)}_{i+1} \Vert^2 / 2 (\rho^2_i)^2  -  d/(2\rho^2_i) \right\rbrace \right]$},
\ENDFOR
\STATE{Output: $\overline{\theta}_m^{(j)} = \sum_{k=0}^{m} w_k \theta_k^{(j)} / \sum_{k=0}^{m} w_k$ for $j \in \{1,\ldots,d^{\prime}\}$, $\overline{\rho}^2_m = \sum_{k=0}^{m} w_k \rho^2_k / \sum_{k=0}^{m} w_k$}.
\end{algorithmic}
\end{algorithm}

\begin{figure}[htb]
	\centering
	\subfloat[]{
		\includegraphics[scale=.4]{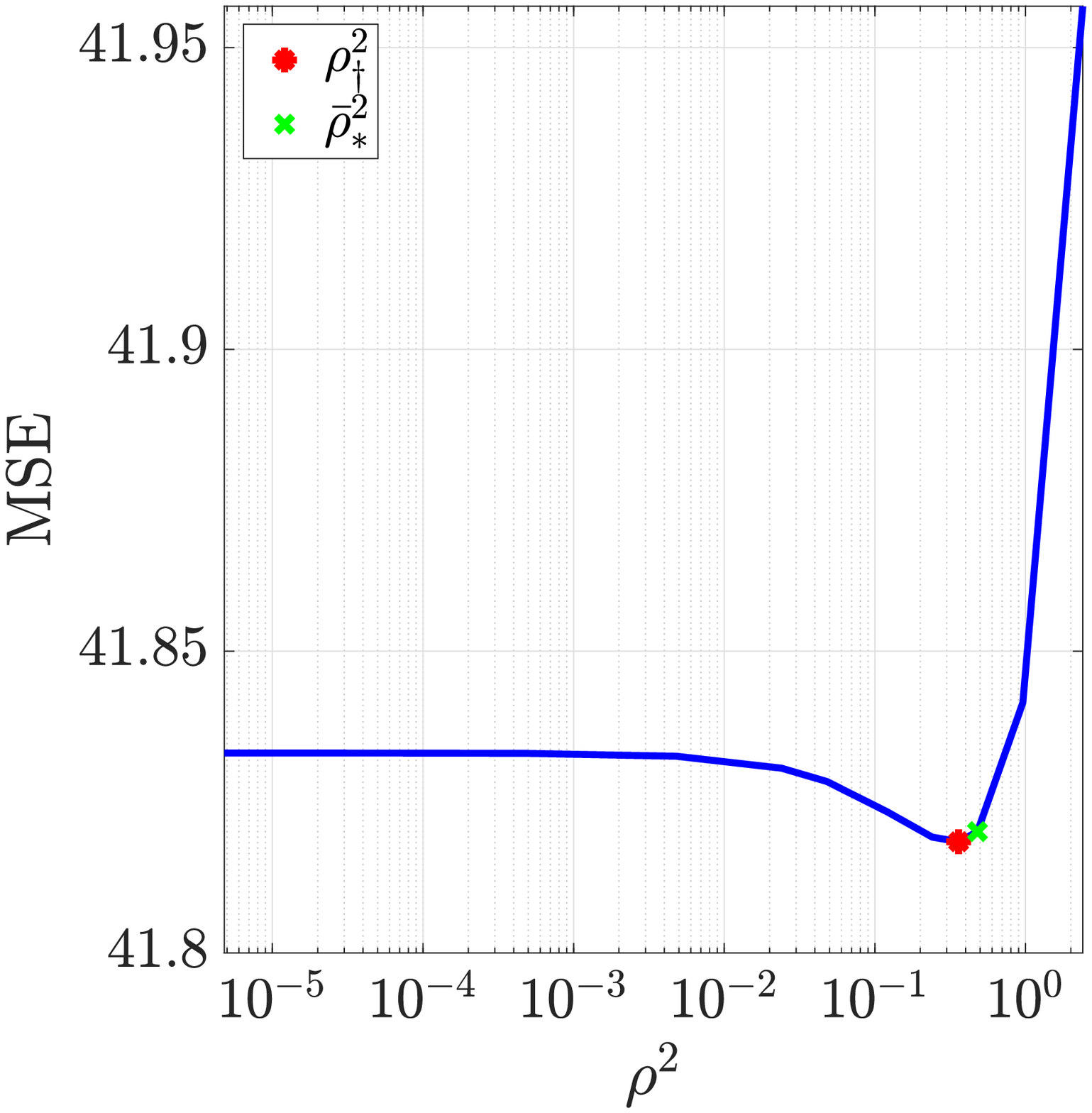}
	}
	\subfloat[]{
		\includegraphics[scale=.4]{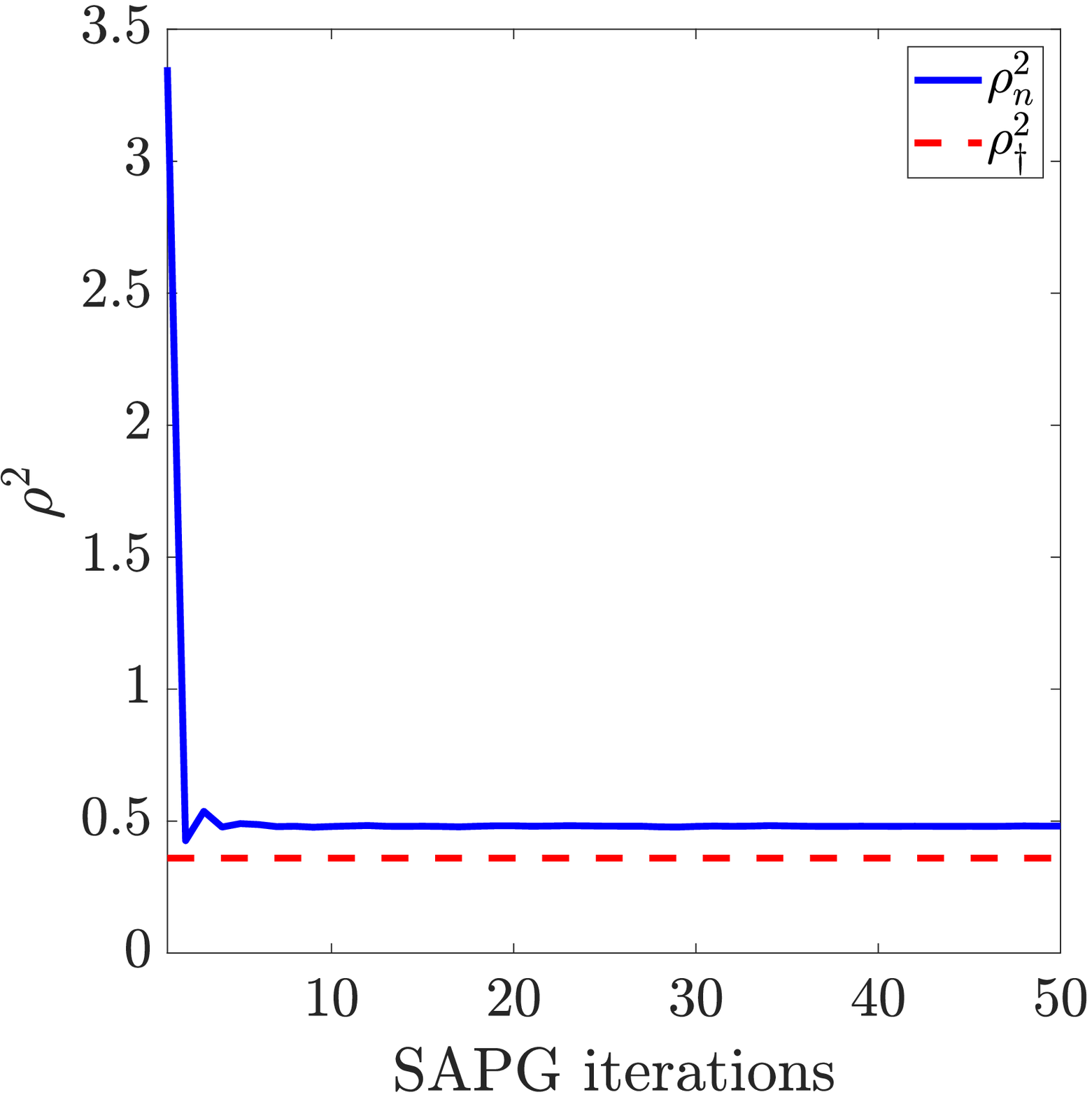}
	}
	\caption{Image deblurring experiment: {\normalfont(a)} MSE between the true image and the posterior mean estimated using Algorithm \ref{alg:ls_SK_ROCK}, for some values of $\rho^2$. In red, the optimal value of $\rho^2$ that minimises the MSE, and in green, the value of $\rho^2$ found by Algorithm \ref{alg:SAPG_augm_model}. {\normalfont(b)} Iterations of SAPG algorithm to estimate $\rho^2$.}
	\label{fig:cam_deb_mse_vs_rho2_w_SAPG}
\end{figure}

To illustrate Algorithm \ref{alg:SAPG_augm_model} in action, Figure \ref{fig:cam_deb_mse_vs_rho2_w_SAPG} shows the value of $\rho^2$ estimated by the algorithm for the image deblurring problem. Observe that the MMLE estimate is close to the value that produces the best estimation MSE in this case. This is in agreement with the results reported in \cite{vidal2019maximum} for other problems. Lastly, notice that Step 10 of Algorithm \ref{alg:SAPG_augm_model} involves the original prior, with terms $\{g_j\}_{j=1}^{d^\prime}$, and not the smooth approximations $\{g^\lambda_j\}_{j=1}^{d^\prime}$, as the SAPG scheme to estimate $\theta$ and $\rho^2$ is not affected by the non-smoothness $p(z|\theta)$ w.r.t $z$ (instead, it requires some smoothness of the Markov kernels w.r.t. $\theta$, see \cite{DeBortoli2021, doi:10.1137/20M1339842} for technical details). The approximation $\{g^\lambda_j\}_{j=1}^{d^\prime}$ is used within the MYULA step of Step 7 Algorithm \ref{alg:SAPG_augm_model}, which does require a smooth prior. This mismatch introduces some bias, which is controlled by using a small value of $\lambda $ \cite{DeBortoli2021}. One could use $\{g^\lambda_j\}_{j=1}^{d^\prime}$ instead of $\{g_j\}_{j=1}^{d^\prime}$ within Step 10. However, doing so would prevent the SAPG scheme from exploiting the homogeneity of $\{g_j\}_{j=1}^{d^\prime}$, resulting in a significantly more expensive SAPG scheme (see \cite[Algorithm 3]{vidal2019maximum}).

\section{Reinterpretation of SGS as noisy MYULA \& new MCMC methods}
\label{sec:SGS_noisy_MYULA_new_MCMC}
\subsection{Noisy MYULA}
We now proceed to show that the SGS algorithm \ref{alg:SP} can be viewed as a noisy version of MYULA. This link will be crucial in allowing us to write it as a noisy discretisation of an SDE, which will help us to propose more efficient MCMC methods for sampling (\ref{eqn:postDist_exp_augm}).

First, note that the marginal of $z$ computed from \eqref{eqn:MY_augmented} can be written as follows
\begin{equation*}
p^{\lambda}(z|y,\theta,\rho^2) = \int_{\mathbb{R}^d} p^{\lambda} (x,z|y,\theta,\rho^2) \, \mathrm{d}x \propto p(y|z,\rho^2)p^{\lambda}(z|\theta),
\end{equation*}
where
\begin{equation*}
p(y|z,\rho^2) \propto \int_{\mathbb{R}^d} \exp \left[ -f_{y}(x) -\frac{1}{2 \rho^2} \Vert x-z \Vert^2 \right]\mathrm{d}x, \; \; p^{\lambda}(z|\theta) \propto \exp [- \theta^{\mathsf{T}} g^{\lambda}(z)].
\end{equation*}
Notice  that in the case where $f_{y}(x)$ is quadratic, $p(y|z,\rho^2)$ is Gaussian with eigenvalues in its covariance matrix shifted by $\rho^2$, when compared with the covariance of $f_{y}(x)$.
Now applying the MYULA to $p^{\lambda}(z|y,\theta,\rho^2)$, we have that
\begin{equation}\label{eqn:ULA_z}
Z_{n+1} = Z_n + \delta \nabla_{z} \log p^{\lambda}(Z_n|\theta) + \delta \nabla_{z} \log p(y | Z_n,\rho^2 ) + \sqrt{2 \delta} \zeta_{n+1},
\end{equation}
where $(\zeta_{n+1})_{n \geq 0}$ is an i.i.d. sequence of $d$-dimensional standard Gaussian random vectors. Due to the complexity of the model, it is difficult to compute $\nabla_{z} \log p(y | Z_n,\rho^2 )$, however, we can express it as an expectation by using Fisher's identity \cite[Proposition D.4]{DoucRandal2014NTST} as follows
\begin{equation*}
\begin{split}
\nabla_{z} \log p(y|z,\rho^2) & = \int_{\mathbb{R}^d} p(x | y,z,\rho^2) \nabla_{z} \log p(x,y|z,\rho^2) \mathrm{d}x \\
& = \mathbb{E}_{x | y,z,\rho^2} \left[ \nabla_{z} \log p(x,y|z,\rho^2) \right].
\end{split}
\end{equation*}
As $p(x,y|z,\rho^2) = p(y|x)p(x|z,\rho^2)$, we have
\begin{equation*}
\begin{split}
\nabla_{z} \log p(x,y|z,\rho^2) & =\mathbb{E}_{x | y,z,\rho^2} \left[ \nabla_{z} \log p(x|z,\rho^2) \right] \\  & = \frac{1}{\rho^2} \mathbb{E}_{x | y,z,\rho^2} \left( x - z  \right).
\end{split}
\end{equation*}
Using this expression in (\ref{eqn:ULA_z}) we obtain
\begin{equation}\label{eqn:ULA_z_with_expectation_1}
Z_{n+1} = Z_n - \delta \nabla_{z} p^{\lambda}(Z_n|\theta) - \frac{\delta}{\rho^2} \mathbb{E}_{x | y,z,\rho^2} \left( Z_n - x \right) + \sqrt{2 \delta} \zeta_{n+1},
\end{equation}
We are now ready to explicitly establish the connection to SGS. SGS stems from dealing with the presence of the expectation in this algorithm by replace it by a Monte Carlo empirical average, i.e.,
\begin{equation}\label{eqn:empiricalAvgGrad_SGS}
\mathbb{E}_{x | y,z,\rho^2} \left( Z_n - x  \right) \approx Z_n -  \frac{1}{N} \sum_{i=1}^N X^{(i)}, \; \text{where} \; X^{(i)} \sim  p(x|y,Z_n ; \rho).
\end{equation}
More precisely, to recover SGS we take $N=1$ and substitute in  (\ref{eqn:ULA_z_with_expectation_1}) to obtain
\begin{equation}
\label{eqn:noisy_MYULA_z}
Z_{n+1} = Z_n - \delta \nabla_{z} p^{\lambda}(Z_n|\theta) - \frac{\delta}{\rho^2} (Z_n - X^{(1)}) + \sqrt{2 \delta} \zeta_{n+1}.
\end{equation}
Since $X^{(1)}$ is an exact sample from $p(x|y,Z_n , \rho^2)$, \eqref{eqn:noisy_MYULA_z} corresponds to the fourth line of Algorithm \ref{alg:SP}.

This establishes that SGS is equivalent to a noisy version of MYULA that relies on one sample from $p(x | y,z,\rho^2)$ to compute a stochastic estimate of the gradient $\nabla_{z} \log p(y|z,\rho^2)$ via (\ref{eqn:empiricalAvgGrad_SGS}).  Using multiple samples from $p(x|y,Z_n ; \rho)$ would improve the estimation of the expectation (\ref{eqn:empiricalAvgGrad_SGS}) and hence the behaviour of the algorithm. Alternatively, in the experiments considered in this paper $p(x|y,Z_n ; \rho)$ is Gaussian, and hence this expectation can be calculated exactly. This is exploited in the MCMC methods proposed below.

\subsection{Latent space MYULA}
We established that SGS is equivalent to MYULA targeting  the marginal of $z$  with an inexact (i.e., stochastic) estimate of the gradient. Replacing this stochastic estimate with its exact value in Algorithm \ref{alg:SP} produces the following recursion
\begin{equation}
\label{eqn:alg_lsMYULA_postMean}
\begin{split}
X_{\mathrm{grad}}^{i+1} & = \mathbb{E}_{x | y,Z_i,\rho^2} [x], \\
Z_{i+1} & = Z_i - \frac{\delta}{\lambda} \sum_{k=1}^{d^{\prime}} [ Z_i -\mathrm{prox}_{\theta_k g_k}^{\lambda}(Z_i) ] - \delta ( Z_i - X_{\mathrm{grad}}^{i+1} ) / \rho^2 +\sqrt{2 \delta} \zeta_{i+1}, \\
\end{split}
\end{equation}
where $\zeta_{i+1} \sim \mathcal{N}(0,\mathbb{I}_d)$. 

We now discuss how to use samples $\{Z_{i}\}_{i\geq 1}^m$ to compute expectations w.r.t. the marginal of interest $x|y,\theta,\rho^2$. More precisely, consider the computation of an expectation $\mathbb{E}_{x|y,\theta,\rho^2} [h(x)]$ for some function $h$ w.r.t. the posterior distribution $p^{\lambda}(x|y,\theta,\rho^2)$ defined in (\ref{eqn:marginal_x_augm}) by using \eqref{eqn:alg_lsMYULA_postMean}. Formally,
\begin{equation*}
\mathbb{E}_{x|y,\theta,\rho^2} [h(x)] = \int_{\mathbb{R}^d} h(x) \int_{\mathbb{R}^d} p^{\lambda} (x,z|y,\theta,\rho^2) \, \mathrm{d}z \, \mathrm{d}x.
\end{equation*}
Using the fact that $p^{\lambda}(x,z|y,\theta,\rho^2) = p(x|y,z,\rho^2)p^{\lambda}(z|\theta)$ we have that
\begin{equation}
\label{eqn:expectation_augmented_lsAlg}
\begin{split}
\mathbb{E}_{x|y,\theta,\rho^2} [h(x)] & = \int_{\mathbb{R}^d} \int_{\mathbb{R}^d} h(x)  p(x|y,z,\rho^2) p^{\lambda}(z|\theta) \, \mathrm{d}z \, \mathrm{d}x \\ & = \int_{\mathbb{R}^d} \int_{\mathbb{R}^d} h(x)  p(x|y,z,\rho^2) \, \mathrm{d}x \, p^{\lambda}(z|\theta) \, \mathrm{d}z \\ & = \mathbb{E}_{z|\theta} \left[ \mathbb{E}_{x|y,z,\rho^2} \left( h(x) \right) \right].
\end{split}
\end{equation}
In cases where $\mathbb{E}_{x|y,\theta,\rho^2} [h(x)]$ is available analytically, we suggest using a Rao-Blackwellised estimator of the form \cite{RobertChristianP.2004MCsm}
\begin{equation*}
	\mathbb{E}_{x|y,\theta,\rho^2} [h(x)] \approx \frac{1}{m} \sum_{i=1}^m \mathbb{E}_{x|y,Z_i,\rho^2} \left[ h(x) \right].
\end{equation*}
The computation of $\mathbb{E}_{x|y,Z_i,\rho^2} \left[ h(x) \right]$ can be done as a postprocessing step, or alternatively within the iterations of the sampler. If $\mathbb{E}_{x|y,\theta,\rho^2} [h(x)]$ is not available analytically, we would draw samples from the conditional $x|y,Z_i,\theta,\rho^2$ and apply a standard Monte Carlo estimator.

We are now ready to present our first new MCMC method, summarised in Algorithm \ref{alg:ls_MYULA} below. We henceforth refer to this method as \textit{latent space MYULA} (ls-MYULA), since it corresponds to MYULA applied to the marginal of the latent variable $z$.

\begin{algorithm}
\caption{ls-MYULA}
\label{alg:ls_MYULA}
\begin{algorithmic}[1]
\STATE{Input: $X_0, Z_0 \in \mathbb{R}^d$, $\lambda, \rho > 0$, $m \in \mathbb{N}$}.
\FOR{$i = 0:m-1$}
\STATE{Compute $X_{\mathrm{grad}}^{i+1} = \mathbb{E}_{x | y,Z_i,\rho^2} [x], $}
\STATE{Compute $Z_{i+1} = Z_i - \delta \sum_{k=1}^{d^{\prime}} [ Z_i -\mathrm{prox}_{\theta_k g_k}^{\lambda}(Z_i) ] / \lambda - \delta ( Z_i - X_{\mathrm{grad}}^{i+1} ) / \rho^2 +\sqrt{2 \delta} \zeta_{i+1}$; where $\zeta_{i+1} \sim \mathcal{N}(0,\mathbb{I}_d)$},
\STATE{Compute $\hat{h}_{i+1} = \mathbb{E}_{x | y,Z_{i+1},\rho^2} [h(x)] $},
\ENDFOR
\STATE{Output: an estimator of $\mathbb{E}_{x|y,\theta,\rho^2} [h(x)]$ given by $\{ \sum_{k=1}^m \hat{h}_k \} / m$}.
\end{algorithmic}
\end{algorithm}

\begin{remark}
The underlying assumption in Algorithm \ref{alg:ls_MYULA} is that one can explicitly calculate $\mathbb{E}_{x | y,Z_i,\rho^2} [x]$ which, for example, is the case when the expectation represents the first moment of a Gaussian distribution, which corresponds to the likelihood models we consider in our experiments. In cases where $\mathbb{E}_{x | y,Z_i,\rho^2} [x]$ is intractable, we recommend to replace the expectation by its corresponding MCMC estimation, i.e.,
\begin{equation*}
\mathbb{E}_{x | y,Z_i,\rho^2} [x] \approx \frac{1}{M} \sum_{i=1}^M X_i \, \, \text{where} \, \, X_i \sim p(x|y,Z_i,\rho^2).
\end{equation*}
\end{remark}

\begin{remark}
    Notice that Algorithm \ref{alg:ls_MYULA} relies implicitly on two forms of smoothing, which operate differently and provide complementary benefits. On the one hand, through the MYULA construction, $\{g_k\}_{k=1}^{d^{\prime}}$ is replaced by the (smooth) Moreau-Yosida envelopes $\{g^\lambda_k\}_{k=1}^{d^{\prime}}$. On the other, the use of the auxiliary variable $z$ introduces smoothing on the marginal prior $p(x|\rho^2) = \int p(x|z,\rho^2)p(z) \textrm{d}z$, where $p(x|z,\rho^2)$ acts as a Gaussian smoothing kernel. It can thus appear that there is some redundancy and that a single smoothing mechanism would suffice. However, Algorithm \ref{alg:ls_MYULA} operates on the latent space, and from that perspective, the smoothing related to $p(x|z,\rho^2)$ acts on the likelihood function of $z$ given $y$, and not to the prior of $z$. This can lead to significant benefits in terms of convergence speed, as illustrated in Section \ref{sec:experiments}. This effect can be analysed in detail in the case of the Gaussian likelihood function, where the smoothing introduced by $p(x|z,\rho^2)$ shifts the eigenvalues of the likelihood covariance matrix by $\rho^2$. As a result, the likelihood of $y$ w.r.t. $z$ is by construction strongly log-concave. The same remark holds for the latent space SK-ROCK algorithm described below. Also note that the bias introduced by this additional smoothing is undone exactly when the samples are mapped from the latent space of $z$ to the canonical space of $x$.
\end{remark}

\subsection{Latent space SK-ROCK}
\label{subsec:ls_SK_ROCK}
In the same way that an exact MYULA discretization is more beneficial than the stochastic MYULA discretization used in SGS, we can further improve results by using an exact SK-ROCK discretization which, as we described in Section \ref{subsec:SK_ROCK}, has many important advantages compared to MYULA. In particular, we present this method in Algorithm \ref{alg:ls_SK_ROCK}, and  we will refer to it as \textit{latent space SK-ROCK} (ls-SK-ROCK). The main difference between this algorithm and Algorithm \ref{alg:SKROCK} is  that the conditional expectation $\mathbb{E}_{x | y,\text{\~{Z}}_j,\rho^2} [x]$ is computed on each internal stage $s$.

\begin{algorithm}
\caption{ls-SK-ROCK}
\label{alg:ls_SK_ROCK}
\begin{algorithmic}[1]
\STATE{Input: $X_0, Z_0 \in \mathbb{R}^d$, $\lambda, \rho > 0$, $m, s \in \mathbb{N}$, $\eta = 0.05$}.
\STATE{Compute $l_s=(s-0.5)^2 (2 - 4/3 \eta) -1.5$},
\STATE{Compute $\omega_0=1 + \eta / s^2$, $\omega_1=T_s(\omega_0) / T_s' (\omega_0)$},
\STATE{Compute $\mu_1=\omega_1 / \omega_0$, $\nu_1=s \omega_1 /2$, $k_1 = s \omega_1 / \omega_0$},
\STATE{Choose $\delta  \in (0,\delta_s^{\max} ]$, where $\delta_s^{\max}=l_s / ( 1 / (\rho^2 + L_f^{-1}) + 1/\lambda)$},
\FOR{$i = 0:m-1$}
\STATE{Set $\text{\~{X}}_{\mathrm{grad}}^0  =  X_{\mathrm{grad}}^i$, $\text{\~{Z}}_0  =  Z_i$},
\STATE{Sample $\xi_{i+1} \sim \mathcal{N} (0,2 \delta \mathbb{I}_d)$},
\STATE{Compute $\text{\~{X}}_{\mathrm{grad}}^1 = \mathbb{E}_{x | y,\text{\~{Z}}_0 + \nu_1  \xi_{i+1},\rho^2} [x]$},
\STATE{Compute $\Lambda (\text{\~{Z}}_0) = \sum_{k=1}^{d^{\prime}} [\text{\~{Z}}_0 +\nu_1 \xi_{i+1} - \text{prox}_{\theta_k g_k}^{\lambda}(\text{\~{Z}}_0 +\nu_1 \xi_{i+1}) ] / \lambda  + (\text{\~{Z}}_0 +\nu_1 \xi_{i+1} - \text{\~{X}}_{\mathrm{grad}}^1)/\rho^2$},
\STATE{Compute $\text{\~{Z}}_1 = \text{\~{Z}}_0 - \mu_1 \delta \Lambda (\text{\~{Z}}_0) +  k_1^2 \xi_{i+1} $},
\FOR {$j = 2:s$}
\STATE{Compute $\mu_j = 2\omega_1 T_{j-1}(\omega_0) / T_j(\omega_0)$, $\nu_j = 2\omega_0 T_{j-1}(\omega_0) / T_j(\omega_0)$, $k_j = 1-\nu_j$},
\STATE{Compute $\text{\~{X}}_{\mathrm{grad}}^j = \mathbb{E}_{x | y,\text{\~{Z}}_{j-1},\rho^2} [x]$},
\STATE{Compute $\Lambda (\text{\~{Z}}_{j-1}) = \sum_{k=1}^{d^{\prime}} [\text{\~{Z}}_{j-1} - \text{prox}_{\theta_k g_k}^{\lambda}(\text{\~{Z}}_{j-1} ) ]/ \lambda + (\text{\~{Z}}_{j-1} - \text{\~{X}}_{\mathrm{grad}}^j)/\rho^2$},
\STATE{Compute $\text{\~{Z}}_j  =  -\mu_j \delta \Lambda (\text{\~{Z}}_{j-1})  + \nu_j \text{\~{Z}}_{j-1} + k_j \text{\~{Z}}_{j-2}$},
\ENDFOR
\STATE{Set $X_{\mathrm{grad}}^{i+1} = \text{\~{X}}_{\mathrm{grad}}^s$, $Z_{i+1} = \text{\~{Z}}_s$, $\hat{h}_{i+1} = \mathbb{E}_{x | y,Z_{i+1},\rho^2} [h(x)]$},
\ENDFOR
\STATE{Output: an estimator of $\mathbb{E}_{x|y,\theta,\rho^2} [h(x)]$ given by $\{\sum_{k=1}^m \hat{h}_k\} / m$}.
\end{algorithmic}
\end{algorithm}

\subsection{Implementation guidelines}
\label{subsec:implementation_guides}
\subsubsection*{Setting $\lambda$}
As the priors of the experiments performed in this work are non-differentiable, we will use the Moreau-Yosida envelope defined in (\ref{eqn:moreauYosida}) with $\lambda \in [L_f^{-1},10 L_f^{-1}]$. We chose $\lambda = L_f^{-1}$ in our numerical experiments, however, we have found numerically that values of $\lambda = 5 L_f^{-1}$ or $\lambda = 10 L_f^{-1}$ lead to faster convergence at the cost of additional bias.

\subsubsection*{Setting $\gamma_i^{(j)}$, $\gamma_i'$ and $n$}
With respect to Algorithm \ref{alg:SAPG_augm_model}, it is suggested in \cite{vidal2019maximum} to set $\gamma_i^{(j)} = C_0^{(j)} i^{-p}$ and $\gamma_i' = C_0' i^{-p}$ where $p \in [0.6, 0.9]$ (in the experiments performed in this paper, we have set $p=0.8$), $C_0^{(j)}$ and $C_0'$ starting with $(\theta_0^{(j)} d)^{-1}$ and $(\rho^2_0 d)^{-1}$ respectively, and then readjusting it if necessary. With respect to $n$, we have followed the recommendation in \cite{vidal2019maximum} and used  a single sample (i.e., $n=1$) on each iteration (we did not observe significant difference for larger values of $n$). For more details about these parameters and their effect on the convergence properties of the algorithms please see \cite[Section 4.2]{doi:10.1137/20M1339842} and \cite[Theorem 1]{DeBortoli2021}.

\subsubsection*{Setting $w_n$}
Following \cite{vidal2019maximum}, we recommend setting $w_n$ as follows
\begin{equation*}
  w_n =
    \begin{cases}
      0 & \text{if } n < N_0 , \\
      1 & \text{if } N_0 \leq n \leq N_1 , \\
      \gamma_n & \text{otherwise} , \\
    \end{cases}       
\end{equation*}
where $N_0$ is the number of initial iterations to be discarded (when $n < N_0$, the values of $\rho^2_n$ and $\theta_n$ are still bouncing and not stabilized), $n \in [N_0,N_1]$ corresponds to the averaging estimation phase, in which the values of $\rho^2_n$ and $\theta_n$ have stabilized and start converging, and $ n > N_1$ is known as the refinement phase where we use decreasing weights to enhance the accuracy of the estimator (see \cite[Section 3.3.1]{vidal2019maximum} for details).

\subsubsection*{Setting an stopping criteria}
It is recommended to supervise the evolution of $| \overline{\theta}_{m+1} - \overline{\theta}_m | / \overline{\theta}_m  $ and $| \overline{\rho}^2_{m+1} - \overline{\rho}^2_m | / \overline{\rho}^2_m  $ in the execution of Algorithm \ref{alg:SAPG_augm_model} until they reach a tolerance level $\beta$ to stop the algorithm execution. In our imaging experiments, we set $\beta = 10^{-4}$ but we have observed that $\beta = 10^{-3}$ is often enough to reach an acceptable estimate of the hyper-parameters in small computational times.

\subsubsection*{Other implementation considerations}
In the implementation of the SAPG method, it is important to update the step-size of the MCMC method to sample $X_i$ and $Z_i$ within each iteration of the SAPG scheme, as the maximum step-size depends on the value of $\rho_i^2$.

Regarding the Lipschitz constant $L$ one needs to compute the step-size for MYULA and SK-ROCK algorithms, the model (\ref{eqn:postDist_w_MY}) has $L = \lambda^{-1} + L_f$. With respect to the augmented model \eqref{eqn:marginal_x_augm}, the Lipschitz constant is $L_a = \lambda^{-1} + (\rho^2 + L_f^{-1})^{-1}$ which we use to implement SGS, ls-MYULA and ls-SK-ROCK. Therefore, we set the step-size of the MCMC methods to $1/L$ for MYULA, to $1/L_a$ for SGS and ls-MYULA, and to $\delta_s^{\max} $ for SK-ROCK and ls-SK-ROCK, where $\delta_s^{\max} $ can be found in Algorithms \ref{alg:SKROCK} and \ref{alg:ls_SK_ROCK}, respectively. Regarding the SK-ROCK and ls-SK-ROCK algorithms, their most important parameters are the number of internal stages $s \in \mathbb{N}$ and the maximum step-size $\delta_s^{\max}$. Note the increase in the step-size of Algorithm \ref{alg:ls_SK_ROCK} compared to Algorithm \ref{alg:SKROCK}, since $\delta_s^{max}=l_s / ( 1 / (\rho^2 + L_f^{-1}) + 1/\lambda) = l_s / L_a $ in ls-SK-ROCK, compared to $\delta_s^{max}=l_s / ( L_f + 1/\lambda) = l_s / L $ in SK-ROCK (with $l_s=[(s-0.5)^2 (2 - 4/3 \eta) -1.5]$ and $\eta = 0.05$ in both algorithms). This leads to a shift of $\rho^2$ in $\delta_s^{max}$ that allows ls-SK-ROCK to converge faster. We have empirically observed that a good bias-variance trade-off is achieved by taking $\delta \in [\delta_{max}/2, \delta_{max})$ and $s \in \{5,\ldots,15\}$.

It is worth highlighting at this point that the improvement in Lipschitz constant experiences by the latent-space model allows ls-MYULA and ls-SK-ROCK to take larger step-sizes than MYULA and SK-ROCK, respectively. This leads to an improvement in convergence speed and therefore to higher computational efficiency, without noticeable additional bias.

\section{Numerical experiments}
\label{sec:experiments}
We now illustrate the improvement that can be obtained by sampling the augmented model (\ref{eqn:postDist_exp_augm}) using Algorithms \ref{alg:ls_MYULA} and \ref{alg:ls_SK_ROCK} together with an optimal estimate of $\rho^2$ using Algorithm \ref{alg:SAPG_augm_model}. To evaluate the performance of the methods in a variety of situations, we perform two imaging experiments related to \emph{image deblurring} (whose model is strongly log-concave) and \emph{image inpainting} (whose model is weakly log-concave). We implement these algorithms as described in the implementation guidelines (see Section \ref{subsec:implementation_guides}).

\begin{figure}[htb]
	\centering
	\subfloat[\textit{cameraman}, true image $x$]{
		\includegraphics[scale=.47]{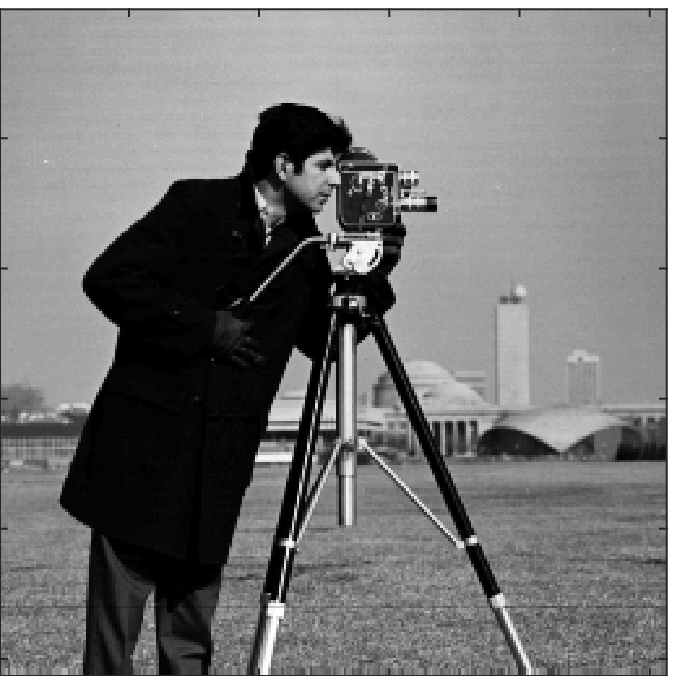}
	}
	\subfloat[\textit{skier}, true image $x$]{
		\includegraphics[scale=.47]{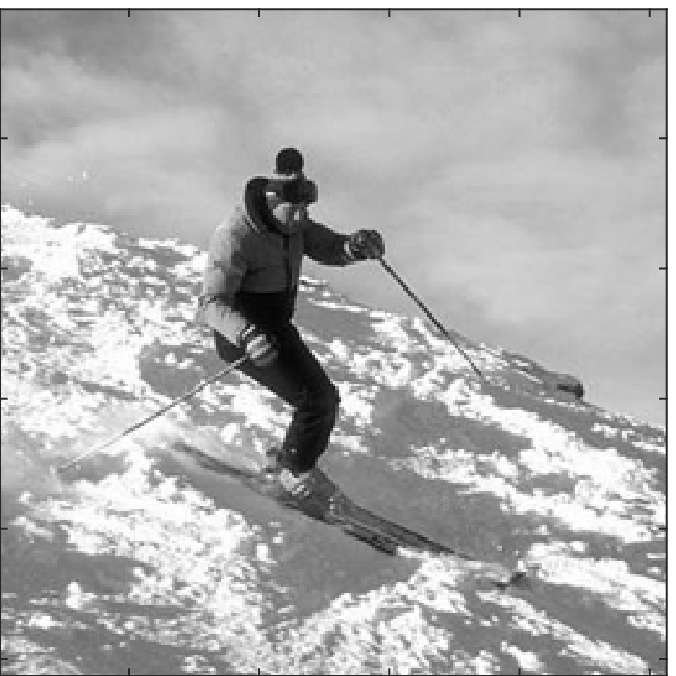}
	}
 \\
	\subfloat[\textit{cameraman}, observation $y$]{
		\includegraphics[scale=.47]{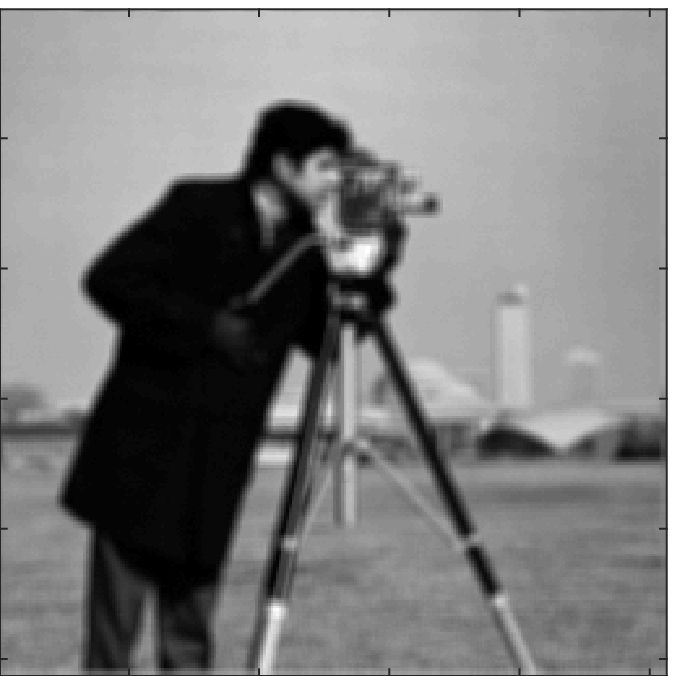}
	}
	\subfloat[\textit{skier}, observation $y$]{
		\includegraphics[scale=.47]{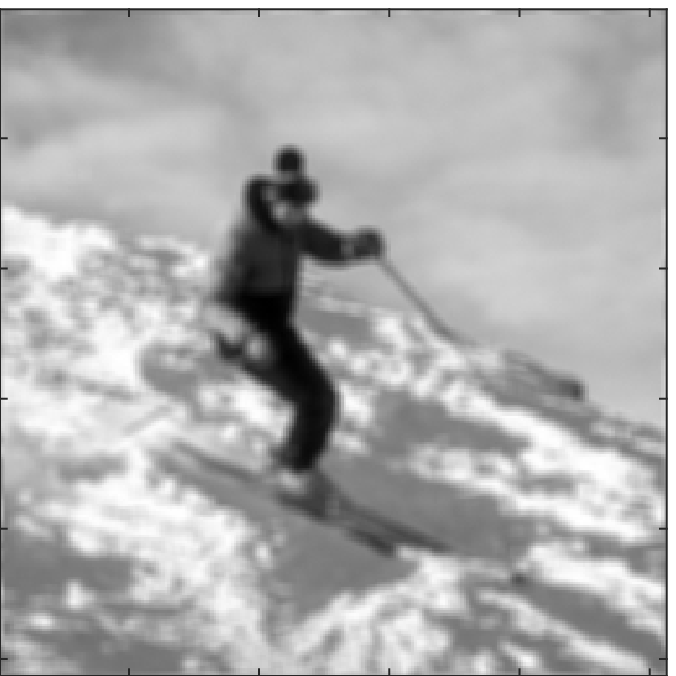}
	}
	\caption{Image deblurring experiments: Test images $x$ and their corresponding noisy and blurred observations $y$.}
	\label{fig:test_images_deconvolution}
\end{figure}

For a fair comparison the results we show have been plotted as a function of the number of gradient evaluations, i.e., the number of times $\nabla \log p^{\lambda}(x|,y,\theta)$ and $\nabla_{z}\log p^{\lambda}(z|y,\theta,\rho^2)$ are computed in our algorithms. The plots we show include the evolution of the MCMC samples in the burn-in stage using the scalar statistic $\log p (X_n|y,\theta)$ for MYULA and SK-ROCK, and $\log p (X_n^{\mathrm{grad}}|y,\theta)$ for SGS, ls-MYULA and ls-SKROCK. We have also plotted the progression of the mean-squared error (MSE) between the posterior mean and the true image, when all the algorithms have reached stationarity, including the MAP estimate defined in (\ref{eqn:MAP_estimate}) and computed using a highly efficient optimization algorithm called SALSA \cite{5445028,5570998} for the \emph{image deblurring} and \emph{image inpainting} experiments.

We also provide pixel-wise standard deviation plots as a way of quantifying the uncertainty in the delivered solution. We have also computed standard deviation plots performing downsampling by averaging the samples by a factor of $2 \times j$ where $j = \{1,2,4\}$, which allows us to observe the uncertainty in image structures at different scales. Finally, we also show autocorrelation plots of the slowest component of the samples produced by each of the methods, applying a 1-in-$s$ thinning to the MYULA, SGS and ls-MYULA chains to equal the number of gradient evaluations between the mentioned methods (one gradient evaluation per iteration) and SK-ROCK/ls-SKROCK methods ($s$ gradient evaluations per iteration). The chain's slowest component was identified by computing the approximated eigenvalue decomposition of the posterior covariance matrix and projecting the samples onto the leading eigenvector.
Now using the slowest component, we have also computed effective sample sizes (ESS) of the five algorithms discussed in this paper, where the sum is truncated at lag $k$ when the lag-$k$ autocorrelation reaches a value less than $0.05$.

For completeness, in Table \ref{tab:times_experiments} we have also provided computing times of all the experiments. These results have been obtained on an Intel core i5-8350U@1.70GHz workstation running MATLAB R2018a.

\subsection{Image deblurring}
\label{subsec:imageDeconvolution}
To examine the performance of the MCMC methods in different scenarios, we consider a deblurring problem with two test images: \textit{cameraman}, and the training image \#$61060$ from the Berkeley Segmentation Dataset and Benchmark \cite{MartinFTM01}, we henceforth refer to this image as \texttt{skier}. Both images have a size of $d=256 \times 256$ pixels. An uniform blur operator $H$ of size $5 \times 5$ is applied to the true image $x \in \mathbb{R}^d$ and then additive Gaussian noise $\eta$ is added with a sigma-to-noise level of $40$dB, to produce an observation $y \in \mathbb{R}^d$ related to the true image by $y= Hx + \eta$. As $H$ is nearly singular, the problem becomes ill-conditioned. So, to promote regularity, we have used the isotropic TV pseudonorm as a prior, given by $\mathrm{TV}(x) =  \sum_i \sqrt{(\Delta_i^h x)^2 + (\Delta_i^v x)^2}$, where $\Delta_i^h$, $\Delta_i^v$ denote horizontal and vertical first-order local diference operators. This leads to the following posterior distributions
\begin{gather} \label{eqn:deconvolution_posterior_dist_nonAugm}
    p (x|y,\theta) \propto \exp \left[ -\| y - Hx \|^2 / 2\sigma^2 - \theta \mathrm{TV}(x) \right] \\
\label{eqn:deconvolution_posterior_dist_augm}
    p (x,z|y,\theta,\rho^2) \propto \exp \left[ -\| y - Hx \|^2 / 2\sigma^2 - \theta \mathrm{TV}(z) - \Vert x - z \Vert^2 / 2 \rho^2 \right] ,
\end{gather}
where $f_{y}(x)=\| y - Hx \|^2 / 2\sigma^2$ and $g(x) = \mathrm{TV}(x)$. Figures \ref{fig:test_images_deconvolution}(a), (b) show the test images for this experiment and Figures \ref{fig:test_images_deconvolution}(c), (d) show the corresponding observations $y$ for each image.

\begin{table}[htb]
{\footnotesize
  \caption{Values for $\theta$ and $\rho^2$ estimated using Algorithm \ref{alg:SAPG_augm_model} for (\ref{eqn:deconvolution_posterior_dist_nonAugm}) and (\ref{eqn:deconvolution_posterior_dist_augm}) in the image deblurring experiments, together with the corresponding Lipschitz constants $L$ and $L_a$.} \label{tab:parameterValues_deconvolution}
\begin{center}
  \begin{tabular}{|c|c|c|c|c|c|} \hline
   \bf Experiment & \bf $\theta$ & \bf $\rho^2$ & \bf $\sigma^2$ & \bf $L = 1/\lambda + 1/\sigma^2$ & \bf $L_{a} = 1/\lambda + (\sigma^2 + \rho^2)^{-1}$ \\ \hline
    cameraman & $0.044$ & $0.480$ & $0.335$ & $5.959$ & $4.205$ \\ 
    skier 	& $0.044$ & $0.250$ & $0.175$ & $11.440$ & $8.078$ \\
		\hline
  \end{tabular}
\end{center}
}
\end{table}

\begin{figure}[htb]
	\centering
	\subfloat[\scriptsize{cameraman: $\log p(X_n|y,\theta)$}]{
		\includegraphics[scale=.16]{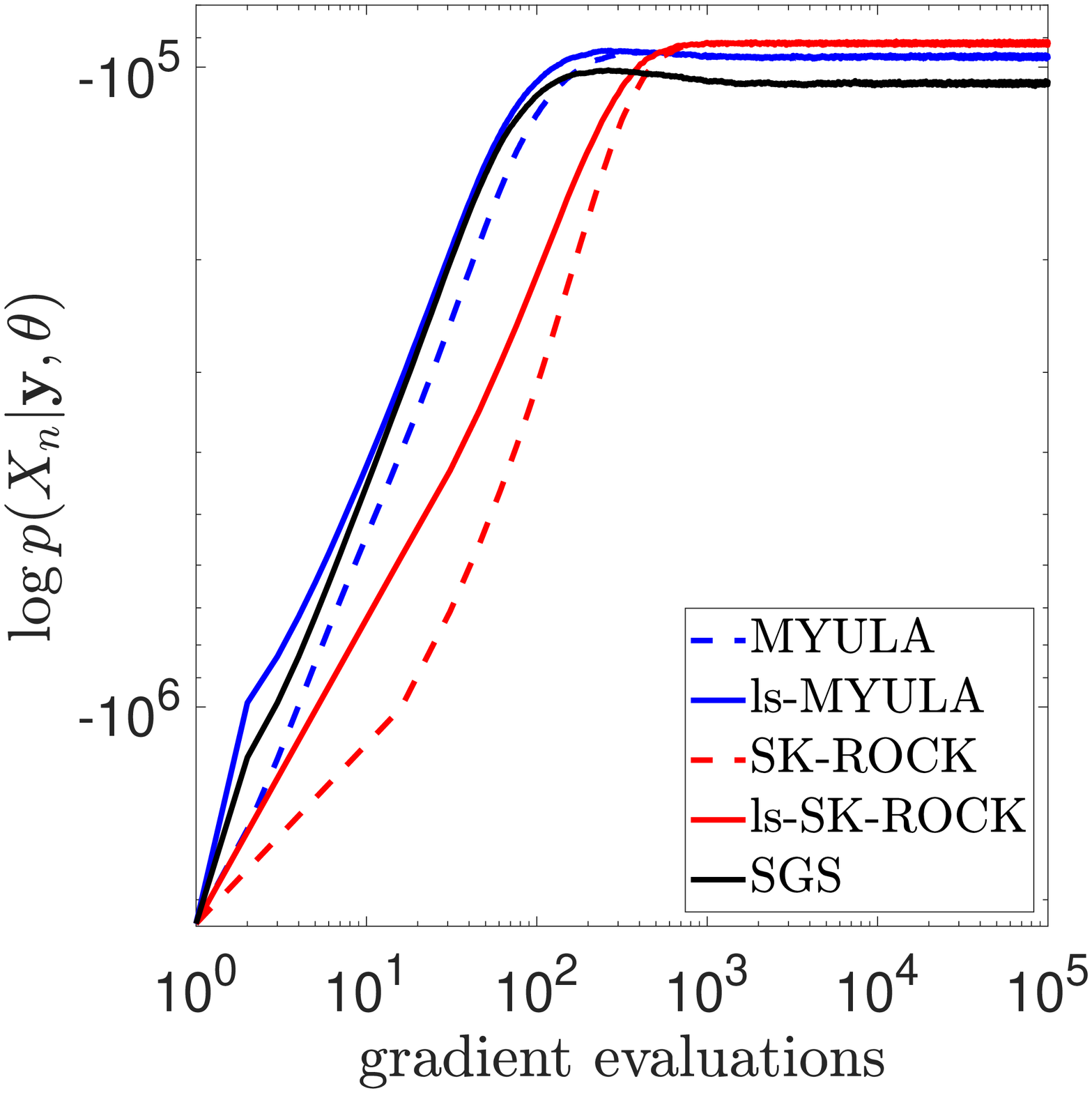}
	}
	\subfloat[\scriptsize{cameraman: MSE}]{
		\includegraphics[scale=.16]{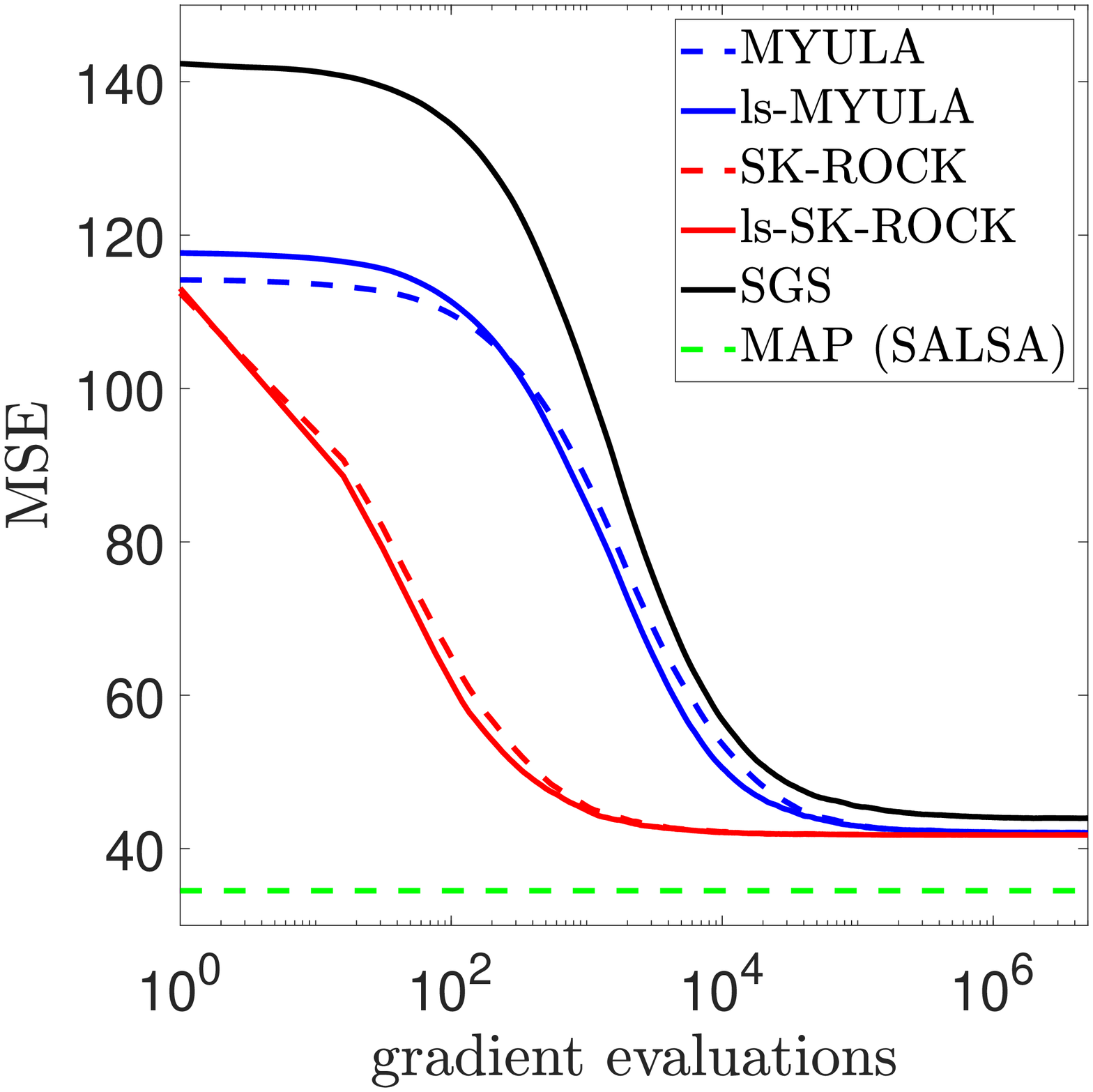}
	}
	\subfloat[\scriptsize{cameraman: ACF}]{
		\includegraphics[scale=.16]{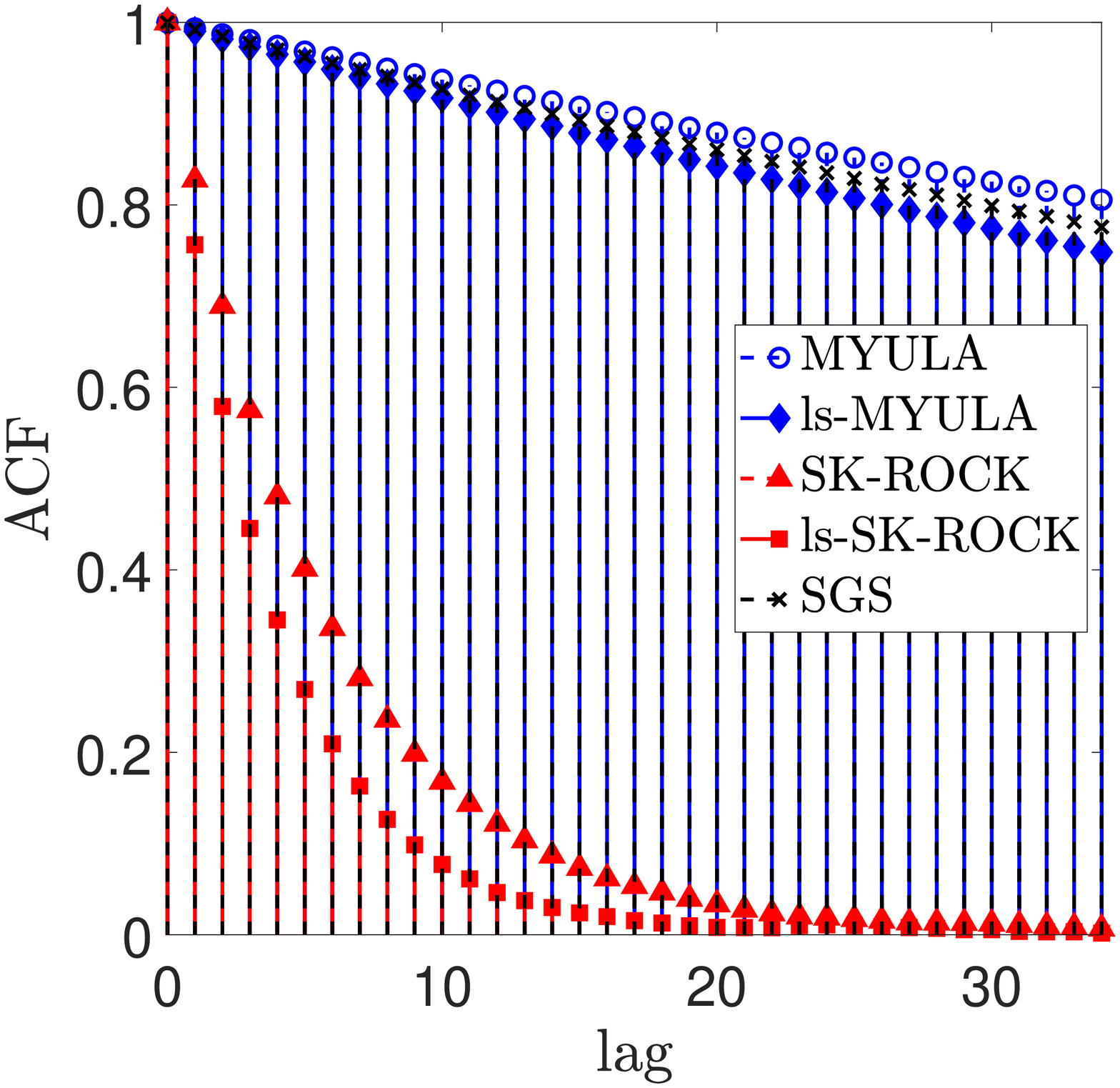}
	} \\
	\subfloat[\scriptsize{skier: $\log p(X_n|y,\theta)$}]{
		\includegraphics[scale=.16]{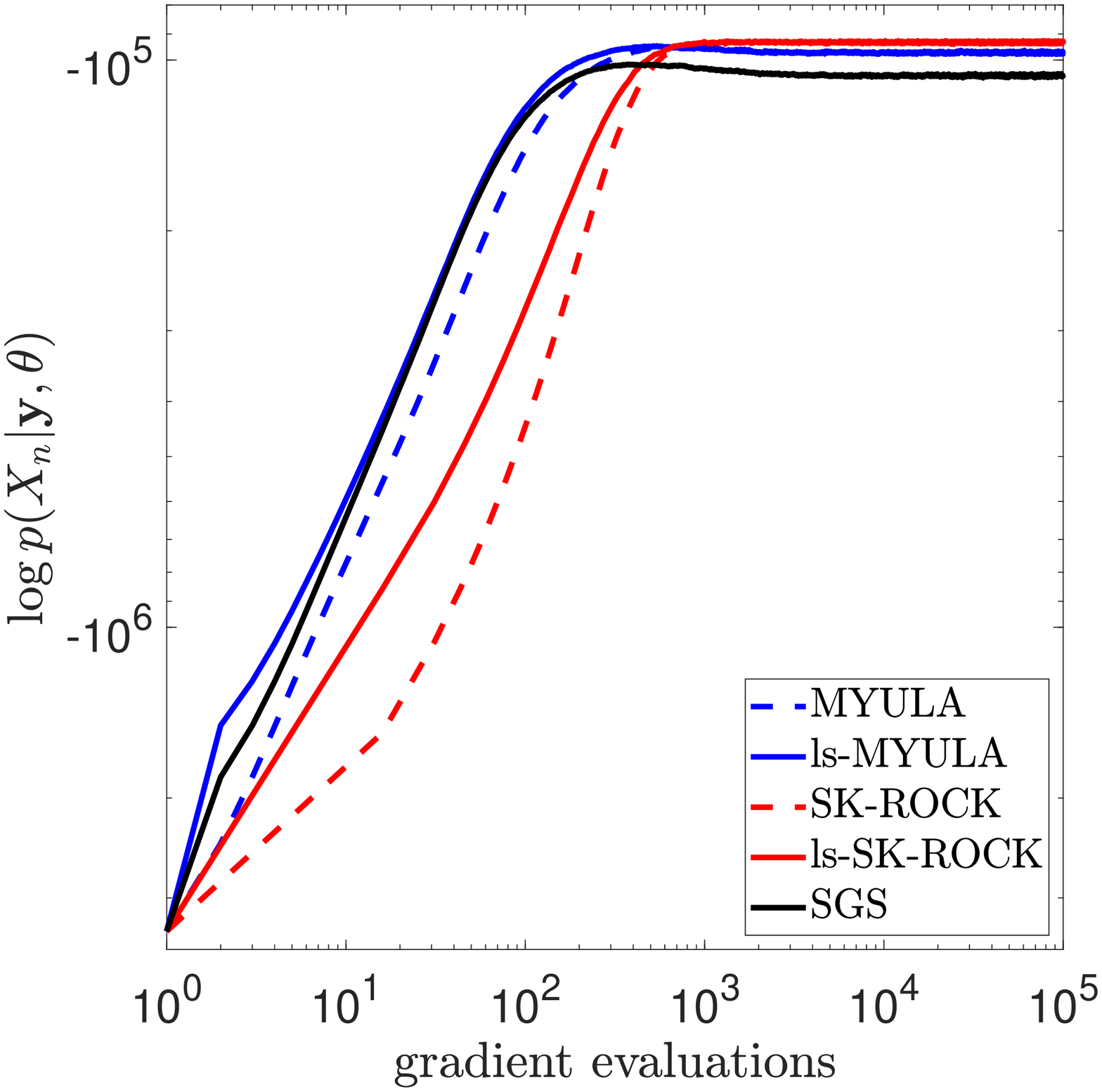}
	}
	\subfloat[\scriptsize{skier: MSE}]{
		\includegraphics[scale=.16]{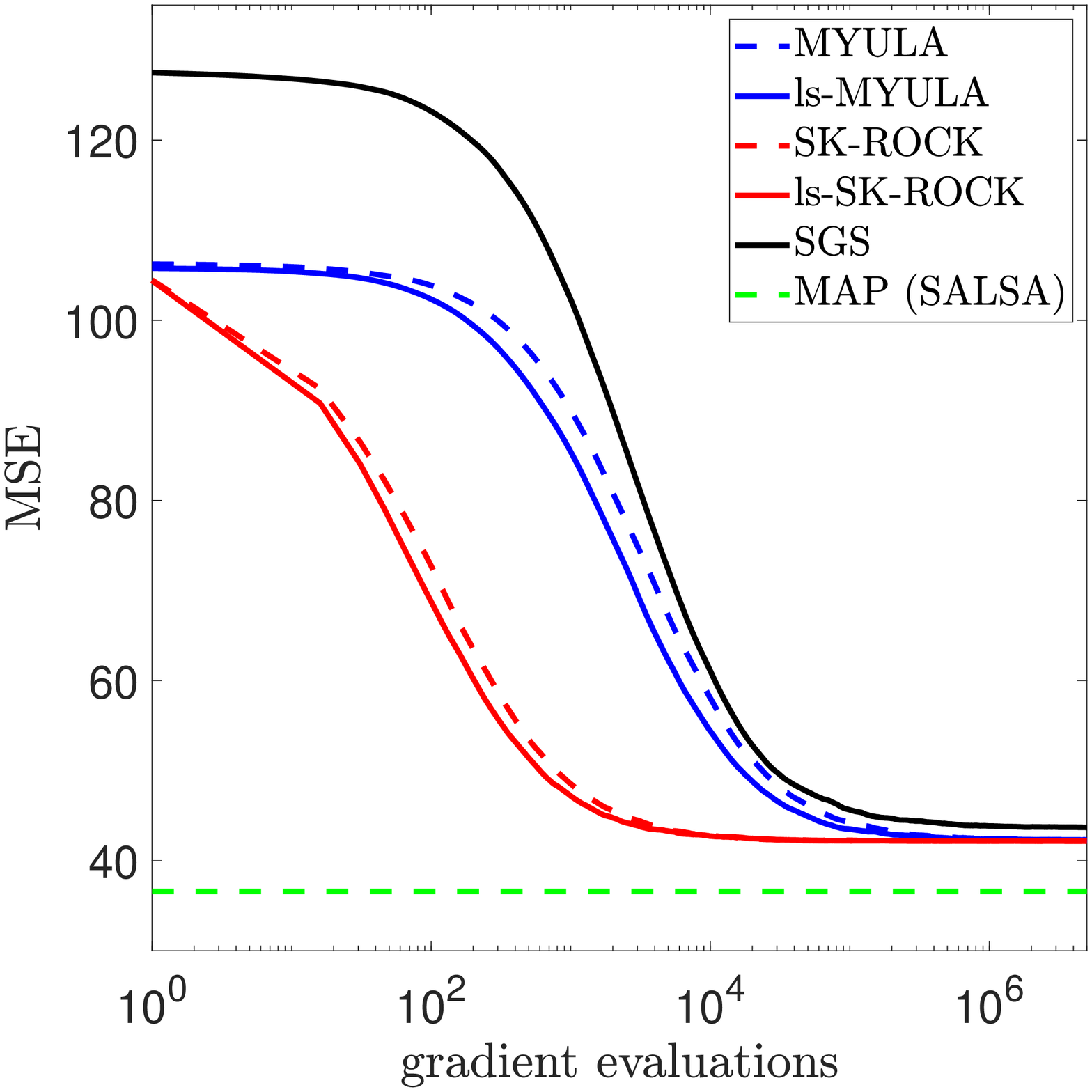}
	}
	\subfloat[\scriptsize{skier: ACF}]{
		\includegraphics[scale=.16]{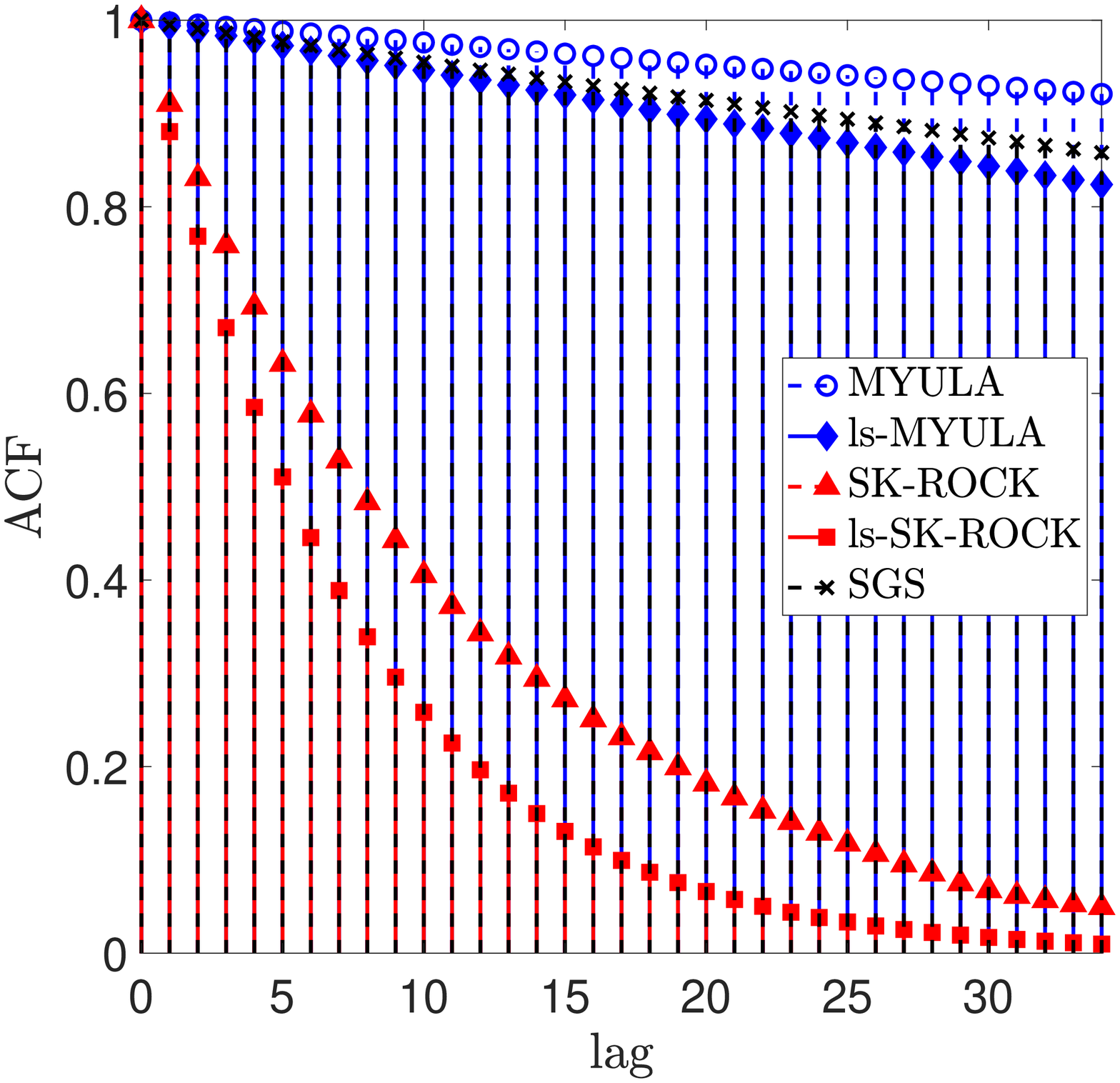}
	} 
	\caption{Image deblurring experiments: {\normalfont(a),(d)} Convergence to the typical set of the posterior distribution (\ref{eqn:deconvolution_posterior_dist_nonAugm}) and (\ref{eqn:deconvolution_posterior_dist_augm}) for the first $10^5$ MYULA, SGS and ls-MYULA samples, and the first $10^5 / s$ SK-ROCK and ls-SK-ROCK samples ($s=15$). {\normalfont(b),(e)} MSE between the mean of the algorithms and the true image, measured using $5 \times 10^6$ MYULA, SGS and ls-MYULA samples, and $5 \times 10^6 / s$ SK-ROCK and ls-SK-ROCK samples ($s=15$), in stationary regime. {\normalfont(c),(f)} Autocorrelation function for the values of the slowest component of the samples.}
	\label{fig:results_deconvolution}
\end{figure}

We begin estimating optimal values for $\theta$ and $\rho^2$ for the given models implementing Algorithm \ref{alg:SAPG_augm_model} setting $\gamma_i = \gamma_i^{\prime} = 10 \times i^{-0.8} / d$, $\theta_0 = 0.04$, $\rho^2_0 = L_f^{-1} = \sigma^2$ and $X_0 = Z_0 = H^{\intercal} y$. The corresponding results for the parameters estimation are given in Table \ref{tab:parameterValues_deconvolution}, together with the Lipschitz constants $L$ and $L_a$ required to sample (\ref{eqn:deconvolution_posterior_dist_nonAugm}) and (\ref{eqn:deconvolution_posterior_dist_augm}) respectively. We then generate  $5 \times 10^6$ samples using MYULA and $5 \times 10^6 / s$ samples using SK-ROCK (with $s=15$) from (\ref{eqn:deconvolution_posterior_dist_nonAugm}), and $5 \times 10^6$ samples using SGS and ls-MYULA and $5 \times 10^6 / s$ samples using ls-SK-ROCK (with $s=15$) from (\ref{eqn:deconvolution_posterior_dist_augm}). The results of these experiments are plotted in \ref{fig:results_deconvolution}. We note from the evolution of the MSE (when the chains have reached the typical set of the target distributions) that ls-MYULA and ls-SK-ROCK outperform SGS in terms of the convergence speed of the posterior mean. The improvement of ls-MYULA w.r.t. SGS illustrates the benefits of using an exact MYULA implementation rather than a noisy one, as shown in Section \ref{sec:SGS_noisy_MYULA_new_MCMC}. The minor improvements between ls-MYULA and MYULA, and between SK-ROCK and ls-SK-ROCK, are due to the effect of $\rho^2 >0$, which does not have a significant impact on the estimation of the posterior mean. The step-sizes used by each method are reported in Table \ref{tab:stepsize_s_deconvolution}.

\begin{table}[htb]
	{\footnotesize
		\caption{Image deblurring experiment: Summary of the values for the step-size $\delta$ for each of the MCMC methods applied to the two imaging experiments: cameraman and skier.}  \label{tab:stepsize_s_deconvolution}
		\begin{center}
			\begin{tabular}{|l|c|c|} \hline
			\bf MCMC method & \bf Cameraman & \bf Skier \\ \hline
			\bf MYULA & $0.167$ & $0.087$ \\
			\bf SK-ROCK ($s=15$) & $67.959$ & $35.402$ \\
			\bf SGS & $0.237$ & $0.124$ \\
			\bf ls-MYULA & $0.237$ & $0.124$ \\
			\bf ls-SK-ROCK ($s=15$) & $96.294$ & $50.161$ \\ \hline
			\end{tabular}
		\end{center}
	}
\end{table}

\begin{table}[htbp]
	{\footnotesize
		\caption{Image deblurring experiments: Effective sample sizes of the slowest component, after generating $15 \times 10^3$ samples using the five algorithms discussed in this work, and the speed increase (i.e., speed-up) achieved by the algorithms w.r.t. MYULA.}  \label{tab:ess_deblurring}
		\begin{center}
			
			\begin{tabular}{|c||c|c||c|c||c|c||c|c||c|c|} \hline
				\multirow{2}{*}{} & \multicolumn{2}{c||}{\bf MYULA} & \multicolumn{2}{c||}{\bf SK-ROCK} & \multicolumn{2}{c||}{\bf SGS} & \multicolumn{2}{c||}{\bf ls-MYULA} & \multicolumn{2}{c|}{\bf ls-SK-ROCK} \\ \cline{2-11}
				& ESS & Speed-up & ESS & Speed-up & ESS & Speed-up & ESS & Speed-up & ESS & Speed-up \\ \hline
				Cam. & $46$ & - & $1175$ & $25.54$ & $49$ & $1.07$ & $54$ & $1.17$ & $1604$ & $34.87$ \\ 
				Skier & $18$ & - & $636$ & $35.33$ & $35$ & $1.94$ & $37$ & $2.06$ & $924$ & $51.33$  \\ \hline
			\end{tabular}
		\end{center}
	}
\end{table}

\begin{figure}[!htb]
	\centering
	\subfloat[MYULA]{
		\includegraphics[scale=.35]{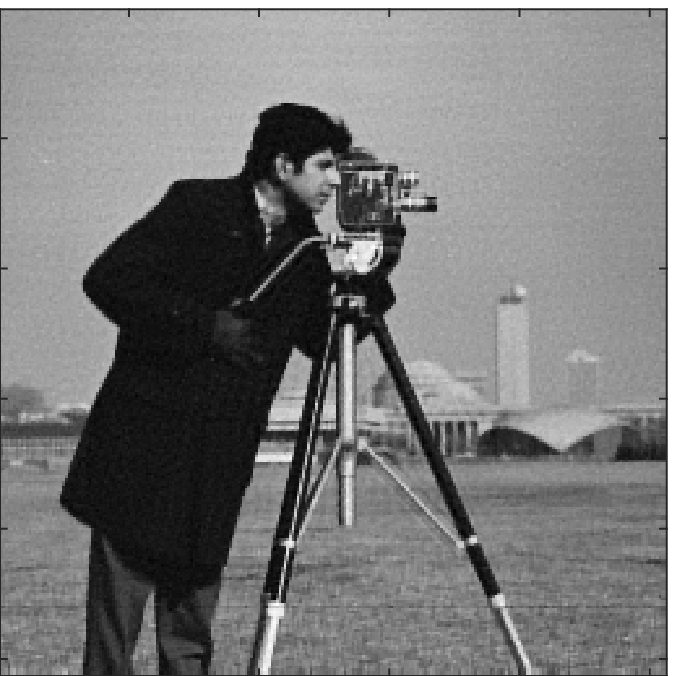}
	}
	\subfloat[SK-ROCK]{
		\includegraphics[scale=.35]{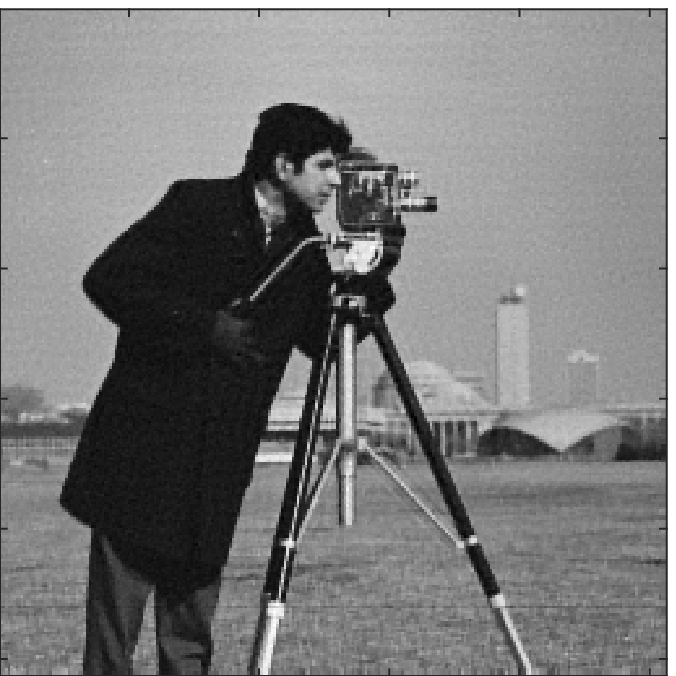}
	}
	\subfloat[SGS]{
		\includegraphics[scale=.35]{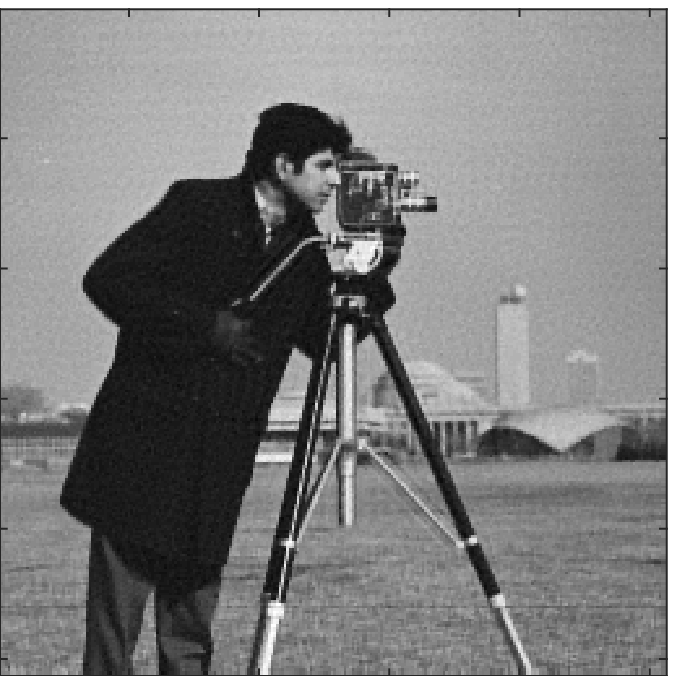}
	}
	\subfloat[ls-MYULA]{
		\includegraphics[scale=.35]{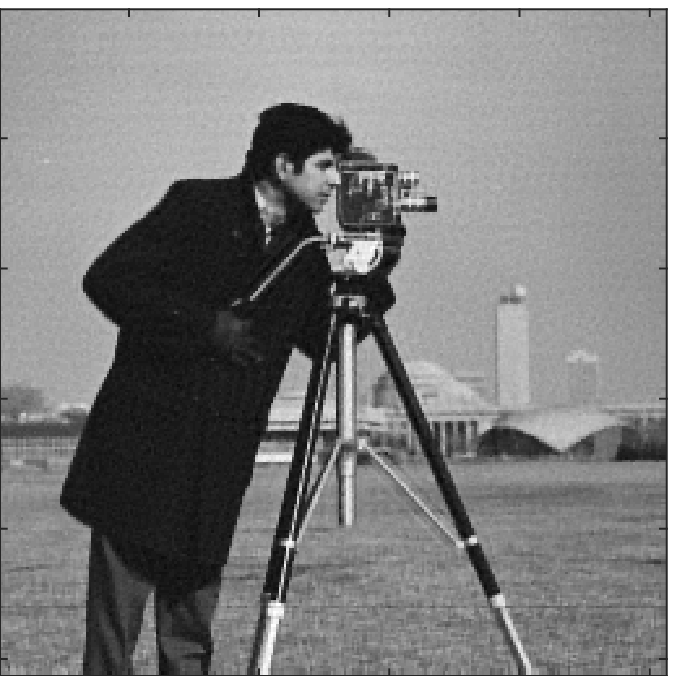}
	}
	\subfloat[ls-SK-ROCK]{
		\includegraphics[scale=.35]{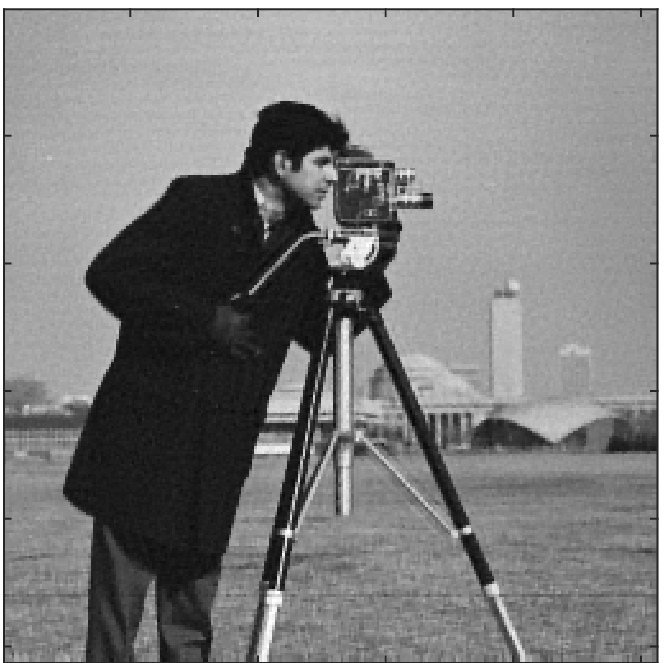}
	} \\
	\subfloat[MYULA]{
		\includegraphics[scale=.35]{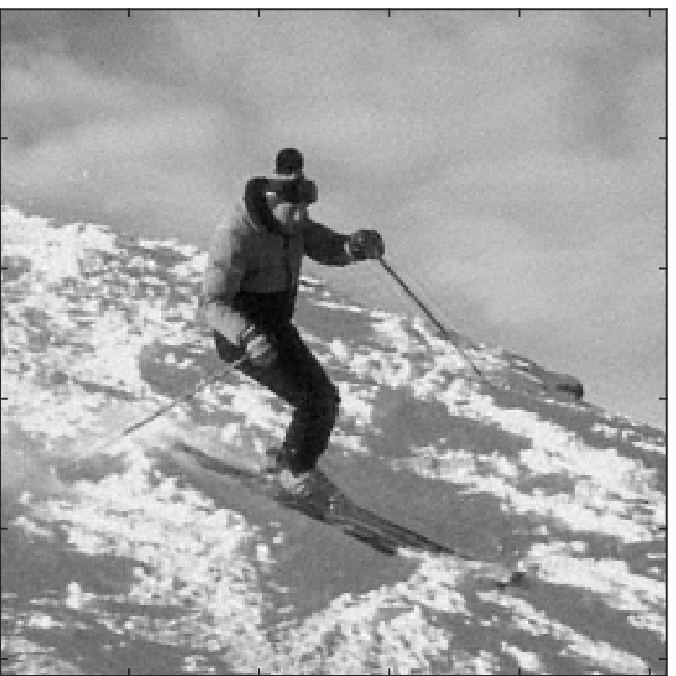}
	}
	\subfloat[SK-ROCK]{
		\includegraphics[scale=.35]{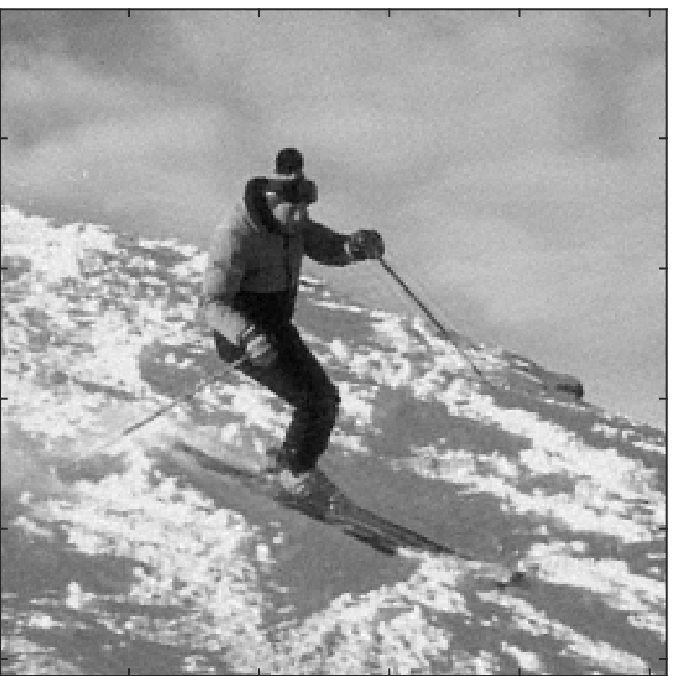}
	}
	\subfloat[SGS]{
		\includegraphics[scale=.35]{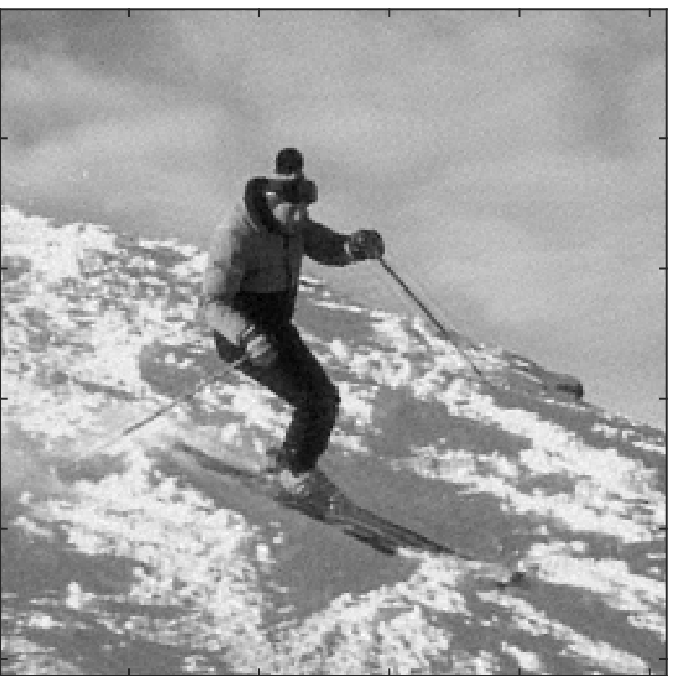}
	}
	\subfloat[ls-MYULA]{
		\includegraphics[scale=.35]{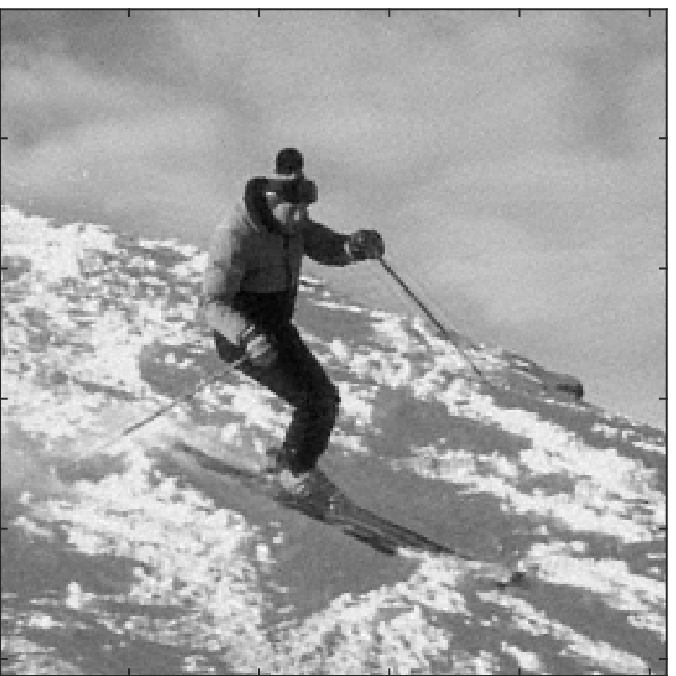}
	}
	\subfloat[ls-SK-ROCK]{
		\includegraphics[scale=.35]{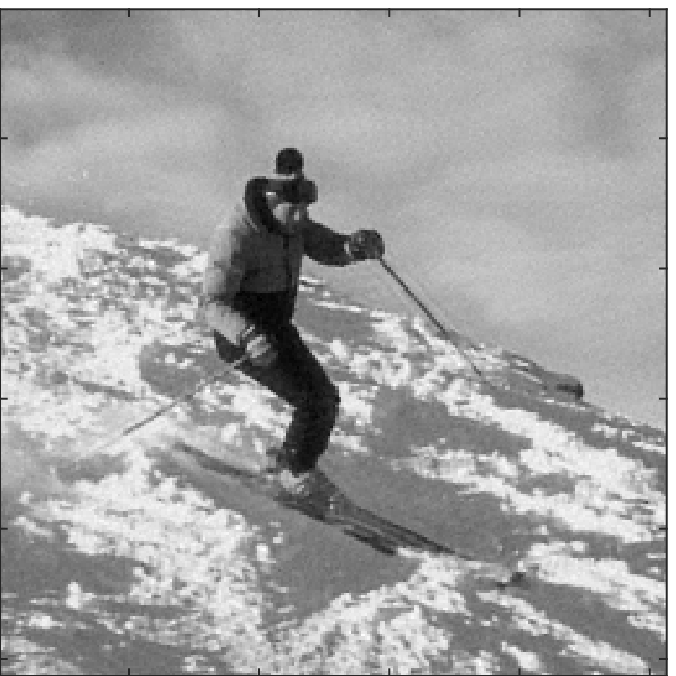}
	} 
	\caption{Image deblurring experiments: MMSE computed using $5 \times 10^6$ MYULA, SGS and ls-MYULA samples, and $5 \times 10^6 / s$ SK-ROCK and ls-SK-ROCK samples, in stationarity.}
	\label{fig:mmse_deconvolution}
\end{figure}

\begin{figure}[htb]
	\centering
	\subfloat[MYULA]{
		\includegraphics[scale=.35]{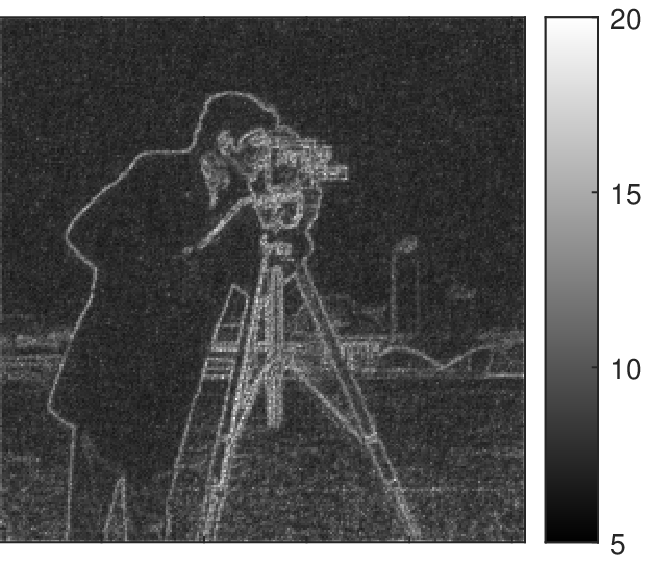}
	}
	\subfloat[SK-ROCK]{
		\includegraphics[scale=.35]{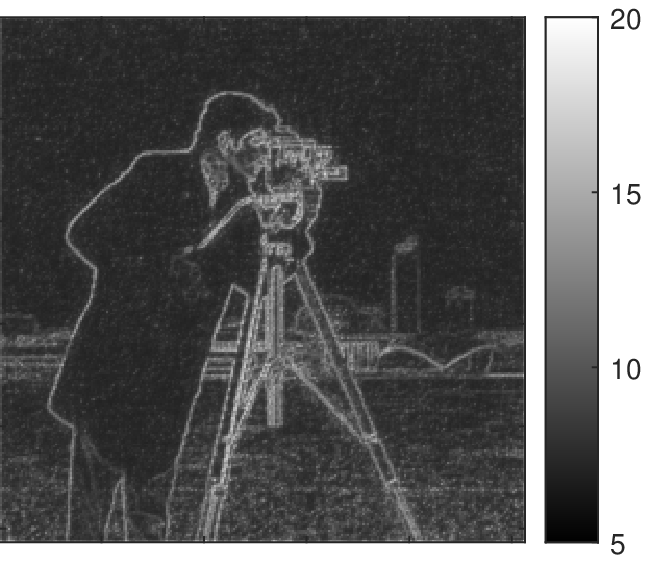}
	}
	\subfloat[SGS]{
		\includegraphics[scale=.35]{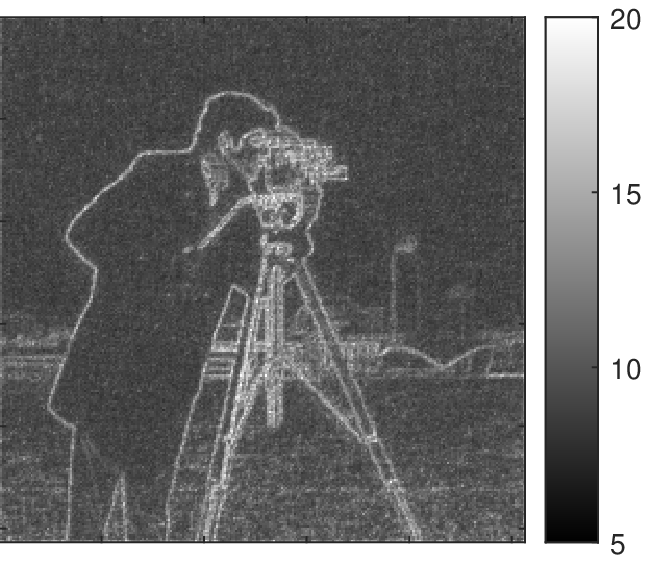}
	}
	\subfloat[ls-MYULA]{
		\includegraphics[scale=.35]{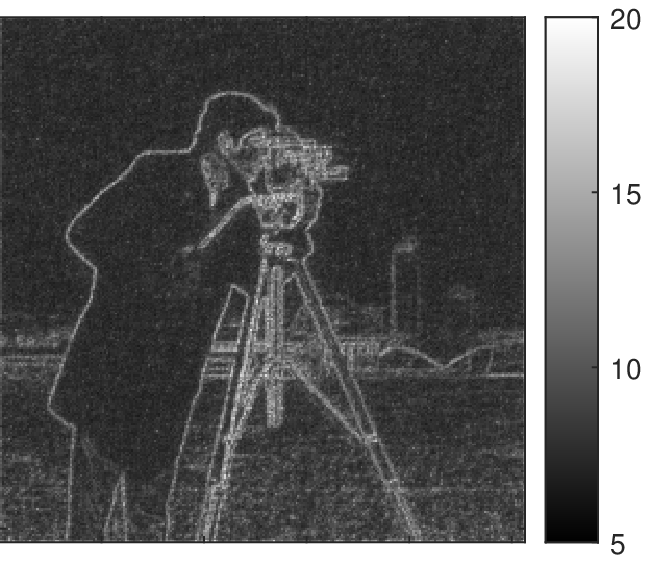}
	}
	\subfloat[ls-SK-ROCK]{
		\includegraphics[scale=.35]{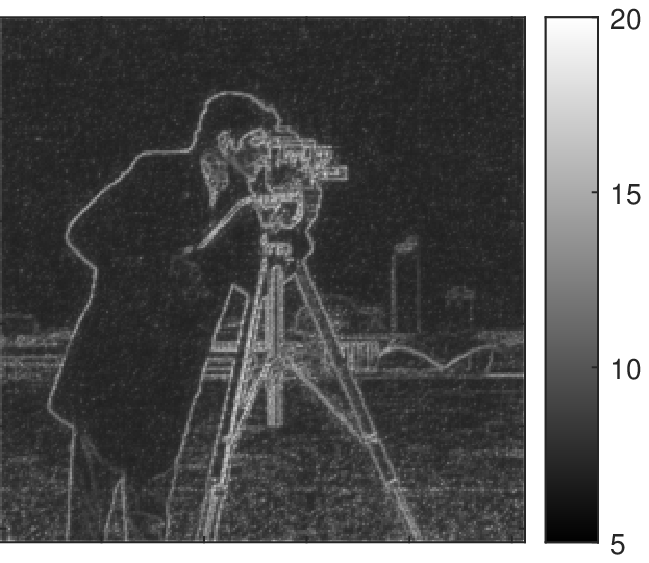}
	} \\
	\subfloat[MYULA]{
		\includegraphics[scale=.35]{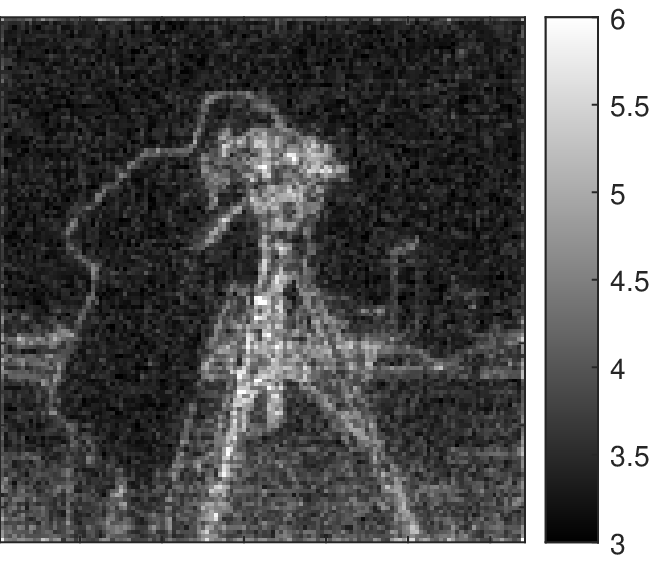}
	}
	\subfloat[SK-ROCK]{
		\includegraphics[scale=.35]{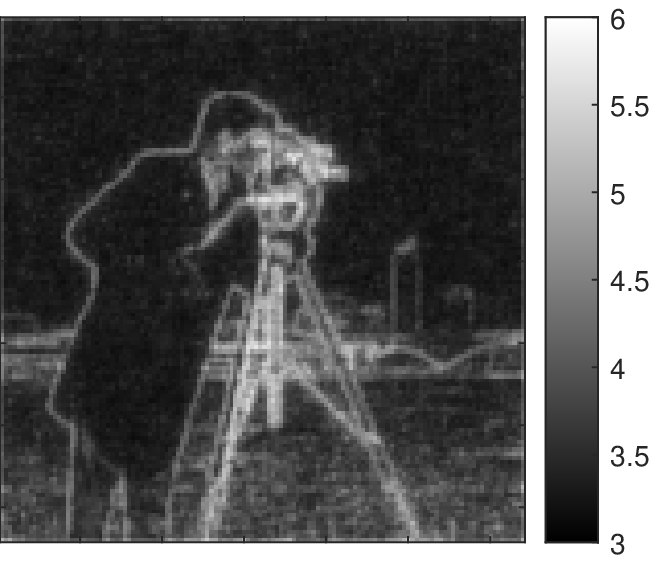}
	}
	\subfloat[SGS]{
		\includegraphics[scale=.35]{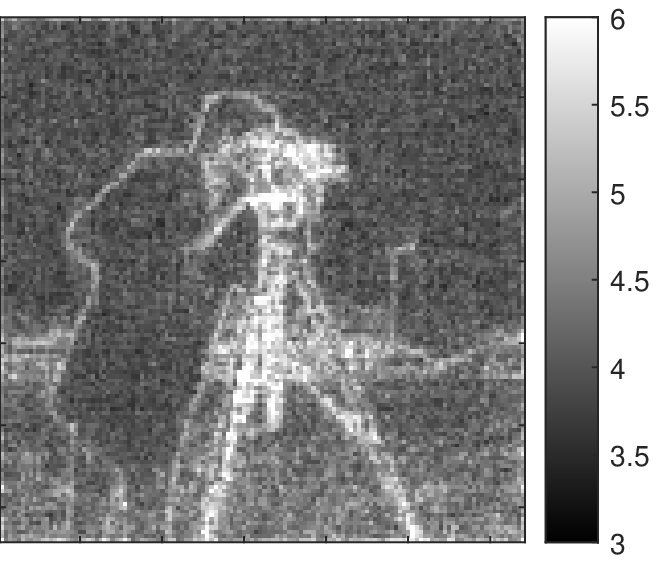}
	}
	\subfloat[ls-MYULA]{
		\includegraphics[scale=.35]{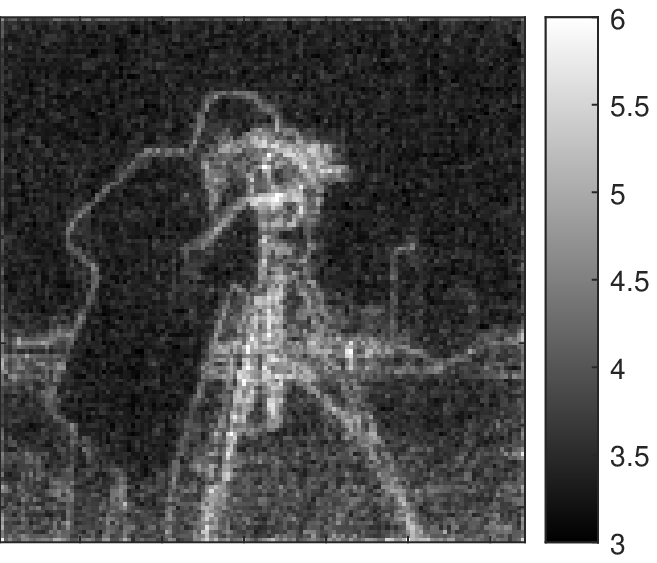}
	}
	\subfloat[ls-SK-ROCK]{
		\includegraphics[scale=.35]{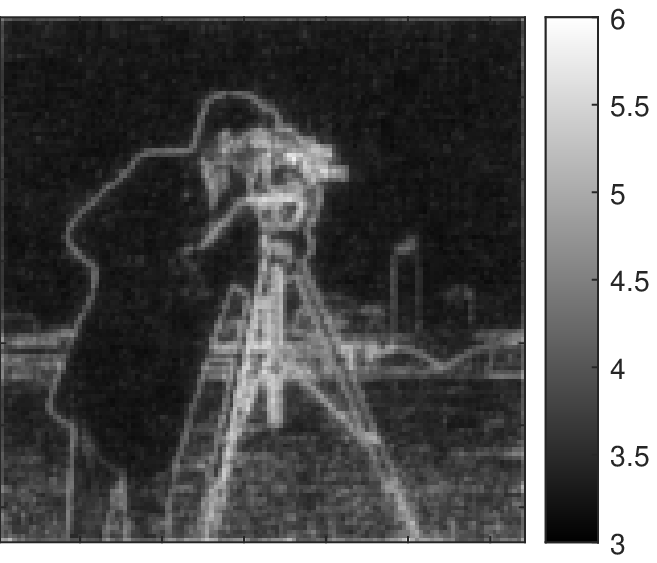}
	} \\
	\subfloat[MYULA]{
		\includegraphics[scale=.35]{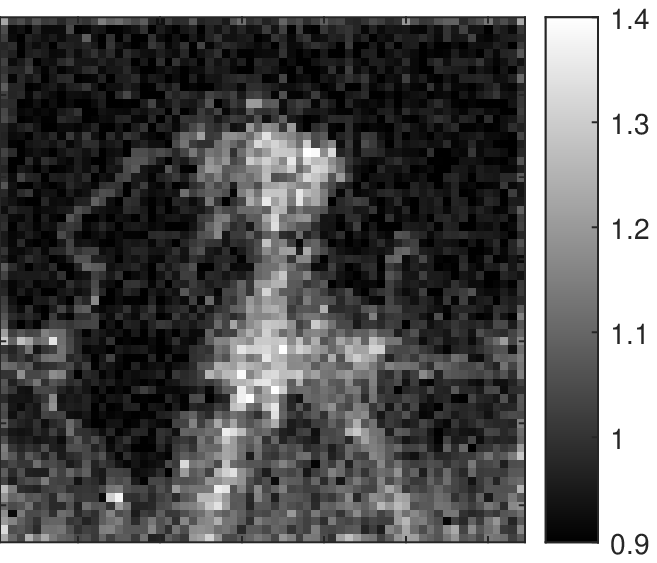}
	}
	\subfloat[SK-ROCK]{
		\includegraphics[scale=.35]{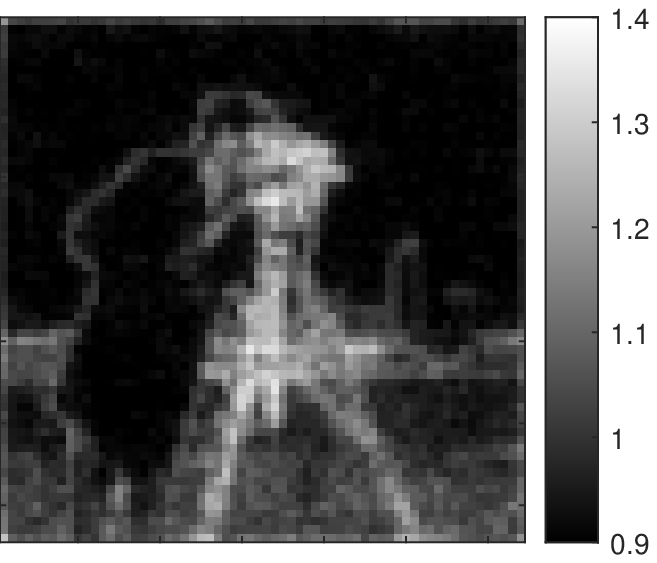}
	}
	\subfloat[SGS]{
		\includegraphics[scale=.35]{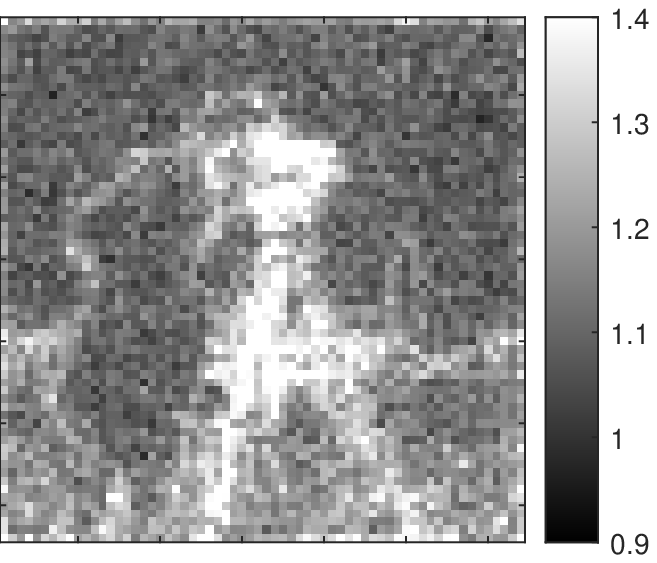}
	}
	\subfloat[ls-MYULA]{
		\includegraphics[scale=.35]{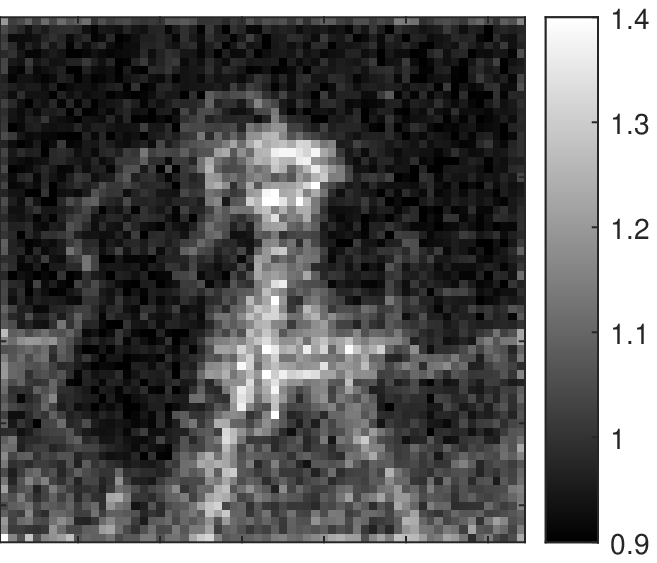}
	}
	\subfloat[ls-SK-ROCK]{
		\includegraphics[scale=.35]{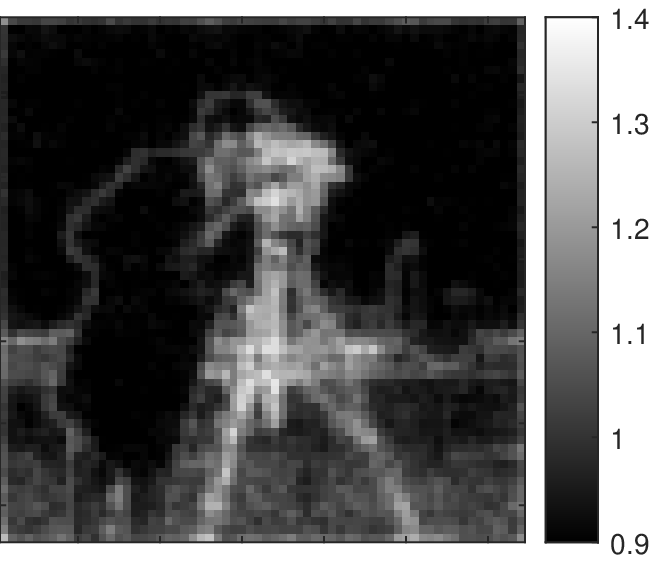}
	} \\
	\subfloat[MYULA]{
		\includegraphics[scale=.35]{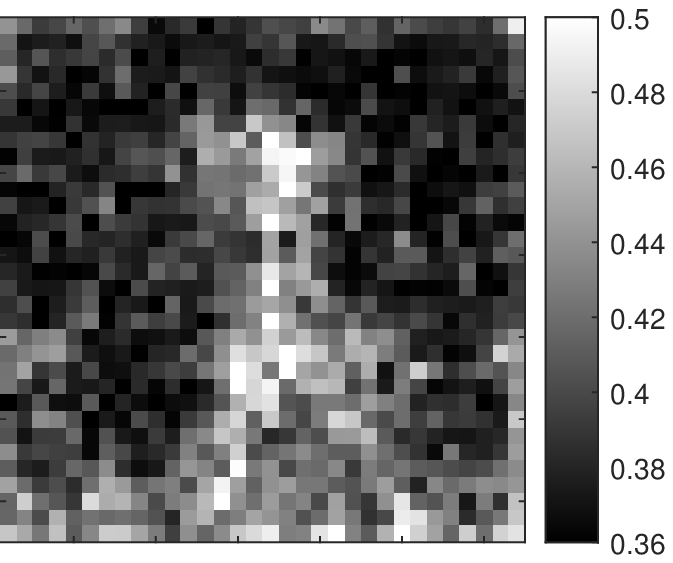}
	}
	\subfloat[SK-ROCK]{
		\includegraphics[scale=.35]{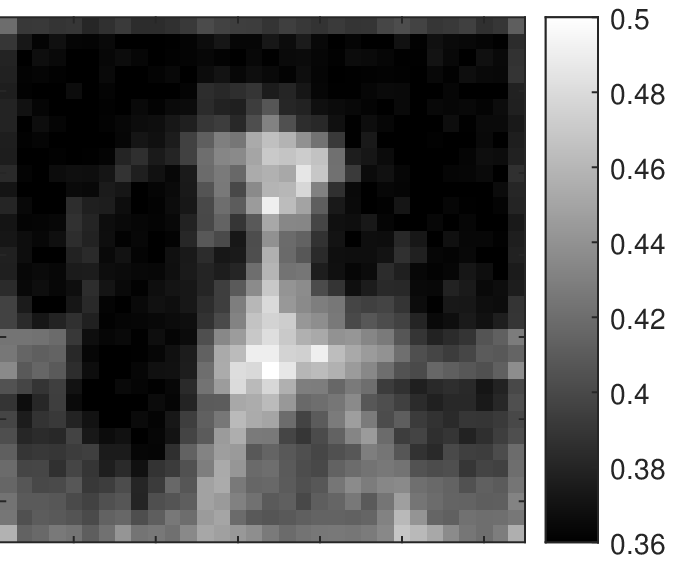}
	}
	\subfloat[SGS]{
		\includegraphics[scale=.35]{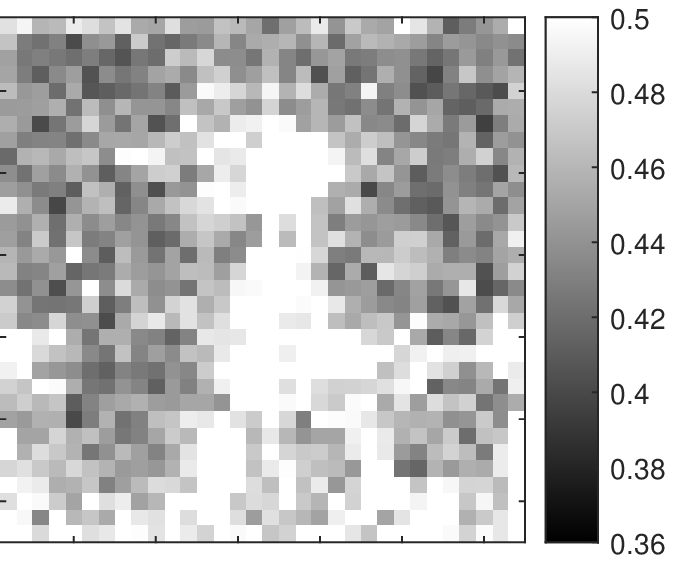}
	}
	\subfloat[ls-MYULA]{
		\includegraphics[scale=.35]{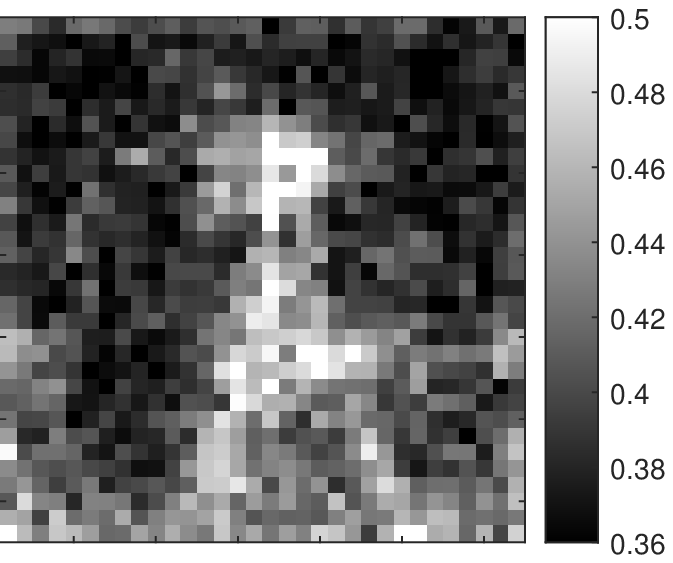}
	}
	\subfloat[ls-SK-ROCK]{
		\includegraphics[scale=.35]{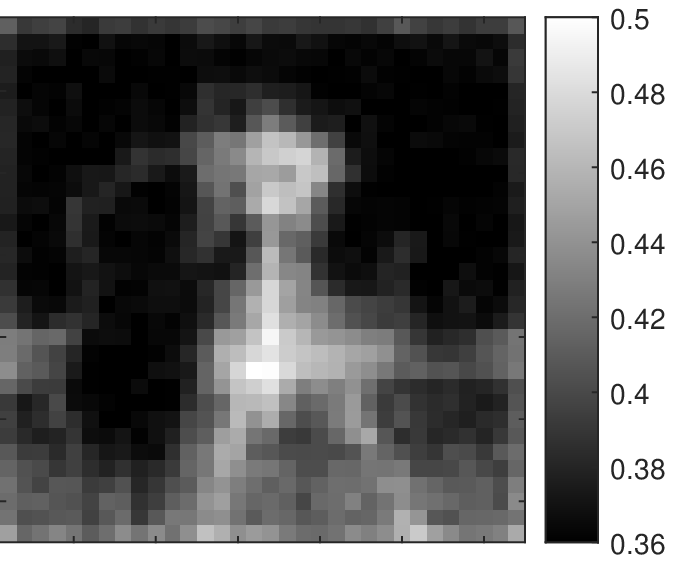}
	}
	\caption{Image deblurring experiments - cameraman: pixel-wise standard deviation computed using $10^4$ MYULA, SGS and ls-MYULA samples with a 1-in-15 thinning, and $10^4$ SK-ROCK and ls-SK-ROCK samples with $s=15$, using (a)-(e) the original sample size ($256 \times 256$) and with downsampling by a factor of (f)-(j) 2, (k)-(o) 4 and (p)-(t) 8.}
	\label{fig:std_dev_deblurring}
\end{figure}

We also plot the autocorrelation function of the slowest component from the chains of the MCMC algorithms, this is shown in Figure \ref{fig:results_deconvolution}(c),(f) and, as can be seen, ls-SK-ROCK presents the fastest decay. In addition, Table \ref{tab:ess_deblurring} shows the effective sample sizes (ESS) associated with these autocorrelation plots, and one can notice that ls-SK-ROCK reaches the largest ESS. The comparison of the autocorrelation function and ESS for the slowest mixing component for MYULA and ls-MYULA illustrates the benefits of operating on the latent space (and the effect of $\rho^2 > 0$). A similar comparison between ls-MYULA and SGS illustrates the efficiency cost that SGS incurs due to the use of an inexact gradient. For completeness, we also illustrated in Figure \ref{fig:mmse_deconvolution} the minimum mean-square estimator (MMSE) of all the MCMC methods for all two deblurring experiments.

Finally, Figure \ref{fig:std_dev_deblurring} shows the marginal posterior standard deviation of the cameraman deblurring experiment at different scales. One can notice that edges show higher uncertainty, which is expected due to the nature of the forward operator. As can be seen, ls-MYULA and, in particular, ls-SK-ROCK outperform SGS in terms of delivering comparable estimates in less computational time, showing the benefit of using these algorithms to sample in a more efficient way the augmented posterior distribution.

\subsection{Image inpainting}
\begin{figure}[htb]
	\centering
	\subfloat[\scriptsize{\textit{cameraman}: observation $y$}]{
		\includegraphics[scale=.45]{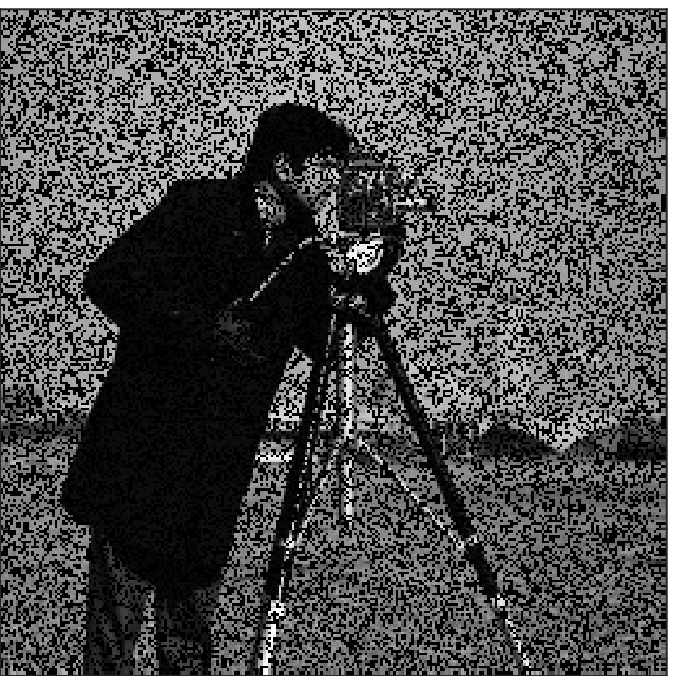}
	}
	\subfloat[\scriptsize{\textit{skier}: observation $y$}]{
		\includegraphics[scale=.45]{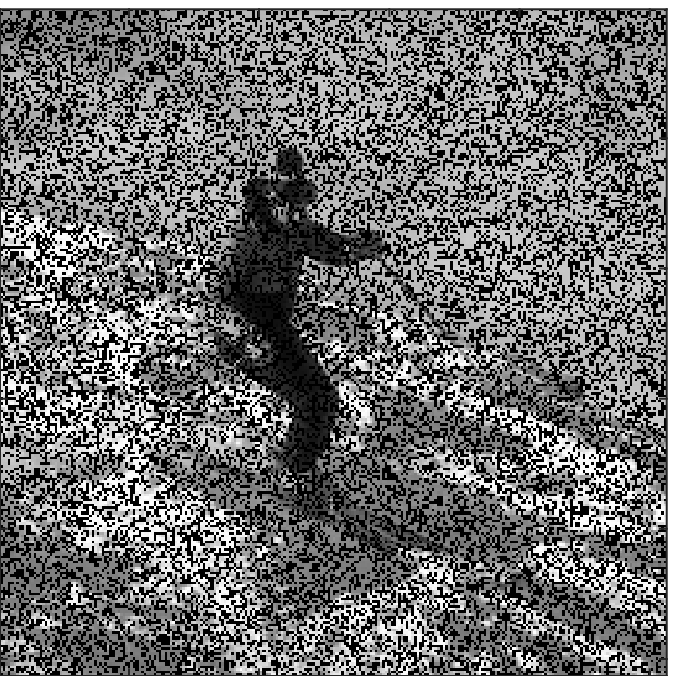}
	}
	\caption{Image inpainting experiments: noisy and incomplete observations $y$ (pixels in black represent unobserved components).}
	\label{fig:obs_images_inpainting}
\end{figure}

We now perform an image inpainting experiment, which consists of randomly selecting $60 \%$ of the image pixels $x \in \mathbb{R}^d$ to form the observation vector $y \in \mathbb{R}^m$ ($m < d$) and then adding Gaussian noise with a SNR level of $40$dB (the observation images $y$ are illustrated in Figure \ref{fig:obs_images_inpainting}). To test the MCMC methods in different regimes, we will use the same two test images given in Section \ref{subsec:imageDeconvolution} (\texttt{cameraman} and \texttt{skier}) and illustrated in Figure \ref{fig:test_images_deconvolution}(a)-(b). For this experiment, we consider the following models
\begin{gather}
\label{eqn:inpainting_posterior_dist_nonAugm}
    p (x|y,\theta) \propto \exp \left[ -\| y - Ax \|^2 / 2\sigma^2 - \theta \mathrm{TV}(x) \right] \\
\label{eqn:inpainting_posterior_dist_augm}
    p (x,z|y,\theta,\rho^2) \propto \exp \left[ -\| y - Ax \|^2 / 2\sigma^2 - \theta \mathrm{TV}(z) - \Vert x - z \Vert^2 / 2 \rho^2 \right] ,
\end{gather}
where $f_{y}(x)=\| y - Ax \|^2 / 2\sigma^2$, $A \in \mathbb{R}^{m \times d}$ is a rectangular matrix obtained by taking a random subset of rows from the identity matrix in dimension $d$, and $g(x) = \mathrm{TV}(x)$, previously defined in Section \ref{subsec:imageDeconvolution}.

We first proceed to estimate optimal hyperparameters $\theta$ and $\rho^2$ for (\ref{eqn:inpainting_posterior_dist_nonAugm}) and (\ref{eqn:inpainting_posterior_dist_augm}) using Algorithm \ref{alg:SAPG_augm_model} setting $\gamma_i = \gamma_i^{\prime} = 10 \times i^{-0.8} / d$, $\theta_0 = 0.5$, $\rho^2_0 = L_f^{-1} / 2 = \sigma^2 / 2$ and $X_0 = Z_0 = A^{\intercal} y$. The estimated parameter values can be seen in Table \ref{tab:parameterValues_inpainting}, together with the Lipschitz constants $L$ and $L_a$ required to sample (\ref{eqn:inpainting_posterior_dist_nonAugm}) and (\ref{eqn:inpainting_posterior_dist_augm}) respectively.

Having obtained our estimates from SAGP algorithm for the values of the hyperparameters, we proceed to generate $5 \times 10^6$ MYULA samples and $5 \times 10^6 / s$ SK-ROCK samples (with $s=15$) from (\ref{eqn:inpainting_posterior_dist_nonAugm}) and $5\times 10^6$ SGS and ls-MYULA samples, and $5\times 10^6 / s$ ls-SK-ROCK samples (with $s=15$) from (\ref{eqn:inpainting_posterior_dist_augm}). The step-sizes for each method are reported in Table \ref{tab:stepsize_s_inpainting}.

\begin{table}[htb]
{\footnotesize
  \caption{Values for $\theta$ and $\rho^2$ estimated using Algorithm \ref{alg:SAPG_augm_model} for (\ref{eqn:inpainting_posterior_dist_nonAugm}) and (\ref{eqn:inpainting_posterior_dist_augm}) in the image inpainting experiments}  \label{tab:parameterValues_inpainting}
\begin{center}
  \begin{tabular}{|c|c|c|c|c|c|} \hline
   \bf Experiment & \bf $\theta$ & \bf $\rho^2$ & \bf $\sigma^2$ & \bf $L = 1/\lambda + 1/\sigma^2$ & \bf $L_{a} = 1/\lambda + (\sigma^2 + \rho^2)^{-1}$ \\ \hline
    cameraman & $0.058$ & $0.65$ & $0.388$ & $5.146$ & $3.530$ \\ 
    skier 	& $0.052$ & $0.37$ & $0.175$ & $9.071$ & $6.220$ \\
		\hline
  \end{tabular}
\end{center}
}
\end{table}

\begin{figure}[!htb]
	\centering
	\subfloat[\scriptsize{cameraman: $\log p(X_n|y,\theta)$}]{
		\includegraphics[scale=.16]{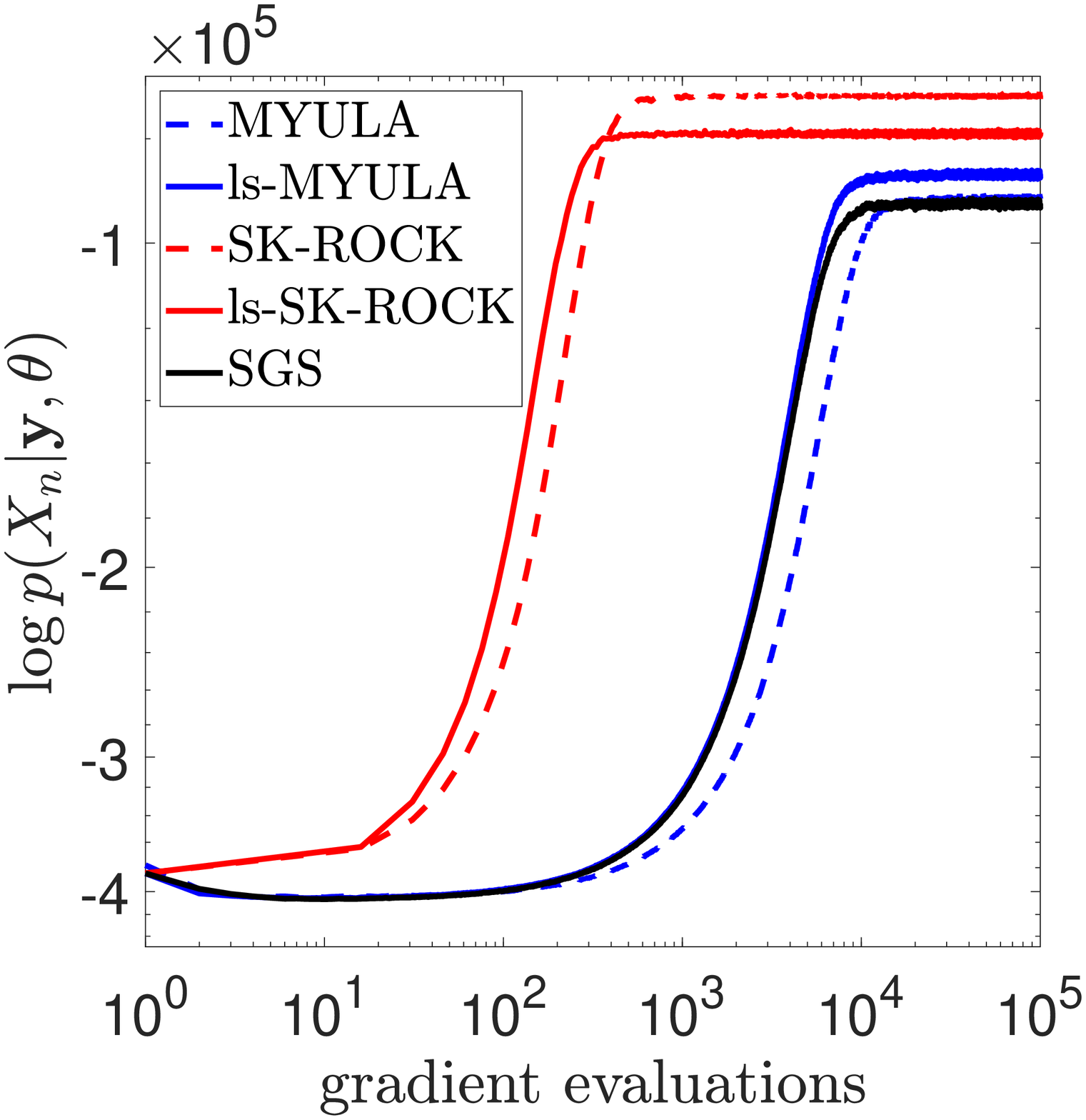}
	}
	\subfloat[\scriptsize{cameraman: MSE}]{
		\includegraphics[scale=.16]{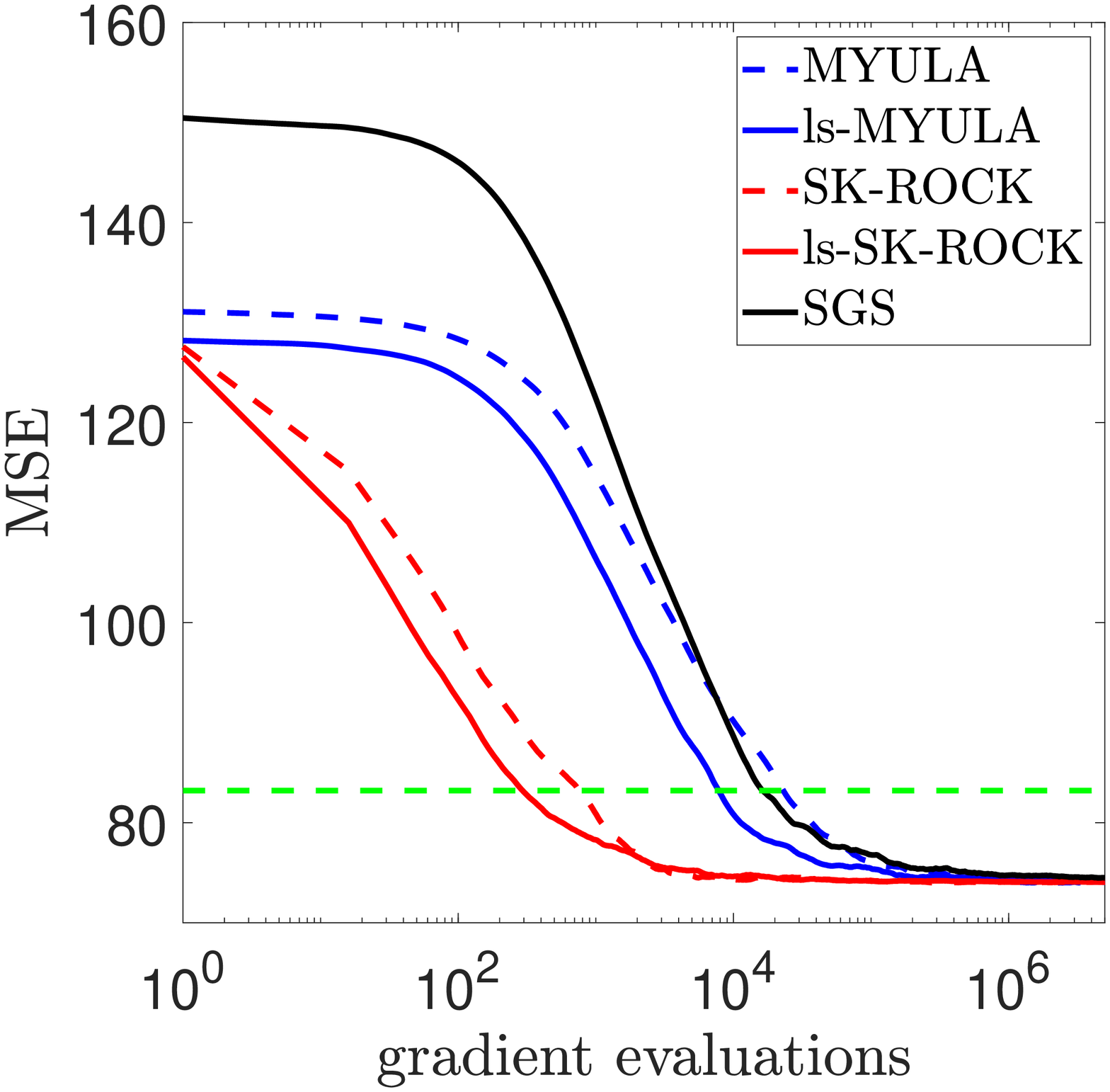}
	}
	\subfloat[\scriptsize{cameraman: ACF}]{
		\includegraphics[scale=.16]{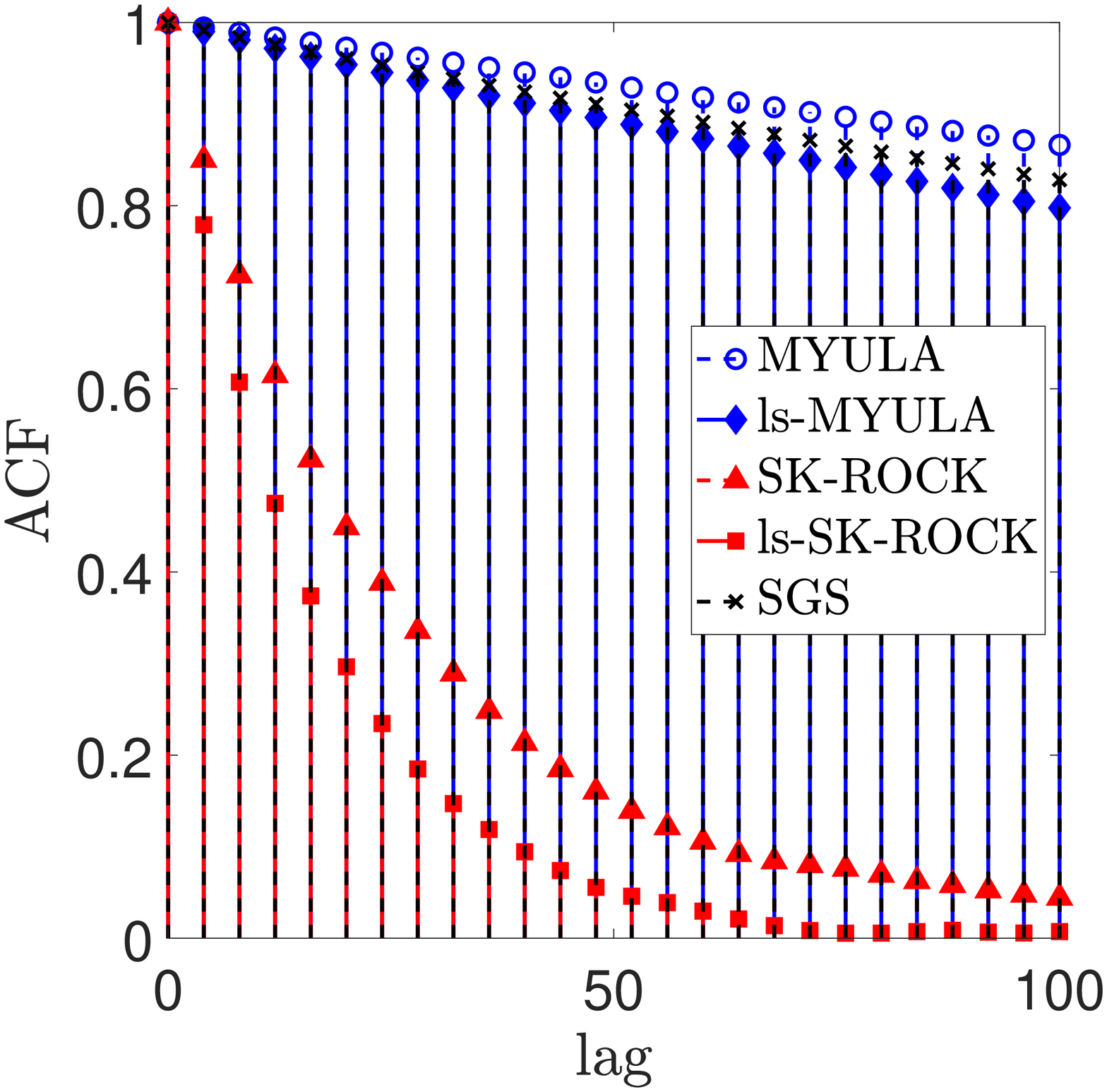}
	} \\
	\subfloat[\scriptsize{skier: $\log p(X_n|y,\theta)$}]{
		\includegraphics[scale=.16]{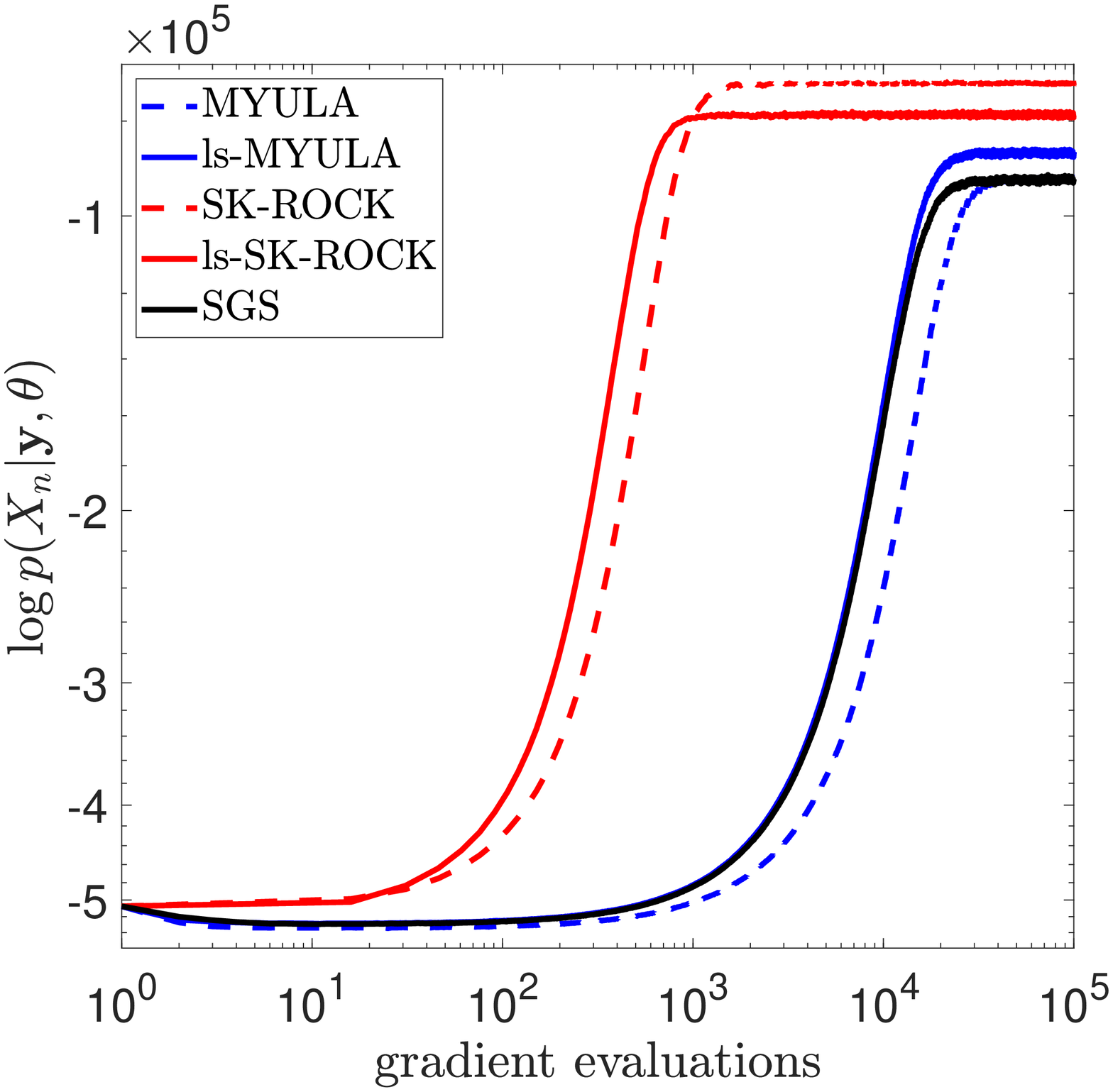}
	}
	\subfloat[\scriptsize{skier: MSE}]{
		\includegraphics[scale=.16]{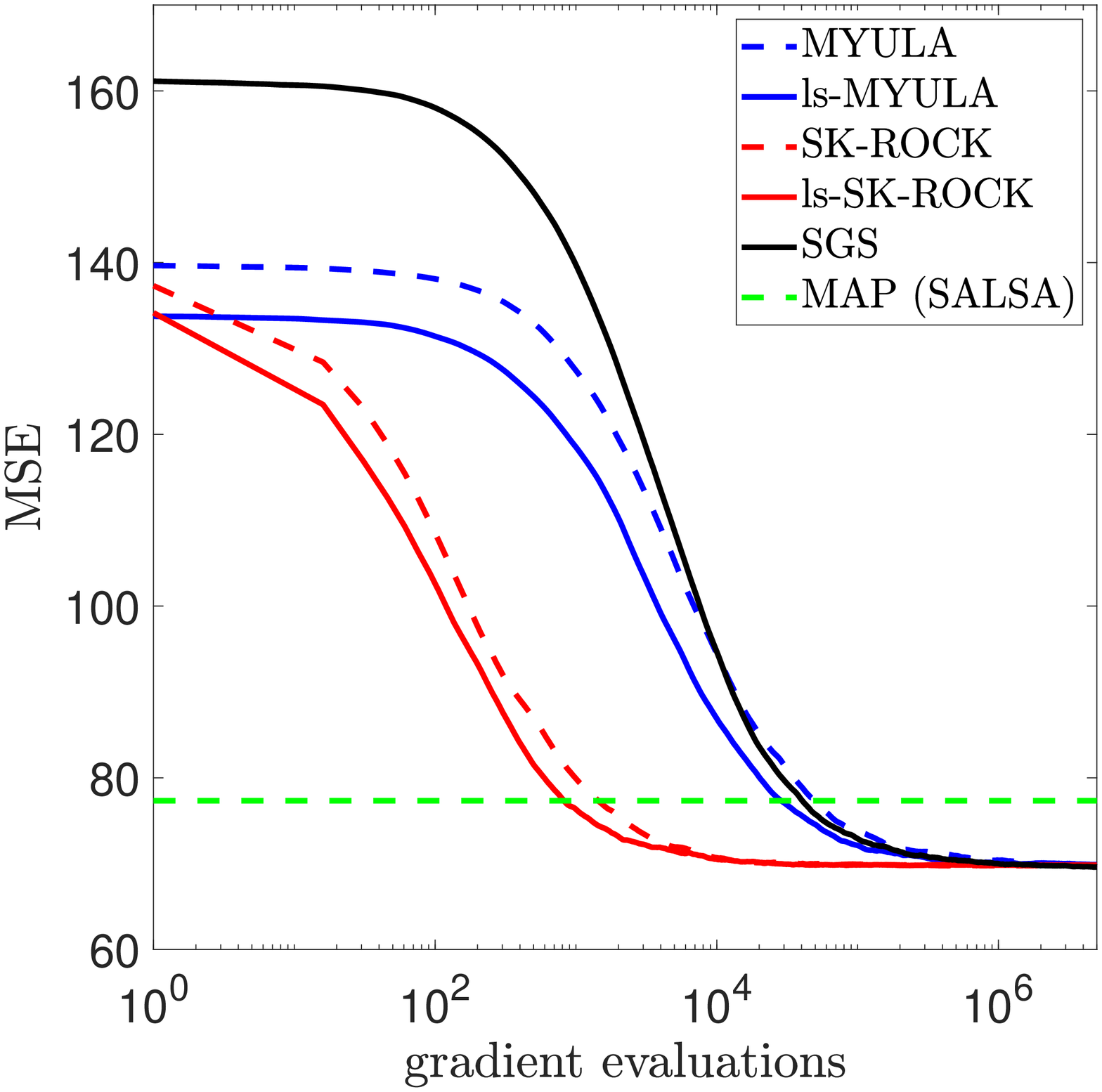}
	}
	\subfloat[\scriptsize{skier: ACF}]{
		\includegraphics[scale=.16]{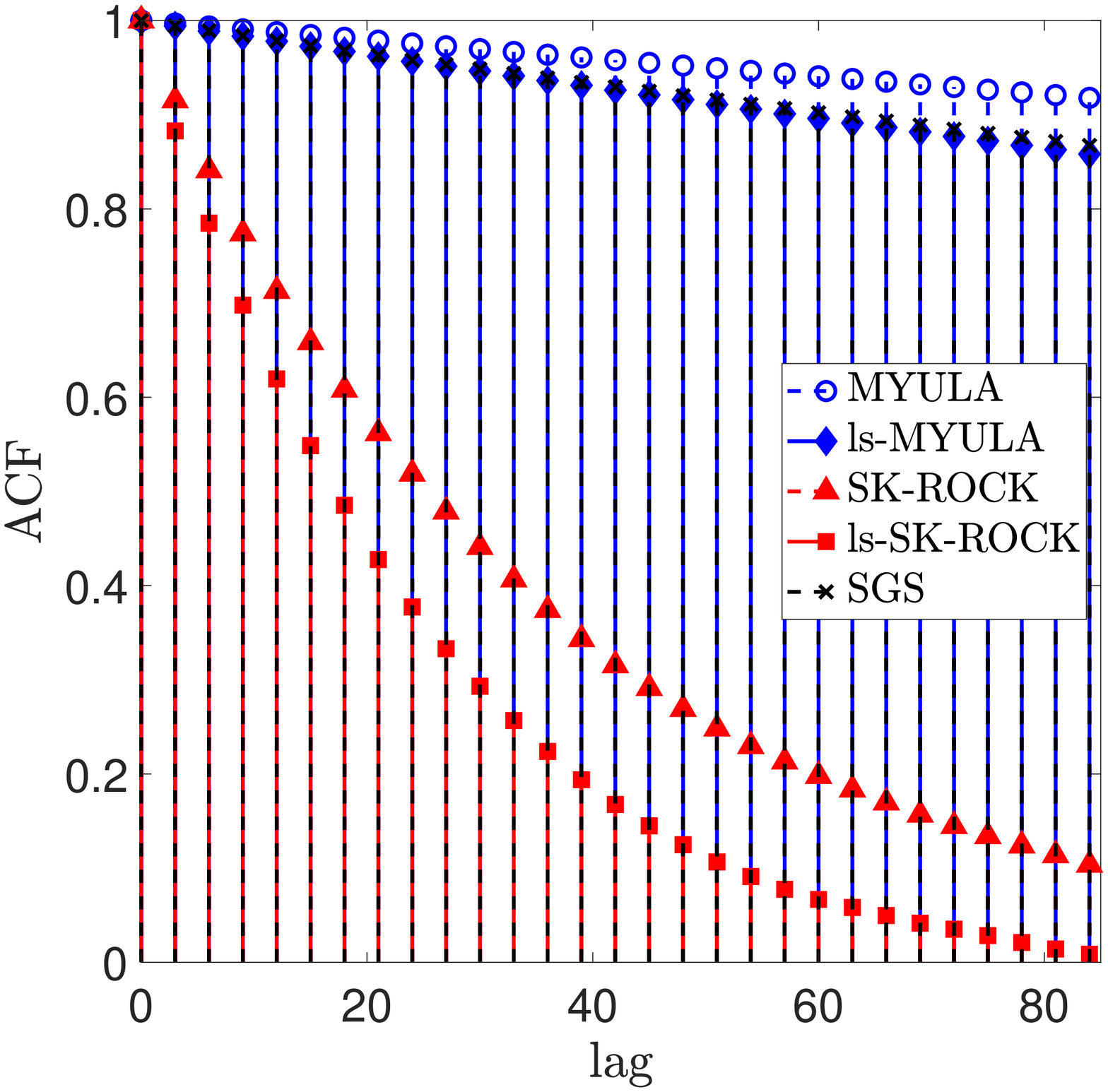}
	} 
	\caption{Image inpainting experiments: {\normalfont(a),(d)} Convergence to the typical set of the posterior distribution (\ref{eqn:inpainting_posterior_dist_nonAugm}) and (\ref{eqn:inpainting_posterior_dist_augm}) for the first $10^5$ MYULA, SGS and ls-MYULA samples, and the first $10^5 / s$ SK-ROCK and ls-SK-ROCK samples ($s=15$). {\normalfont(b),(e)} MSE between the mean of the algorithms and the true image, measured using $5 \times 10^6$ MYULA, SGS and ls-MYULA samples, and $5 \times 10^6 / s$ SK-ROCK and ls-SK-ROCK samples ($s=15$), in stationary regime. {\normalfont(c),(f)} Autocorrelation function for the slowest component of the samples.}
	\label{fig:results_inpainting}
\end{figure}

With the generated samples, we proceed to plot the results of these experiments in Figure \ref{fig:results_inpainting}. We first notice the acceleration one can get from ls-SK-ROCK from the convergence to equilibrium of the MCMC samples in the burn-in stage represented by the evolution of the scalar estimate $\log p (X_n|y,\theta)$. Then, we illustrate the evolution of the mean-squared error (MSE) between the mean of the samples and the true image $x$ in stationarity and, as can be seen, ls-SK-ROCK is computationally efficient in being the fastest method to reach the MMSE in all two experiments, even outperforming the MAP estimate in terms of accuracy in all two experiments; moreover, the improvement of ls-MYULA over SGS, in terms of accuracy is evident similar to our previous results.

\begin{figure}[htb]
	\centering
	\subfloat[MYULA]{
		\includegraphics[scale=.35]{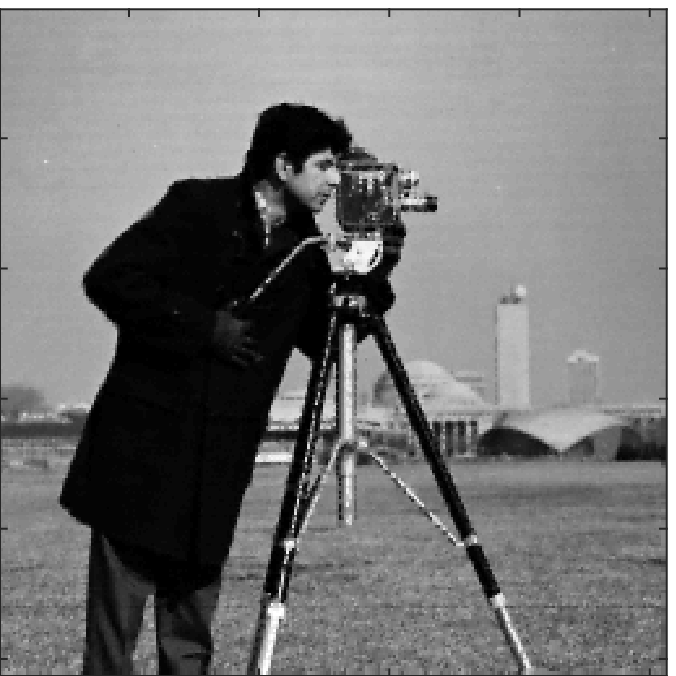}
	}
	\subfloat[SK-ROCK]{
		\includegraphics[scale=.35]{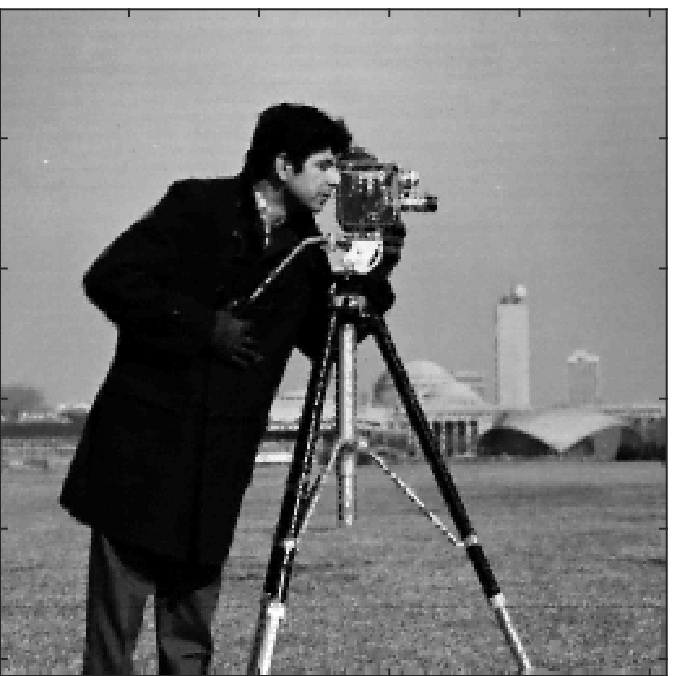}
	}
	\subfloat[SGS]{
		\includegraphics[scale=.35]{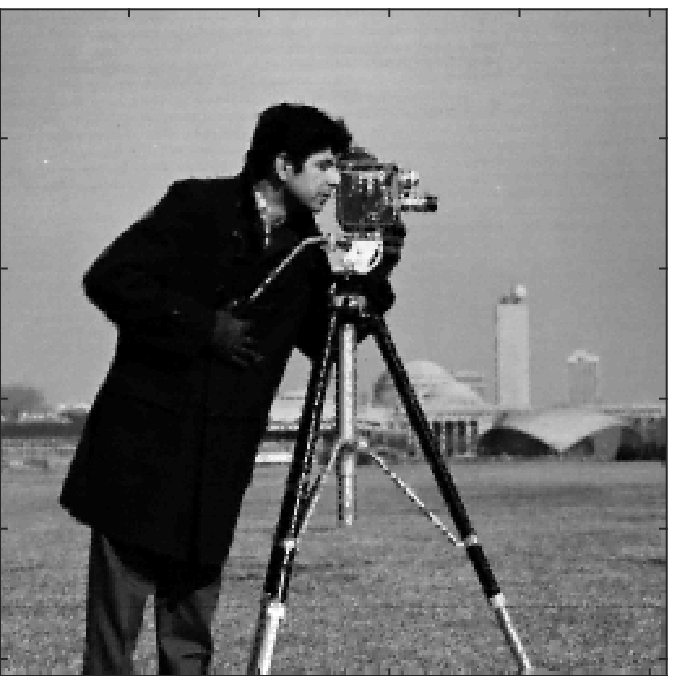}
	}
	\subfloat[ls-MYULA]{
		\includegraphics[scale=.35]{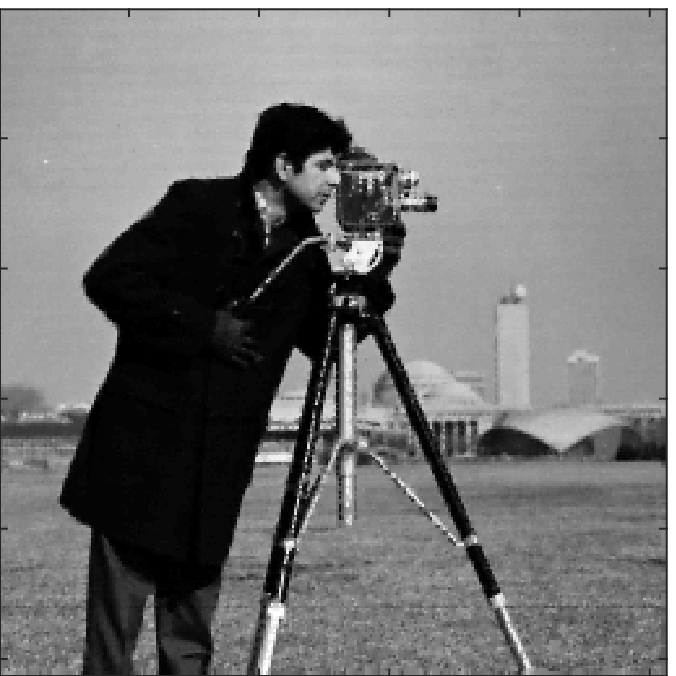}
	}
	\subfloat[ls-SK-ROCK]{
		\includegraphics[scale=.35]{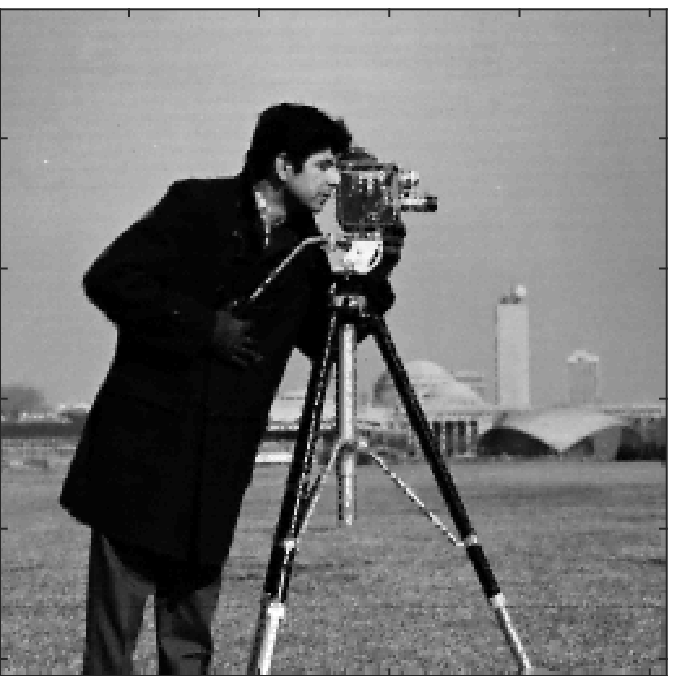}
	} \\
	\subfloat[MYULA]{
		\includegraphics[scale=.35]{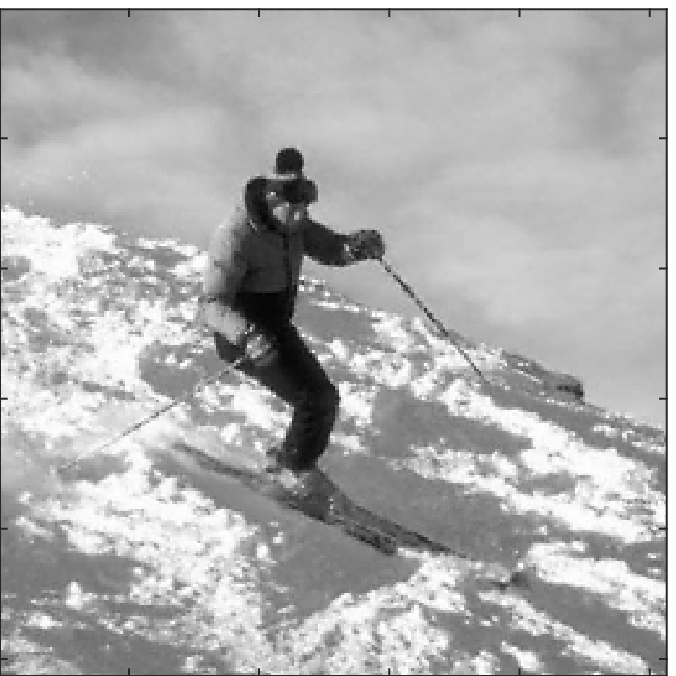}
	}
	\subfloat[SK-ROCK]{
		\includegraphics[scale=.35]{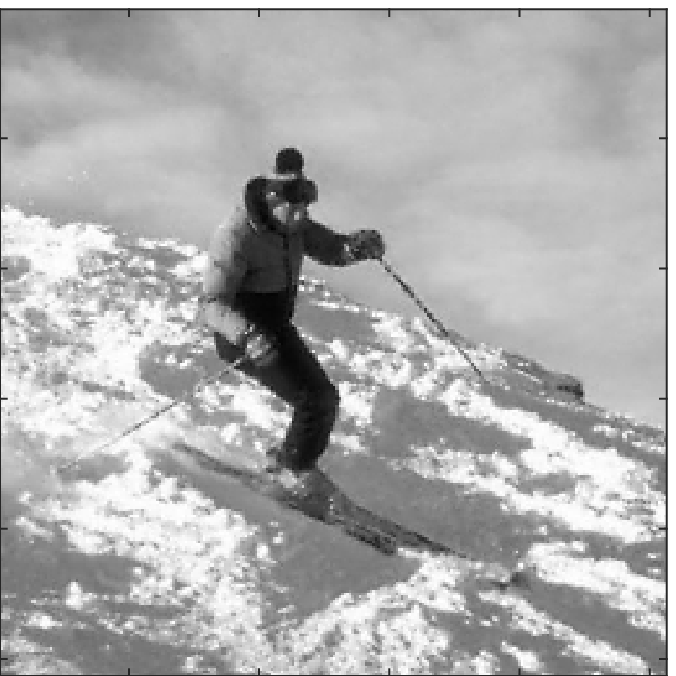}
	}
	\subfloat[SGS]{
		\includegraphics[scale=.35]{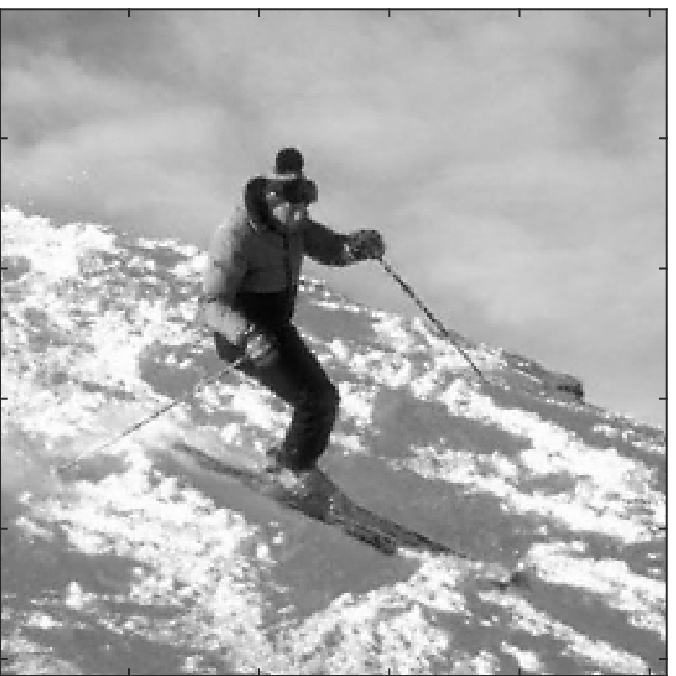}
	}
	\subfloat[ls-MYULA]{
		\includegraphics[scale=.35]{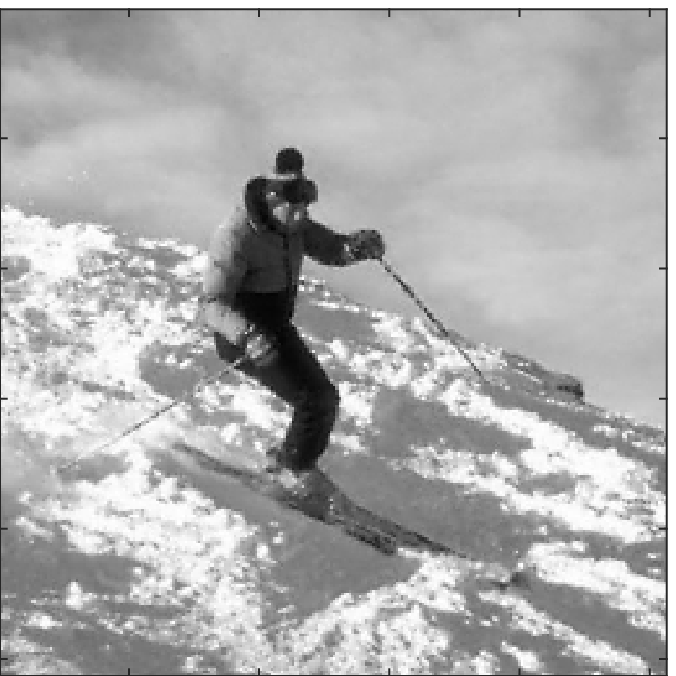}
	}
	\subfloat[ls-SK-ROCK]{
		\includegraphics[scale=.35]{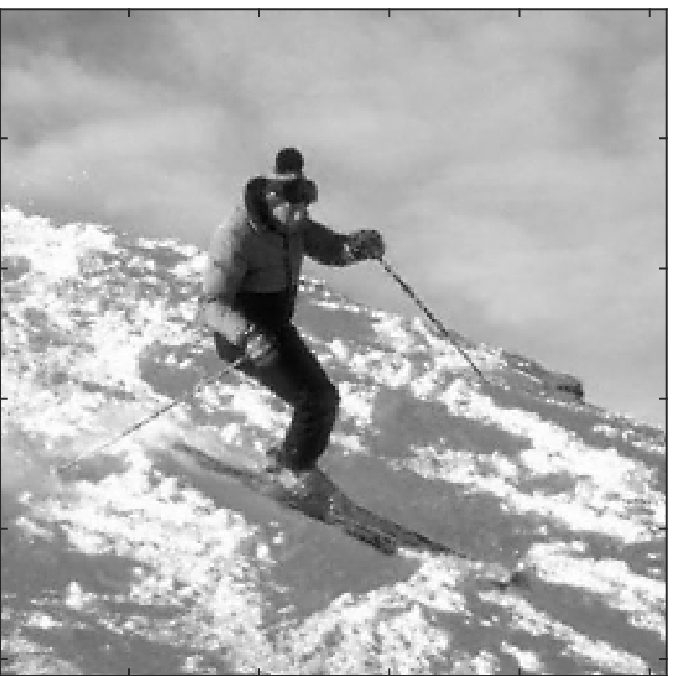}
	} 
	\caption{MMSE for the image inpainting experiment.}
	\label{fig:mmse_inpainting}
\end{figure}

We also plot the autocorrelation function of the pixel values for the slowest component in Figure \ref{fig:results_inpainting}(c),(f) and, as can be seen, the ACF of the ls-SK-ROCK samples decays faster than all the other MCMC methods. Again, a comparison of the autocorrelation function for the slowest mixing component shows the benefits of operating on the latent space (and the effect of $\rho^2 > 0$), as well as the efficiency cost that SGS incurs because of the use of an inexact gradient in this case. For completeness, we also illustrate in Figure \ref{fig:mmse_inpainting} the MMSE of all the MCMC methods for all two inpainting experiments and, as in previous numerical results, we can see in Figure \ref{fig:results_inpainting}(b),(e) that ls-SK-ROCK is the fastest method in compute this estimate.

\begin{table}[htb]
	{\footnotesize
		\caption{Image inpainting experiment: Summary of the values for the step-size $\delta$ for each of the MCMC methods applied to the two imaging experiments: cameraman and skier.}  \label{tab:stepsize_s_inpainting}
		\begin{center}
		\begin{tabular}{|l|c|c|c|} \hline
			\bf MCMC method & \bf Cameraman & \bf Skier \\ \hline
			\bf MYULA & $0.194$ & $0.110$ \\
			\bf SK-ROCK ($s=15$) & $78.698$ & $44.646$ \\
			\bf SGS & $0.283$ & $0.161$ \\
			\bf ls-MYULA & $0.283$ & $0.161$ \\
			\bf ls-SK-ROCK ($s=15$) & $114.717$ & $65.105$ \\ \hline
			\end{tabular}
		\end{center}
	}
\end{table}

\begin{figure}[htb]
	\centering
	\subfloat[MYULA]{
		\includegraphics[scale=.35]{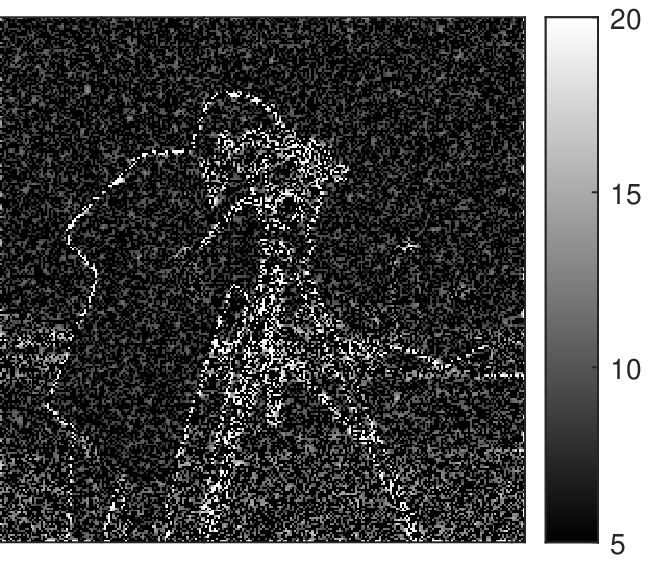}
	}
	\subfloat[SK-ROCK]{
		\includegraphics[scale=.35]{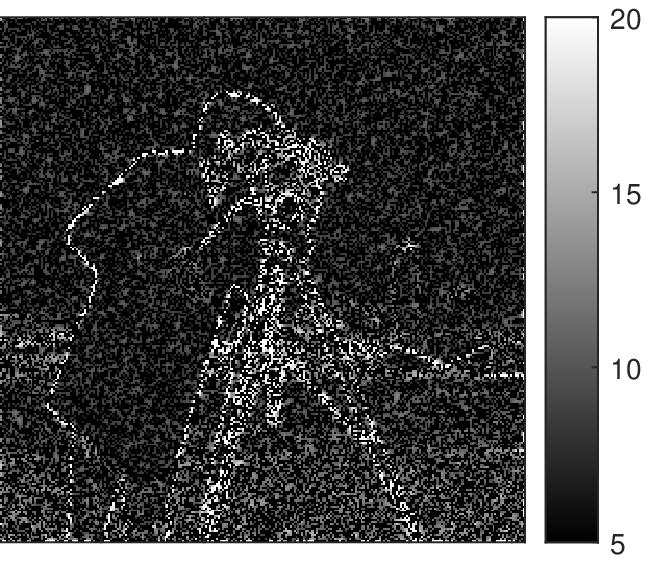}
	}
	\subfloat[SGS]{
		\includegraphics[scale=.35]{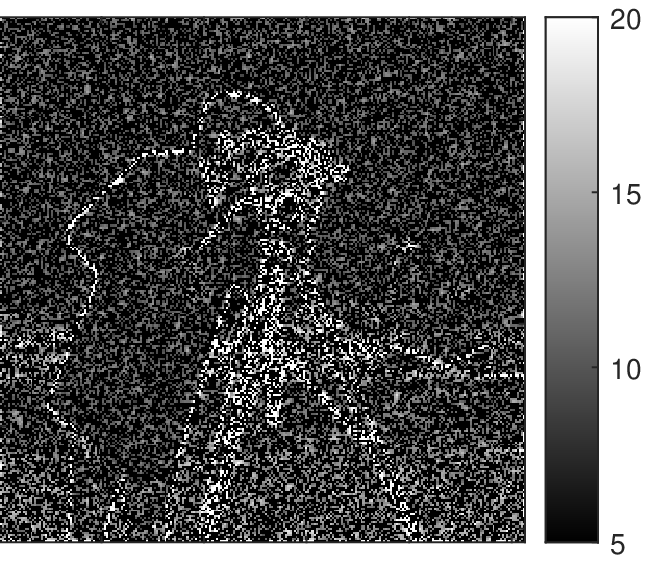}
	}
	\subfloat[ls-MYULA]{
		\includegraphics[scale=.35]{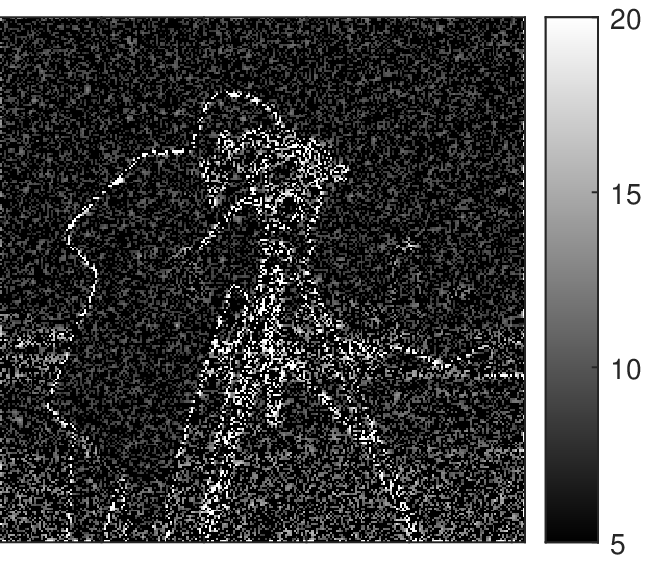}
	}
	\subfloat[ls-SK-ROCK]{
		\includegraphics[scale=.35]{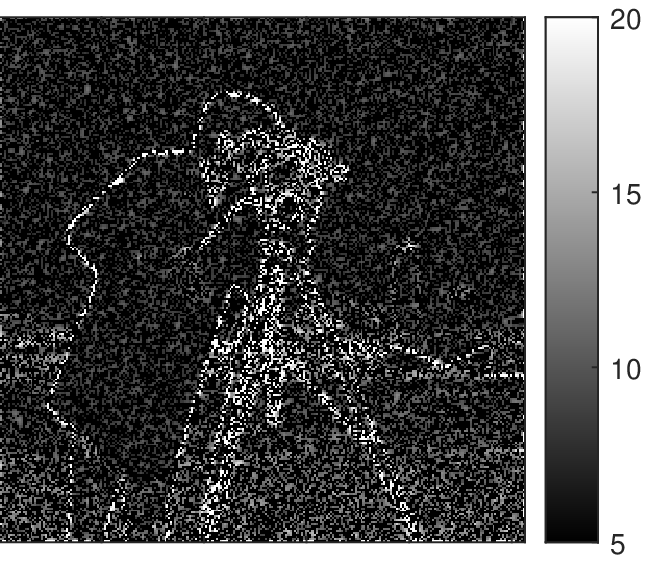}
	} \\
	\subfloat[MYULA]{
		\includegraphics[scale=.35]{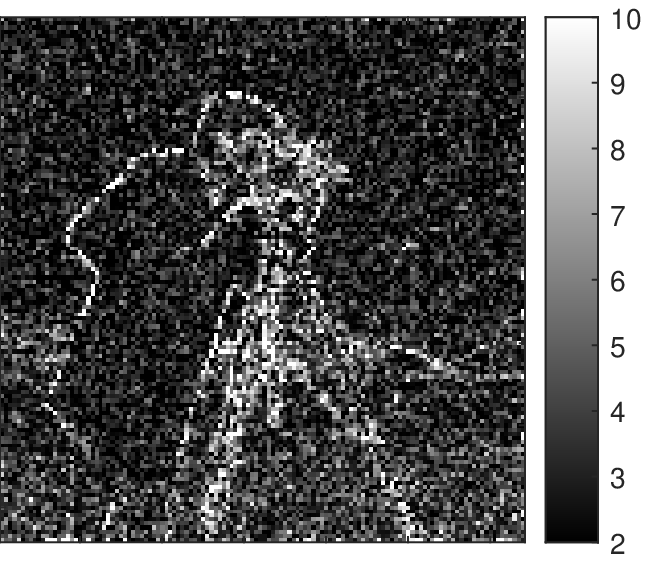}
	}
	\subfloat[SK-ROCK]{
		\includegraphics[scale=.35]{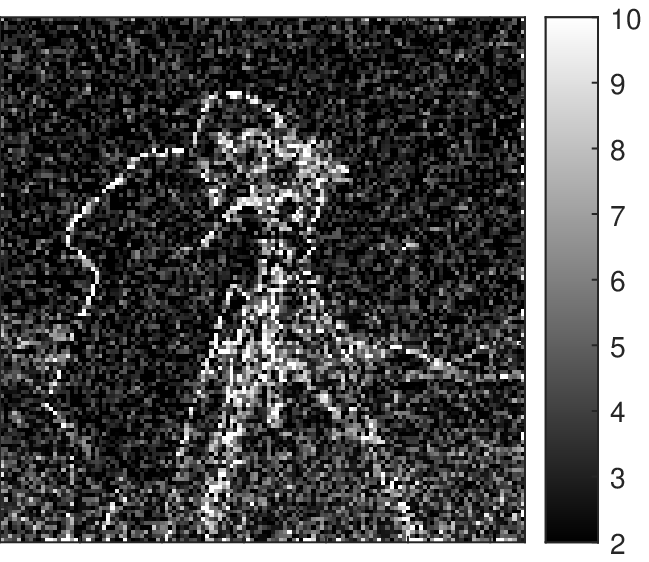}
	}
	\subfloat[SGS]{
		\includegraphics[scale=.35]{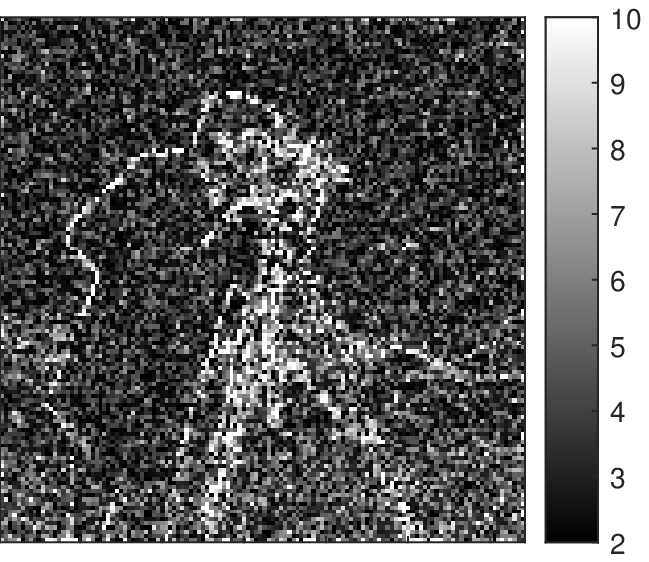}
	}
	\subfloat[ls-MYULA]{
		\includegraphics[scale=.35]{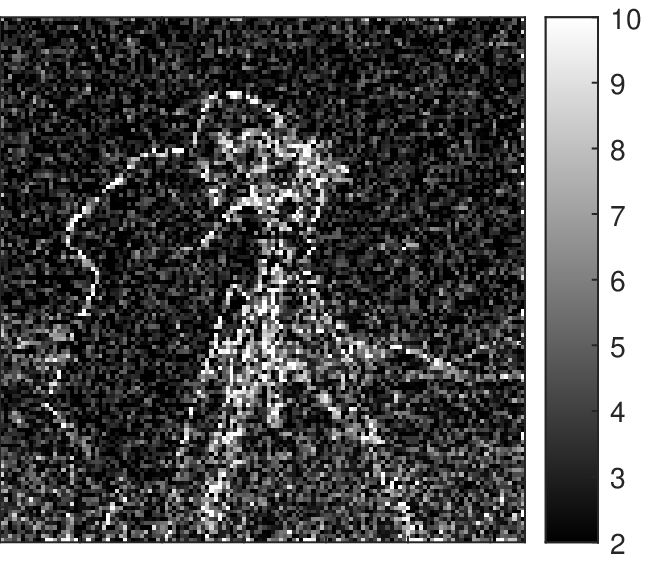}
	}
	\subfloat[ls-SK-ROCK]{
		\includegraphics[scale=.35]{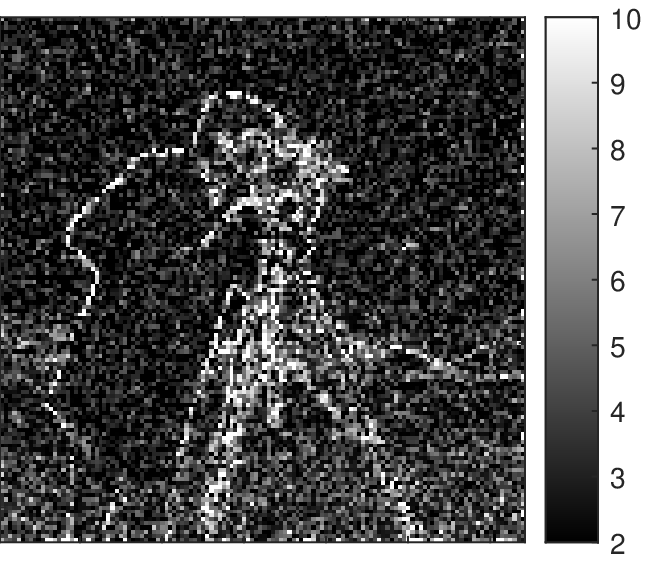}
	} \\
	\subfloat[MYULA]{
		\includegraphics[scale=.35]{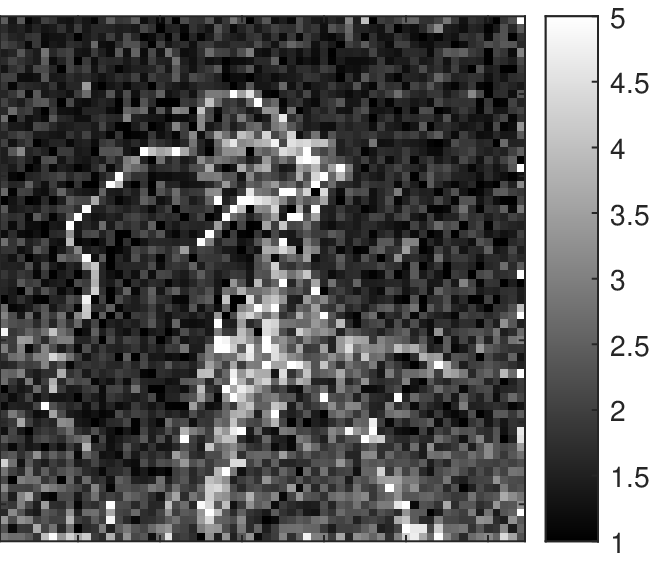}
	}
	\subfloat[SK-ROCK]{
		\includegraphics[scale=.35]{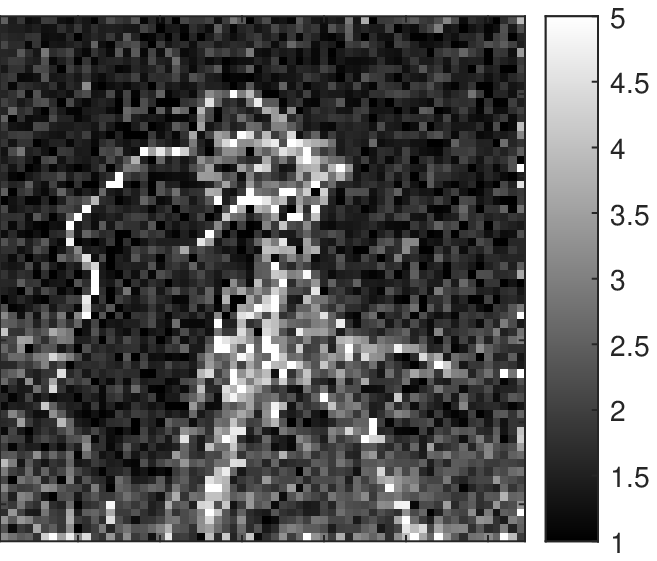}
	}
	\subfloat[SGS]{
		\includegraphics[scale=.35]{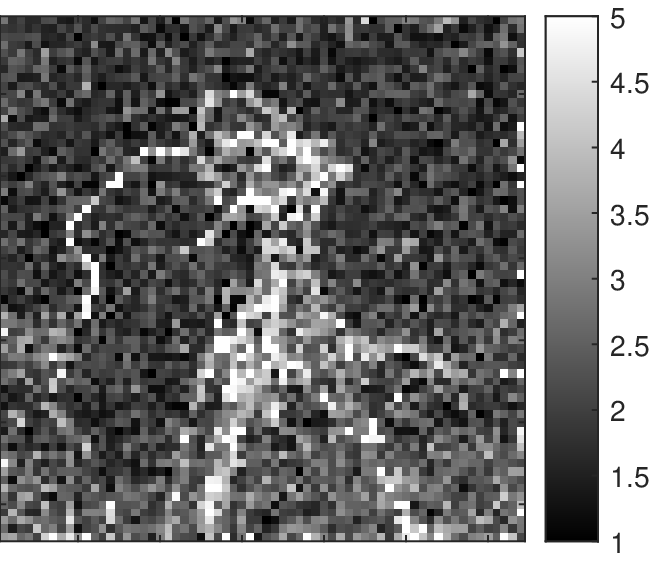}
	}
	\subfloat[ls-MYULA]{
		\includegraphics[scale=.35]{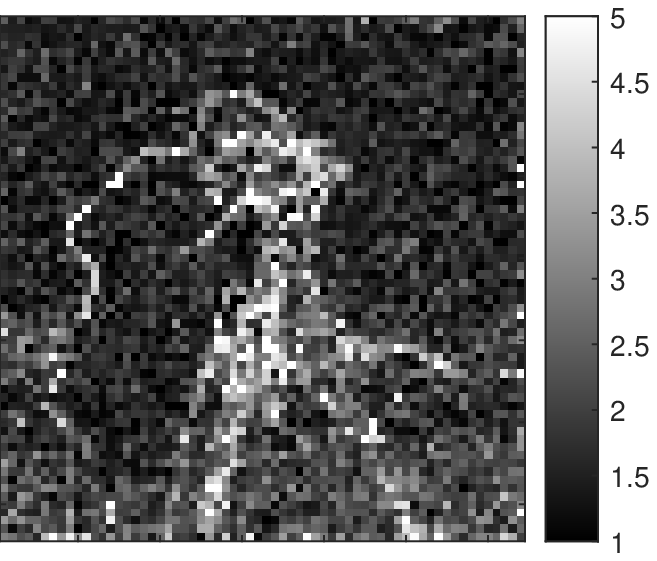}
	}
	\subfloat[ls-SK-ROCK]{
		\includegraphics[scale=.35]{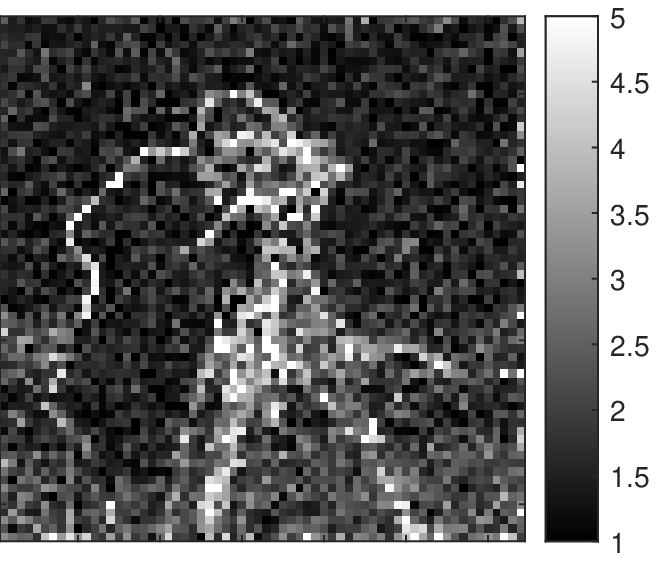}
	} \\
	\subfloat[MYULA]{
		\includegraphics[scale=.35]{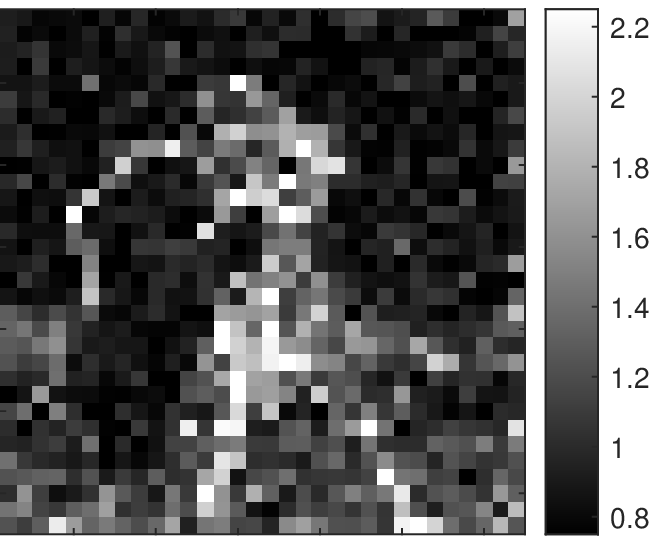}
	}
	\subfloat[SK-ROCK]{
		\includegraphics[scale=.35]{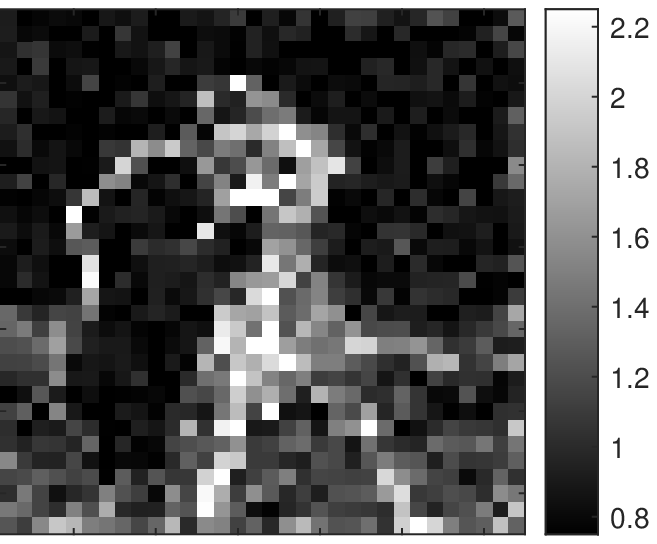}
	}
	\subfloat[SGS]{
		\includegraphics[scale=.35]{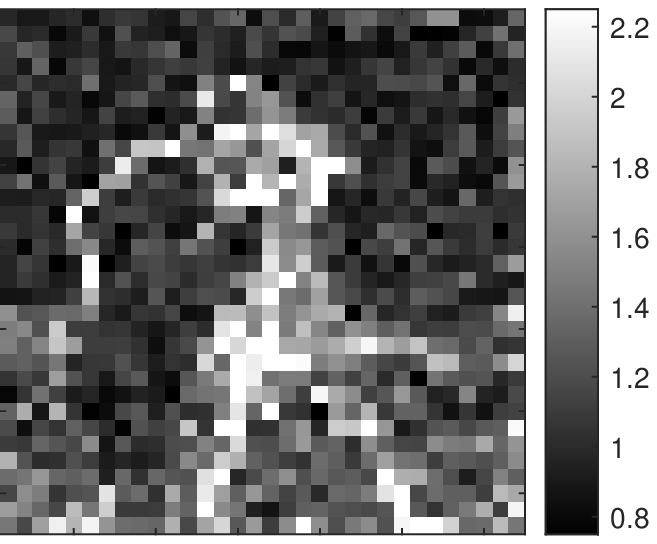}
	}
	\subfloat[ls-MYULA]{
		\includegraphics[scale=.35]{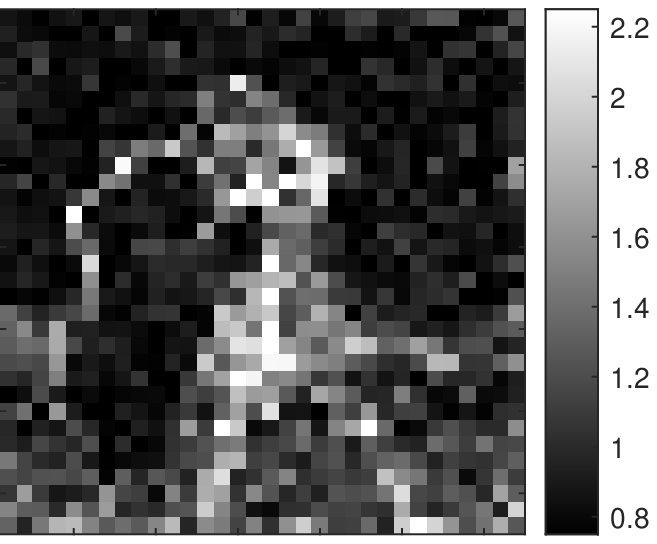}
	}
	\subfloat[ls-SK-ROCK]{
		\includegraphics[scale=.35]{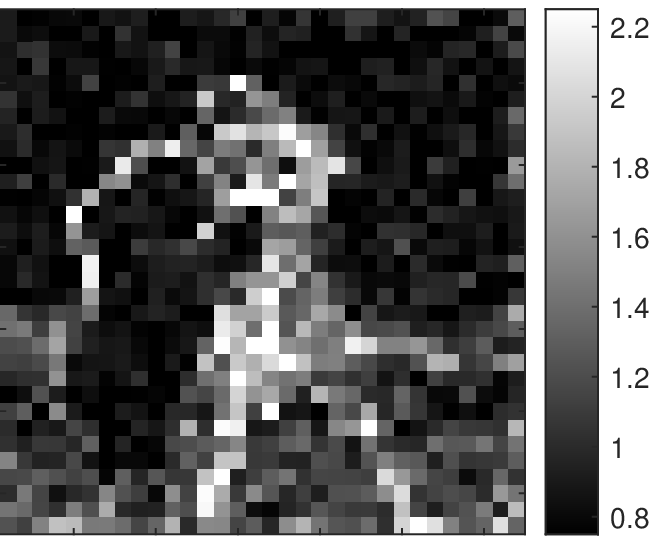}
	}
	\caption{Image inpainting experiments - cameraman: pixel-wise standard deviation computed using $10^4$ MYULA, SGS and ls-MYULA samples with a 1-in-15 thinning, and $10^4$ SK-ROCK and ls-SK-ROCK samples with $s=15$, using (a)-(e) the original sample size ($256 \times 256$) and with downsampling by a factor of (f)-(j) 2, (k)-(o) 4 and (p)-(t) 8.}
	\label{fig:std_dev_inpainting}
\end{figure}

Finally, Figure \ref{fig:std_dev_inpainting} presents uncertainty quantification plots by showing pixel-wise standard deviation estimates for the cameraman inpainting experiment. In this case, the uncertainty is concentrated on the unobserved pixels, which is expected given the nature of the inpainting problem. One can notice that ls-MYULA and ls-SK-ROCK deliver comparable estimates in less computational time than SGS, showing the good performance of these algorithms in sampling the augmented posterior distribution.

\begin{table}[htbp]
	{\footnotesize
		\caption{Image inpainting experiments: Effective sample sizes of the slowest component, after generating $15 \times 10^3$ samples using the five algorithms discussed in this work, and the speed increase (i.e., speed-up) achieved by the algorithms w.r.t. MYULA.}  \label{tab:ess_inpainting}
		\begin{center}
			
			\begin{tabular}{|c||c|c||c|c||c|c||c|c||c|c|} \hline
				\multirow{2}{*}{} & \multicolumn{2}{c||}{\bf MYULA} & \multicolumn{2}{c||}{\bf SK-ROCK} & \multicolumn{2}{c||}{\bf SGS} & \multicolumn{2}{c||}{\bf ls-MYULA} & \multicolumn{2}{c|}{\bf ls-SK-ROCK} \\ \cline{2-11}
				& ESS & Speed-up & ESS & Speed-up & ESS & Speed-up & ESS & Speed-up & ESS & Speed-up \\ \hline
				Cam. & $14$ & - & $281$ & $20.07$ & $14$ & $1$ & $21$ & $1.5$ & $448$ & $32$ \\
				Skier & $8$ & - & $212$ & $26.5$ & $11$ & $1.38$ & $15$ & $1.88$ & $320$ & $40$  \\ \hline
			\end{tabular}
		\end{center}
	}
\end{table}

\begin{table}[htb]
	{\footnotesize
		\caption{Summary of the execution times (in seconds) to produce one sample (i.e., after one iteration) on each of the MCMC algorithms implemented for each experiment.}  \label{tab:times_experiments}
		\begin{center}
			\begin{tabular}{|c|c|c|c|c|c|} \hline
   \bf Imaging & \bf MYULA & \bf SK-ROCK & \bf ls-MYULA & \bf ls-SK-ROCK & \bf SGS \\
   \bf Experiment &  & \bf (s=15) &  & \bf (s=15) &  \\ \hline
    deblurring & $3.8 \times 10^{-2}$ & $6.1 \times 10^{-1}$ & $4.3 \times 10^{-2}$ & $6.1 \times 10^{-1}$ & $4.7 \times 10^{-2}$ \\ 
    inpainting & $4.1 \times 10^{-2}$ & $5.8 \times 10^{-1}$ & $3.8 \times 10^{-2}$ & $5.6 \times 10^{-1}$ & $4.7 \times 10^{-2}$ \\
		\hline
  			\end{tabular}
		\end{center}
	}
\end{table}

\subsection{Image deblurring with a total generalized variation prior}
We conclude this section with an experiment related to image deblurring with a total generalised variation prior. The experiment setup is akin to Section \ref{subsec:imageDeconvolution}, except that the prior is now given by
\begin{equation}
\label{eqn:tgv_prior}
p (x|\theta^{(1)},\theta^{(2)}) \propto \exp \left[ -\mathrm{TGV}^2_{\theta^{(1)},\theta^{(2)}} (x) - \varepsilon \|x\|_2^2  \right],
\end{equation}
where $\varepsilon > 0$ and where $\mathrm{TGV}^2_{\theta^{(1)},\theta^{(2)}} (x)$ is the so-called total generalized variation (TGV) regulariser \cite{bredies2010kunisch,Chambolle1997_2}, defined for every $\theta^{(1)}, \theta^{(2)} \in [0, +\infty)^2$, and $x \in \mathbb{R}^d$ by
\begin{equation*}
    \mathrm{TGV}^2_{\theta^{(1)},\theta^{(2)}} (x) = \min_{u \in \mathbb{R}^{2d}} \{ \theta^{(1)} \| u \|_{1,2} + \theta^{(2)} \| J(\Delta x -u ) \|_{1,\mathrm{Frob.}} \},
\end{equation*}
where $\Delta = (\Delta^v, \Delta^h)$ computes the first-order vertical and horizontal pixel differences, while the second-order information of the image-gradient vector field is computed by the Jacobian matrix $J$. The incorporation of second-order information removes the characteristic stair-casing artefacts commonly associated with (non-generalised) TV regularisation \cite{bredies2010kunisch}.

Image deblurring with a TGV prior is challenging because the results are highly sensitive to the choice of $\theta^{(1)}$ and $\theta^{(2)}$. However, these parameters are difficult to set a priori. Their direct estimation from $y$ by maximum marginal likelihood estimation is also difficult because the TGV prior does not belong to the exponential family. \cite[Section 4.4]{vidal2019maximum} proposes to address this difficulty by using a SAPG scheme to compute a pseudo-maximum marginal likelihood estimator for $\theta^{(1)}$ and $\theta^{(2)}$, which we also adopt in this paper (we refer the reader to \cite[Section 4.4]{vidal2019maximum} for more details). Also note that the term $\varepsilon \|x\|_2^2$ is required so that $p (x|\theta^{(1)},\theta^{(2)})$ defines a proper prior, in practice $\varepsilon$ is very small (we use $\varepsilon = 10^{-10}$).

\begin{figure}[htb]
        \centering
        \subfloat[\textit{skier}, true image $x$]{
		\includegraphics[scale=.47]{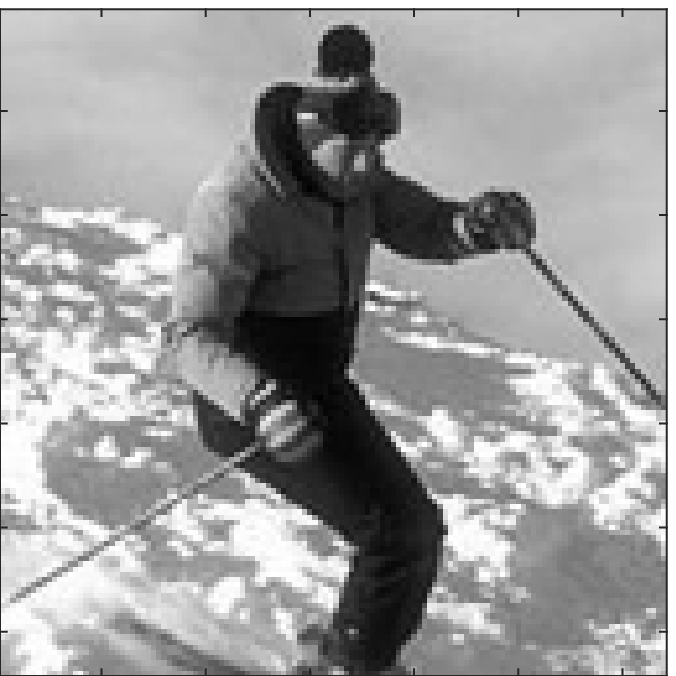}
	}
	\subfloat[\scriptsize{\textit{skier}: observation $y$}]{
		\includegraphics[scale=.47]{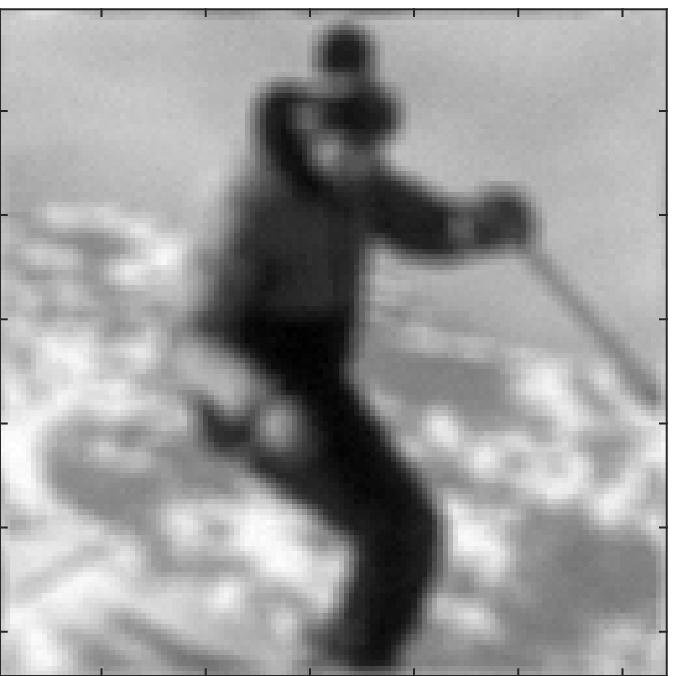}
	}
	\caption{Image deblurring experiment with TGV prior: Test image $x$ (patch from Image \ref{fig:test_images_deconvolution}{\normalfont(b)} of size $128\times128$) and its corresponding noisy and blurred observation $y$.}
	\label{fig:true_obs_skier_deb_tgv}
\end{figure}

In a manner akin to Section \ref{subsec:imageDeconvolution}, we consider a deblurring problem with the \texttt{skier} test images of size $d=128 \times 128$ pixels, which is a patch of Image \ref{fig:test_images_deconvolution}{\normalfont(b)} (a reduced size image is considered in this experiment because the evaluation of the TGV proximal operator is highly computationally expensive). The observation $y$ is generated by applying a uniform blur operator $H$ of size $5 \times 5$ to the true image $x \in \mathbb{R}^d$, followed by additive Gaussian noise with a sigma-to-noise level of $30$dB. This experiment considers the following posterior distributions
\begin{gather} \label{eqn:deb_w_tgv_posterior_dist_nonAugm}
    p (x|y,\theta^{(1)},\theta^{(2)}) \propto \exp \left[ -\| y - Hx \|^2 / 2\sigma^2 - p (x|\theta^{(1)},\theta^{(2)}) \right] \\
\label{eqn:deb_w_tgv_posterior_dist_augm}
    p (x,z|y,\theta^{(1)},\theta^{(2)},\rho^2) \propto \exp \left[ -\| y - Hx \|^2 / 2\sigma^2 - p (z|\theta^{(1)},\theta^{(2)}) - \Vert x - z \Vert^2 / 2 \rho^2 \right] ,
\end{gather}
where $p(x|\theta^{(1)},\theta^{(2)})$ is defined in \eqref{eqn:tgv_prior}, and $p (z|\theta^{(1)},\theta^{(2)})$ is the same TGV prior applied to the latent variable $z$ instead of $x$. Figure \ref{fig:true_obs_skier_deb_tgv}(a) shows the test images for this experiment and Figure \ref{fig:true_obs_skier_deb_tgv}(b) shows the corresponding observations $y$.

\begin{figure}[htb]
	\centering
	\subfloat[\scriptsize{skier: $\log p(X_n|y,\theta_1,\theta_2)$}]{
		\includegraphics[scale=.16]{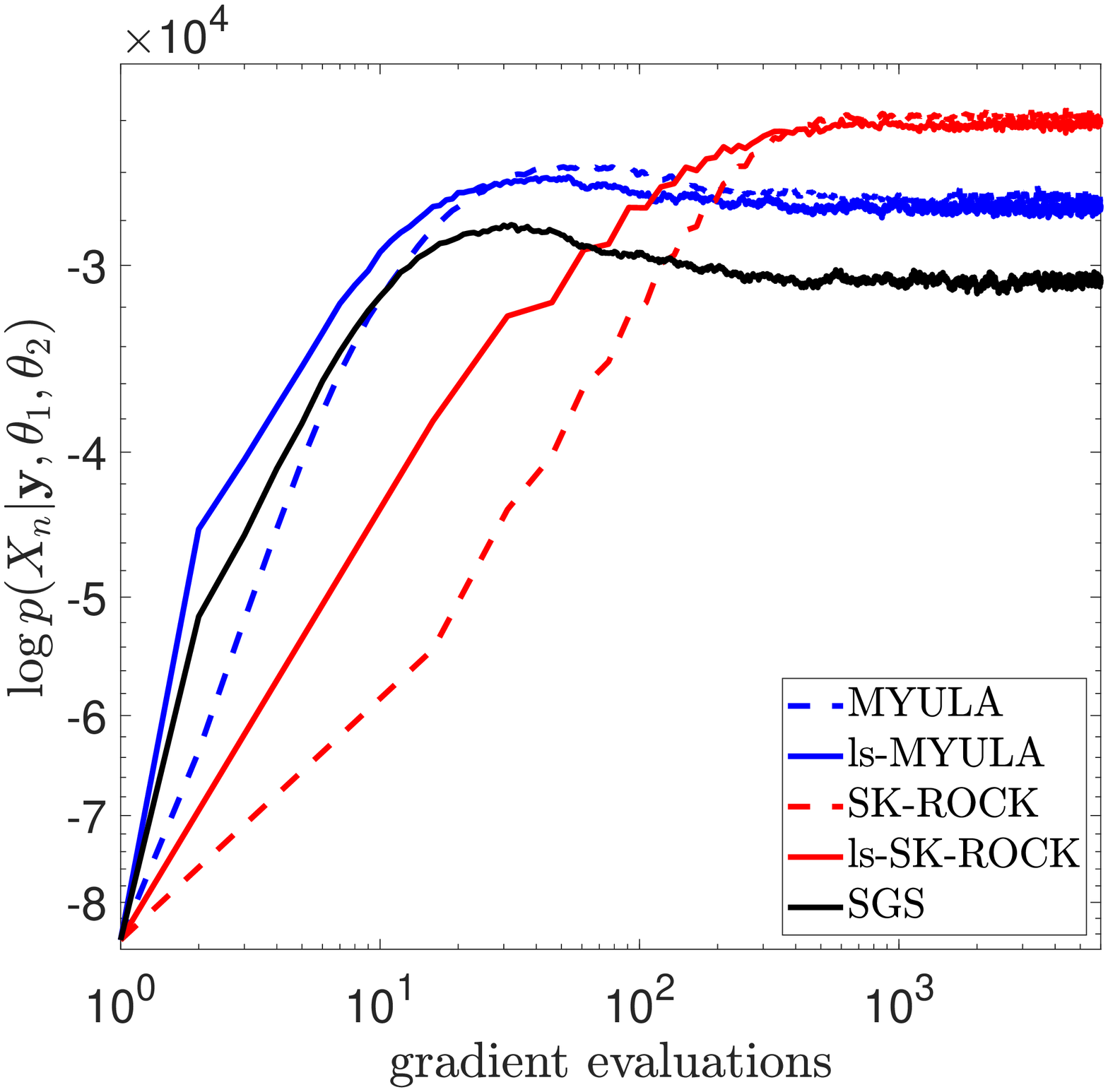}
	}
	\subfloat[\scriptsize{skier: MSE}]{
		\includegraphics[scale=.16]{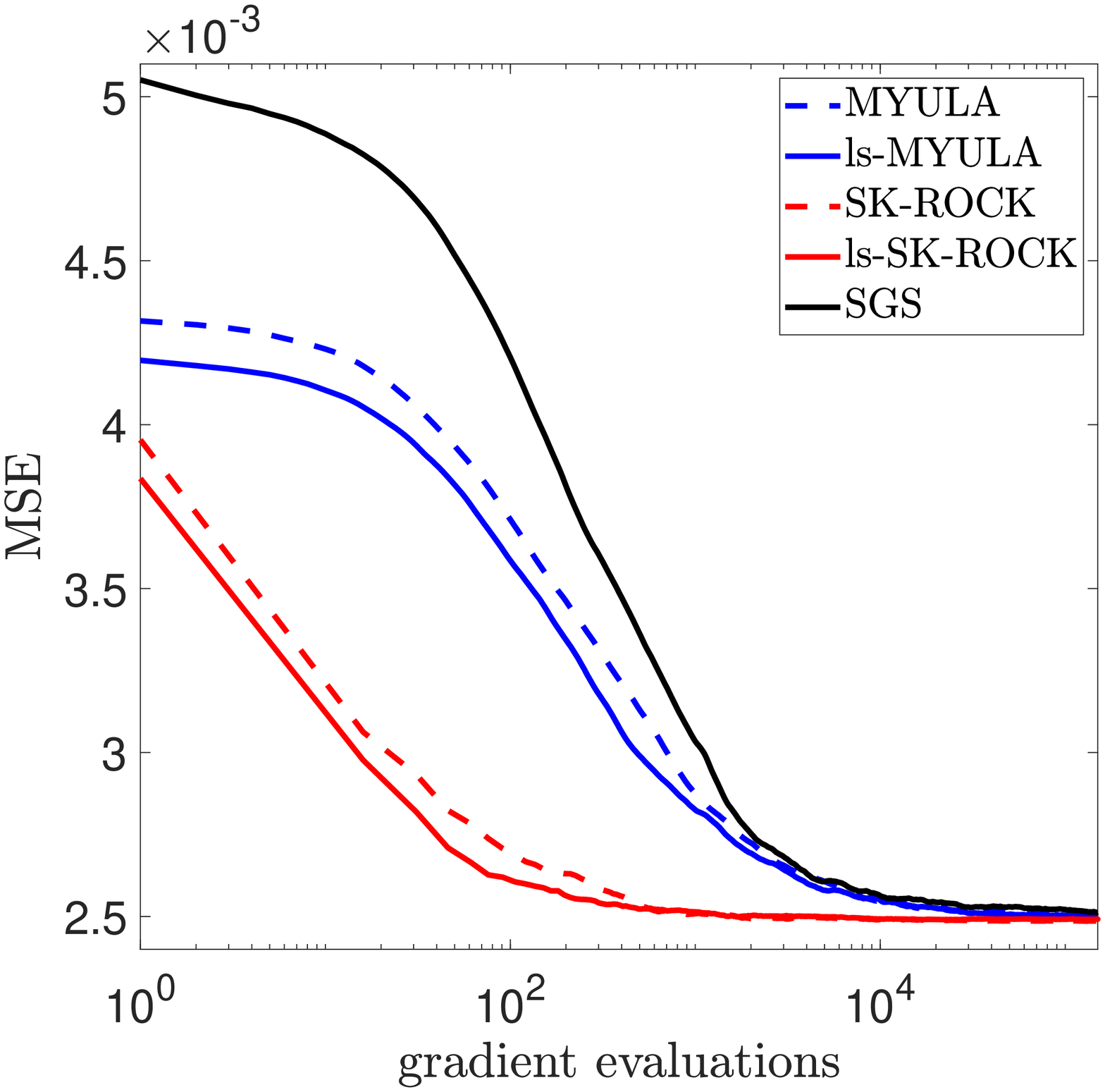}
	}
	\subfloat[\scriptsize{skier: ACF}]{
		\includegraphics[scale=.16]{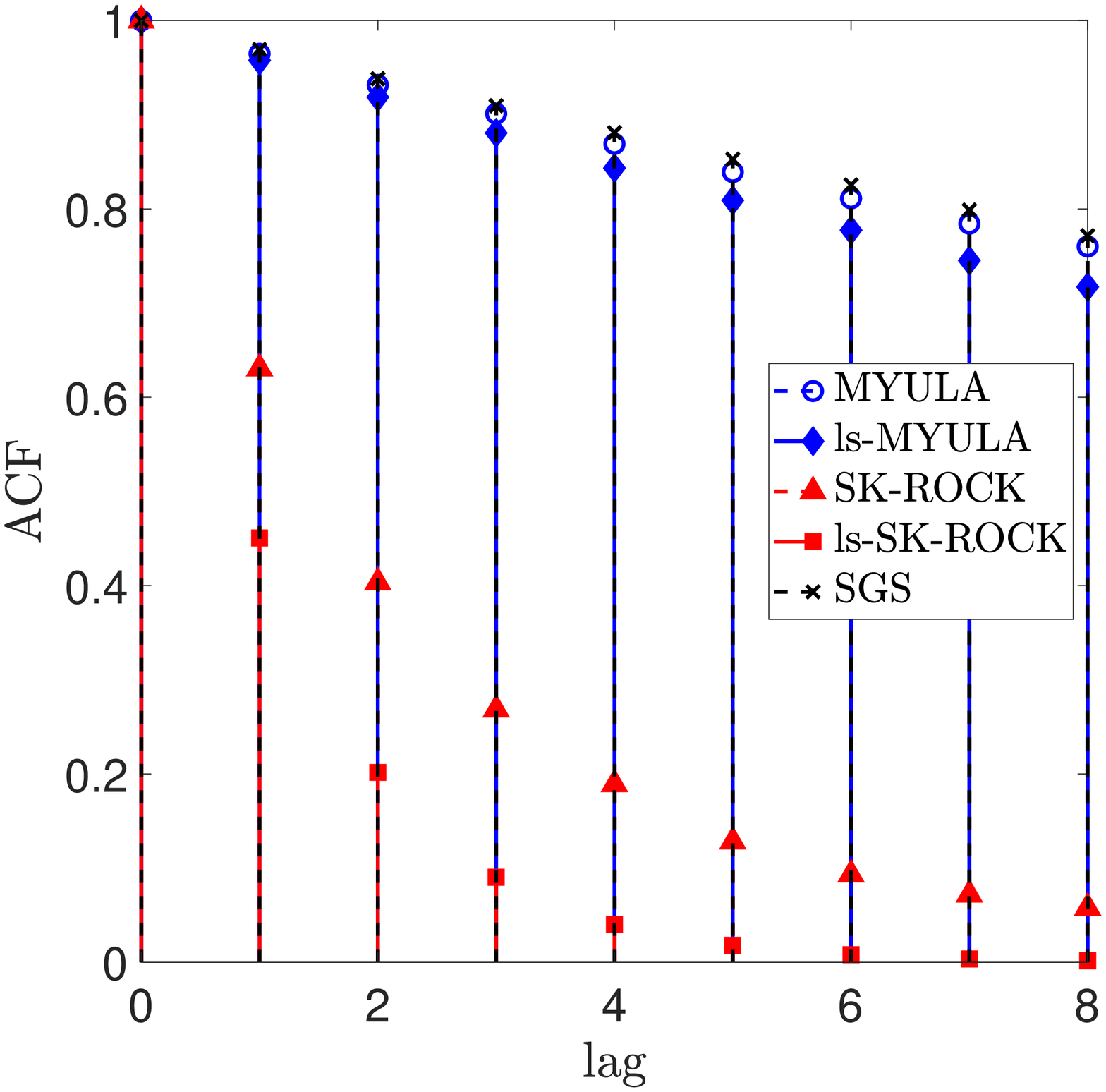}
	}
	\caption{Image deblurring experiment with TGV prior: {\normalfont(a)} Convergence to the typical set of the posterior distribution (\ref{eqn:deb_w_tgv_posterior_dist_nonAugm}) and (\ref{eqn:deb_w_tgv_posterior_dist_augm}) for the first $5 \times 10^3$ MYULA, SGS and ls-MYULA samples, and the first $5 \times 10^3 / s$ SK-ROCK and ls-SK-ROCK samples ($s=15$). {\normalfont(b)} MSE between the mean of the algorithms and the true image, measured using $1.5 \times 10^5$ MYULA, SGS and ls-MYULA samples, and $1.5 \times 10^5 / s$ SK-ROCK and ls-SK-ROCK samples ($s=15$), in stationary regime. {\normalfont(c)} Autocorrelation function for the values of the slowest component of the samples.}
	\label{fig:results_skier_deb_w_tgv}
\end{figure}

We begin estimating optimal values for $\theta^{(1)},\theta^{(2)}$ and $\rho^2$ for the given models implementing a variant of the Algorithm \ref{alg:SAPG_augm_model} for inhomogeneous regularizers (See \cite[Section 3.2.3]{vidal2019maximum} for details), setting $\gamma_i^{(1)} = 100 \times i^{-0.8} / d$, and $\gamma_i^{(2)} = \gamma_i^{\prime} = i^{-0.8} / d$, $\theta_0^{(1)} = \theta_0^{(2)} = 5$, $\rho^2_0 = L_f^{-1} = \sigma^2$ and $X_0 = Z_0 = H^{\intercal} y$. The corresponding results for the estimated parameters are given in Table \ref{tab:parameterValues_skier_deb_tgv}, together with the Lipschitz constants $L$ and $L_a$ required to sample (\ref{eqn:deb_w_tgv_posterior_dist_nonAugm}) and (\ref{eqn:deb_w_tgv_posterior_dist_augm}) respectively. We then generate  $1.5 \times 10^5$ samples using MYULA and $1.5 \times 10^5 / s$ samples using SK-ROCK (with $s=15$ and $\delta = \delta_s^{\max} / 2$) from (\ref{eqn:deb_w_tgv_posterior_dist_nonAugm}), and $1.5 \times 10^5$ samples using SGS and ls-MYULA and $1.5 \times 10^5 / s$ samples using ls-SK-ROCK (with $s=15$ and $\delta = \delta_s^{\max} / 2$) from (\ref{eqn:deb_w_tgv_posterior_dist_augm}). The results of these experiments are plotted in Figure \ref{fig:results_skier_deb_w_tgv}. In particular, we  note from the evolution of the MSE (when the chains have reached the typical set of the target distributions) that ls-MYULA and ls-SK-ROCK outperform SGS, as we are using an exact MYULA implementation rather than a noisy one, as shown in Section \ref{sec:SGS_noisy_MYULA_new_MCMC}. We have also reported the step-size used by each method in Table \ref{tab:stepsize_s_skier_deb_tgv}.

\begin{table}[htb]
{\footnotesize
  \caption{Values for $\theta^{(1)},\theta^{(2)}$ and $\rho^2$ estimated using Algorithm \ref{alg:SAPG_augm_model} for (\ref{eqn:deb_w_tgv_posterior_dist_nonAugm}) and (\ref{eqn:deb_w_tgv_posterior_dist_augm}) in the image deblurring experiment with TGV, together with the corresponding Lipschitz constants $L$ and $L_a$.} \label{tab:parameterValues_skier_deb_tgv}
\begin{center}
  \begin{tabular}{|c|c|c|c|c|c|} \hline
   \bf $\theta^{(1)}$ & \bf $\theta^{(2)}$ & \bf $\rho^2$ & \bf $\sigma^2$ & \bf $L = 1/\lambda + 1/\sigma^2$ & \bf $L_{a} = 1/\lambda + (\sigma^2 + \rho^2)^{-1}$ \\ \hline
    $4.46$ & $7.38$ & $7.15 \times 10^{-5}$ & $5.04 \times 10^{-5}$ & $3.97 \times 10^4$ & $2.81 \times 10^4$ \\ 
		\hline
  \end{tabular}
\end{center}
}
\end{table}

\begin{table}[htb]
	{\footnotesize
		\caption{Image deblurring experiment with TGV: Summary of the values for the step-size $\delta$ for each of the MCMC methods applied to this experiment.}  \label{tab:stepsize_s_skier_deb_tgv}
		\begin{center}
			\begin{tabular}{|l|c|c|c|} \hline
			\bf MCMC method & \bf $\delta$ \\ \hline
			\bf MYULA & $2.51 \times 10^{-5}$ \\
			\bf SK-ROCK ($s=15$) & $5.10 \times 10^{-3}$ \\
			\bf SGS & $3.56 \times 10^{-5}$ \\
			\bf ls-MYULA & $3.56 \times 10^{-5}$ \\
			\bf ls-SK-ROCK ($s=15$) & $7.21 \times 10^{-3}$ \\ \hline
			\end{tabular}
		\end{center}
	}
\end{table}

\begin{figure}[htb]
	\centering
	\subfloat[MYULA]{
		\includegraphics[scale=.35]{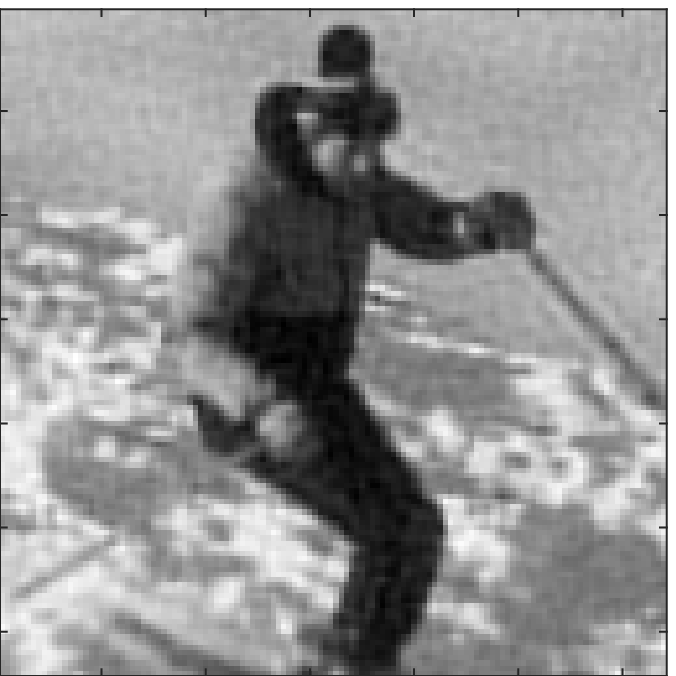}
	}
	\subfloat[SK-ROCK]{
		\includegraphics[scale=.35]{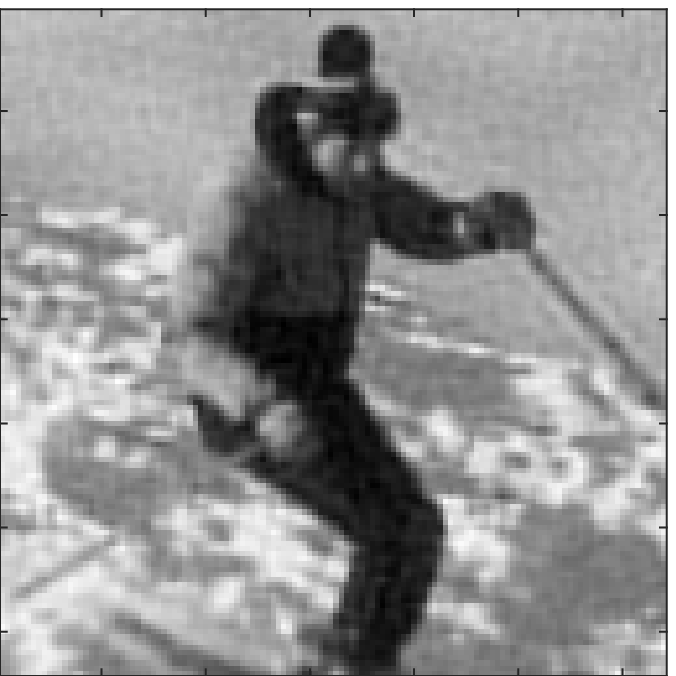}
	}
	\subfloat[SGS]{
		\includegraphics[scale=.35]{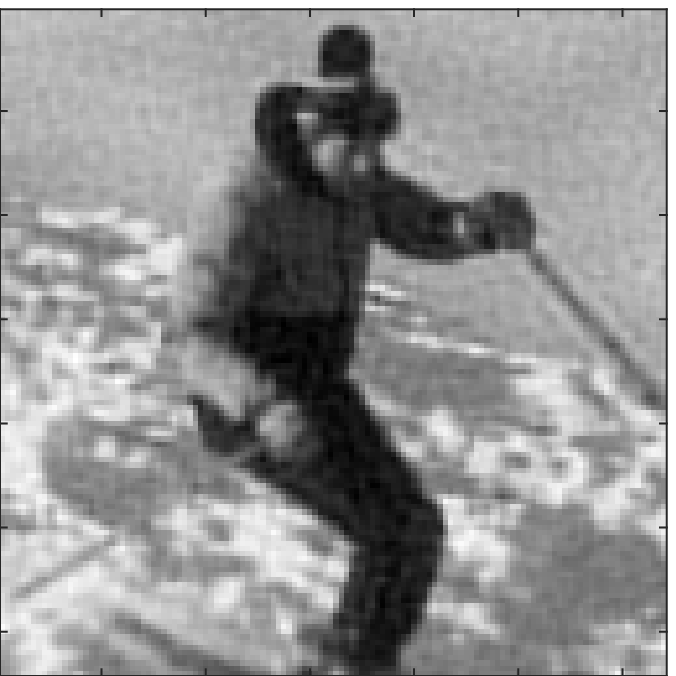}
	}
	\subfloat[ls-MYULA]{
		\includegraphics[scale=.35]{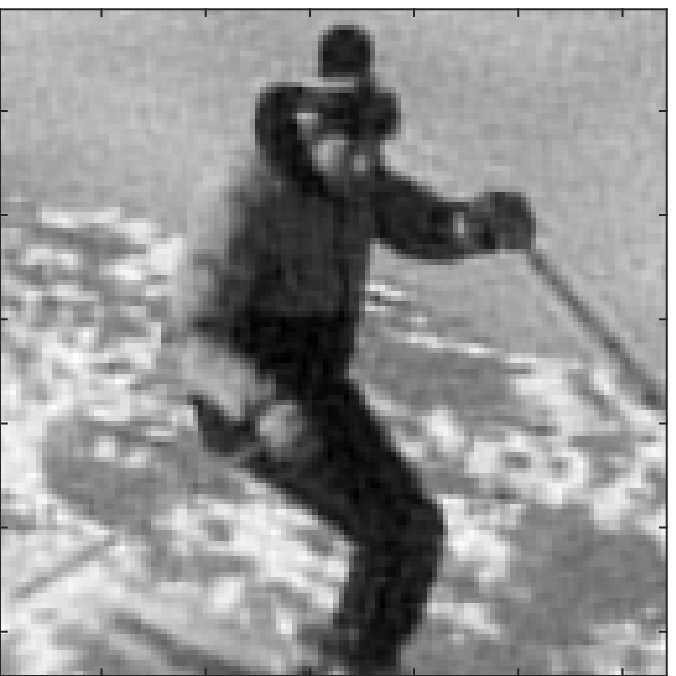}
	}
	\subfloat[ls-SK-ROCK]{
		\includegraphics[scale=.35]{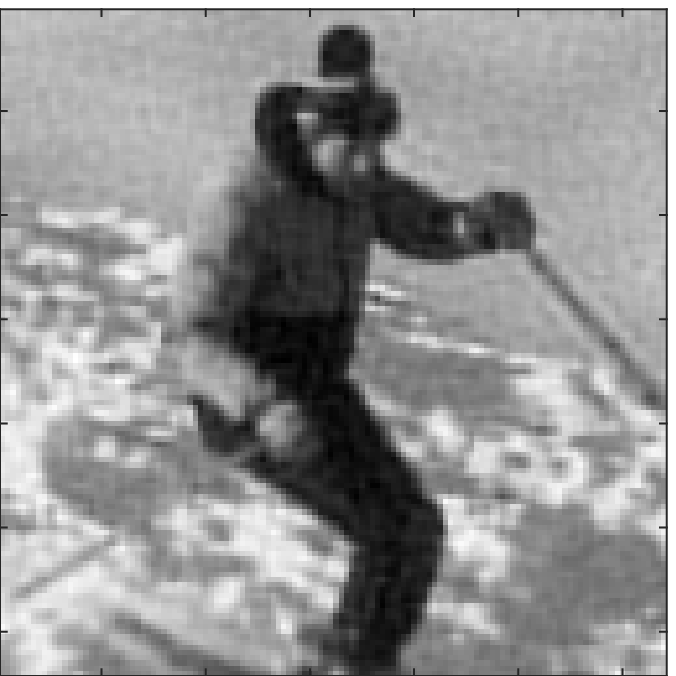}
	}
	\caption{MMSE for the image deblurring experiment with TGV prior.}
	\label{fig:mmse_skier_deb_w_tgv}
\end{figure}

We also plot the autocorrelation function of the slowest component from the chains of the MCMC algorithms, this is shown in Figure \ref{fig:results_skier_deb_w_tgv}(c) and, as can be seen, ls-SK-ROCK presents the fastest decay. In addition, we also illustrated in Figure \ref{fig:mmse_skier_deb_w_tgv} the minimum mean-square estimator (MMSE) of all the MCMC methods for this experiment.

As can be seen, ls-MYULA and, in particular, ls-SK-ROCK outperform SGS in terms of delivering comparable estimates in less computational time, showing the benefit of using these algorithms to sample in a more efficient way the augmented posterior distribution.

\section{Conclusions}
\label{sec:conclusions}
We presented a strategy to combine MYULA or proximal SK-ROCK with augmentation and relaxation in the manner of SGS. This was achieved by first establishing that SGS is equivalent to a noisy ULA scheme applied to the marginal distribution of the latent variable $z$ in an augmented Bayesian model $x,z|y,\theta,\rho^2$. This then naturally led to two new samplers that apply MYULA and SK-ROCK to the latent marginal distribution $z|y,\theta,\rho^2$. Probabilities and expectations w.r.t. the primal marginal $x|y,\theta,\rho^2$ are then straightforwardly computed by using a Rao-Blackwellised Monte Carlo estimator. Moreover, we also observed empirically that there is a range of values for $\rho^2$ for which convergence speed and model quality improve (in the sense of the model evidence). Increasing $\rho^2$ beyond this range leads to improvements in convergence speed at the expense of significant estimation bias. We therefore proposed to adopt an empirical Bayesian approach and set $\rho^2$, together with the regularisation parameter $\theta$, by maximum marginal likelihood estimation from $y$. This was achieved by using an SAPG scheme that convergences in very few iterations. We illustrated the benefits from adopting the proposed methodology with two experiments, image deblurring and image inpainting. The results showed that the new proximal SK-ROCK algorithm that benefits from augmentation and relaxation outperforms the other methods from the state of the art in terms of computational efficiency. Future work will focus on extending the proposed approach to plug-and-play priors encoded by neural network denoisers \cite{doi:10.1137/21M1406349}, and on establishing non-asymptotic convergence results for the methods. Another perspective for future work is to compare the proposed MCMC methods with deterministic Bayesian computation strategies based on approximate message passing and expectation propagation algorithms \cite{7314898}, and to explore ways in which the methods presented in this paper could be used to compute message passing or expectation propagation update steps within these schemes.


\printbibliography

\end{document}